\documentclass[preprint,noshowkeys,amsmath,amssymb,superscriptaddress,nofootinbib,longbibliography]{revtex4-1}

\usepackage{bm}
\usepackage{slashed}
\usepackage{ulem}

\usepackage[pdftex]{graphicx,color}
\usepackage{epstopdf}
\usepackage[bookmarks=true,bookmarksnumbered=true,bookmarkstype=toc, colorlinks=true, citecolor=blue, linkcolor=blue]{hyperref} 
\usepackage{units}

\usepackage{mathtools}

\usepackage{color}

\definecolor{green}{cmyk}{1,0,1,0}

\definecolor{pink}{cmyk}{0,0.5,0,0}

\definecolor{pastelpink}{cmyk}{0,0.25,0,0}

\definecolor{softpink}{cmyk}{0,0.125,0,0}

\definecolor{purple}{cmyk}{0.5,1.0,0.1,0}

\definecolor{violet}{cmyk}{0.75,1,0.25,0}

\newcommand{\nn}{\nonumber}

\usepackage[utf8]{inputenc}

\usepackage[geometry]{ifsym}

\begin{document}
\preprint{YITP-26-29}
\preprint{KYUSHU-HET-357}
\preprint{EPHOU-26-003}
\preprint{UME-PP-030}

\sloppy
\title{
Dark photon and U(1)$_{B-L}$ gauge boson\\
from dark Higgs boson decays at FASER and SHiP
}

\author{Takeshi Araki}
\email{t-araki@den.ohu-u.ac.jp}
\affiliation{%
Faculty of Dentistry, Ohu University, 31-1 Misumidou, Tomita-machi,\\
Koriyama, Fukushima 963--8041, Japan}

\author{Kento Asai}
\email{asai@yukawa.kyoto-u.ac.jp}
\affiliation{%
Yukawa Institute for Theoretical Physics, Kyoto University,\\
Kyoto 606--8502, Japan}
\affiliation{%
Institute for Cosmic Ray Research (ICRR), The University of Tokyo,\\
Kashiwa, Chiba 277--8582, Japan}

\author{Yohei Nakashima}
\email{nakashima.youhei.775@s.kyushu-u.ac.jp}
\affiliation{Department of Physics, Kyushu University, 744 Motooka, Nishi-ku,\\
Fukuoka, 819-0395, Japan}

\author{Osamu~Seto}
\email{seto@particle.sci.hokudai.ac.jp}
\affiliation{%
Department of Physics, Hokkaido University, Sapporo 060--0810, Japan}

\author{Takashi Shimomura}
\email{shimomura@miyazaki-u.ac.jp}
\affiliation{%
Faculty of Education, Miyazaki University, Miyazaki, 889--2192, Japan}

\author{Yoshiki Uchida}
\email{uchida.yoshiki@ccnu.edu.cn}
\affiliation{%
Institute of Particle Physics and Key Laboratory of Quark and Lepton Physics (MOE),\\ Central China Normal University, Wuhan, Hubei 430079, China}
\affiliation{%
State Key Laboratory of Nuclear Physics and Technology,
Institute of Quantum Matter,\\ South China Normal University,
Guangzhou 510006, China}
\affiliation{%
Guangdong Basic Research Center of Excellence for Structure and 
Fundamental\\
Interactions of Matter, Guangdong Provincial Key Laboratory of Nuclear Science,\\
Guangzhou 510006, China}

\begin{abstract}
We study the sensitivity to dark photons and U(1)$_{B-L}$ gauge bosons produced via dark Higgs boson decays at the FASER and SHiP experiments.
In addition to pair production of these vector bosons from both on-shell and off-shell dark Higgs boson decays, a new production process of single vector boson associated with the standard model particles is taken into account.  
Constraints on the parameter space of dark photon are derived with including the latest results from the FASER experiment. The expected sensitivity regions to the dark photon and U(1)$_{B-L}$ gauge boson of the future FASER2 and SHiP experiments are presented.
The sensitivity to the U(1)$_{B-L}$ model with freeze-in sterile neutrino dark matter is also discussed.
\end{abstract}

\date{\today}

\maketitle
\tableofcontents

\section{Introduction}
\label{sec:introduction}
Recently, feebly interacting and light dark sector has attracted broad interest in particle physics. 
A wide range of possibilities has been explored in connection with dark matter (DM), neutrino mass and flavor mixing, the strong CP problem, the baryon asymmetry of the Universe, and so on (for a review, see refs.~\cite{Fabbrichesi:2020wbt,Feng:2022inv,Antel:2023hkf,Alimena:2025kjv}). 
In this framework, portal particles mediate interactions between the Standard Model (SM) sector and the dark sector.
Such portals are typically long-lived and are being probed in present and future experiments such as FASER~\cite{Feng:2017uoz, Feng:2017vli, FASER:2018eoc, FASER:2018bac}, Belle-II~\cite{Belle-II:2018jsg}, NA64~\cite{Gninenko:2013rka,Andreas:2013lya,NA64:2016oww}, EBES~\cite{Ishikawa:2021qna}, FASER2~\cite{FPF:2025bor}, SHiP~\cite{SHiP:2015vad,SHiP:2020vbd,ShipECN3}, FACET~\cite{Cerci:2021nlb}, and CODEX-b~\cite{Gligorov:2017nwh,CODEX-b:2019jve}. 
As many extensions beyond the minimal setup have been proposed, 
detailed studies of portal particle production are important for determining search sensitivities to the dark sectors in these experiments.

Dark photon is one of the most widely studied portals in the dark sector framework.
It is realized by a gauge boson associated with an Abelian dark symmetry U(1)$_D$, which kinetically mixes with the hypercharge gauge boson 
\cite{Okun:1982xi,Galison:1983pa,Holdom:1985ag,Foot:1991kb,Babu:1997st}. Through this kinetic mixing, the dark photon interacts with the SM particles via the electromagnetic current \cite{Pospelov:2007mp,Huh:2007zw,Pospelov:2008zw}. 
Another well studied portal is dark Higgs boson which mixes with the SM Higgs boson. It interacts with the SM fermions and gauge bosons through this scalar mixing~\cite{Patt:2006fw,OConnell:2006rsp,Bezrukov:2009yw,Feng:2017vli}.
Dark photon and dark Higgs boson have been often studied separately in much of the literature. 
Theoretical analyses on the sensitivity to dark photon and dark Higgs boson have been performed independently, and experimental results have been reported respectively. 

When one considers the origin of the dark photon mass, one possibility is the spontaneous breaking of U(1)$_D$ symmetry by dark Higgs fields.  
In refs.~\cite{Araki:2020wkq, Araki:2022xqp, Araki:2024uad}, the dark photon pair production from on-shell and off-shell dark Higgs decays, which is a direct consequence of the mass generation by the spontaneous symmetry breaking, has been studied for the FASER~\cite{Feng:2017uoz, Feng:2017vli, FASER:2018eoc, FASER:2018bac} experiment.
It was shown that the sensitivity to the dark photon parameters can be enlarged by these additional dark photon productions.
Similar studies have been performed for dark photon mediated dark matter~\cite{Cline:2024wja}, inelastic dark matter \cite{Li:2021rzt,Ko:2025drr}, lepton flavor violating gauge bosons~\cite{Araki:2022xqp} and scalar bosons~\cite{Balkin:2024qtf}, dark photon~\cite{Foguel:2022unm,Ferber:2023iso,Felkl:2023nan,Hostert:2023gpk,Cheung:2024oxh,Seto:2025mte,Cheung:2025kmc}, gauged U(1)$_{L_\mu - L_\tau}$~\cite{Nomura:2020vnk}, and U(1)$_{B-L}$ models~\cite{Dev:2021qjj}.
In the previous studies, only the dark photon pair production from dark Higgs boson decays was considered. In this paper, we point out that single dark photon production from on-shell dark Higgs boson decay is important in future long-lived particle search experiments. Even when the pair production of dark photons is kinematically forbidden, single dark photon production in association with SM particles can be allowed for the dark photon lighter than the dark Higgs boson. In such a situation, the dark Higgs boson becomes long-lived due to phase space suppression and can travel long distance. Then, the dark photon can be short-lived, which enables one to explore larger kinetic mixing region in long-lived particle search experiment. 

Light and long-lived particles can also appear in freeze-in dark matter scenario, in particular, in infrared (IR) freeze-in production~\cite{McDonald:2001vt,Hall:2009bx}, in which typical interaction strength is very weak. Sterile neutrino is a good dark matter candidate in gauged U(1)$_{B-L}$ model with the freeze-in mechanism~\cite{Kaneta:2016vkq,Araki:2022xqp,Eijima:2022dec,Seto:2024lik}. Interestingly, the lifetime and mass of the U(1)$_{B-L}$ gauge boson and its symmetry-breaking Higgs boson under certain mass spectra are in the reach of long-lived particle search experiments~\cite{Eijima:2022dec,Seto:2024lik}. Therefore, the above analyses can be applied to this model.

The purpose of this paper is to show that one dark photon or U(1)$_{B-L}$ gauge boson production with SM particles via on-shell dark Higgs boson decay provides a new possibility in long-lived particle search experiments. We include the pair production processes of vector bosons in our analyses of the dark photon model and the gauged U(1)$_{B-L}$ model. The sensitivity regions at future FASER2 and SHiP experiments can be enlarged by the new production process.  Some parameter space of the dark photon can be constrained by the latest result of the FASER experiment~\cite{FASER:2023tle,moriond}.

This paper is organized as follows. In section 2, we introduce dark photon and gauged U(1)$_{B-L}$ models with dark Higgs bosons and give relevant interaction Lagrangians, respectively. In section 3, we explain the production processes of dark photon and U(1)$_{B-L}$ gauge boson from dark Higgs boson decays, and present the decay widths of the dark Higgs boson into a single vector boson with the SM particles. The decay branching ratios and decay lengths of the dark Higgs boson are shown in section 4, and the distributions of the dark photons are shown in section 5. The formulae of the expected number of events from each production process are given in Section 6. Our numerical results are presented in section 7. The last section is devoted to summary. In appendix, the distributions of $B$ meson and dark Higgs bosons are presented.

\section{Models} \label{sec:model}
We start our discussion with introducing two models studied in this paper. 
One is a secluded U(1)$_{D}$ or dark photon model \cite{Foot:1990mn,He:1990pn,He:1991qd,Pospelov:2007mp,Huh:2007zw,Pospelov:2008zw} with a kinetic mixing term \cite{Okun:1982xi,Galison:1983pa,Holdom:1985ag,Foot:1991kb,Babu:1997st}, 
and the other is a U(1)$_{B-L}$ model \cite{Davidson:1978pm,Marshak:1979fm,Mohapatra:1980qe} where $B$ and $L$ denote the baryon and 
the lepton number, respectively. Given spontaneous symmetry breaking 
as the origin of the new gauge boson mass, an extra complex scalar field, namely dark Higgs field, is introduced to both models. 
The dark Higgs field $\Phi$ is assumed to be singlet under the SM gauge group and charged only under the secluded U(1)$_D$ or U(1)$_{B-L}$ symmetry. The vacuum expectation values (vevs) of the SM doublet Higgs field $H$ and the dark Higgs field are written by
\begin{align}
    \langle H \rangle = \frac{1}{\sqrt{2}} 
    \begin{pmatrix}
        0 \\ v
    \end{pmatrix}, ~~~~
    \langle \Phi \rangle = \frac{v_\Phi}{\sqrt{2}}, \label{eq:dp-iggs-vev}
\end{align}
respectively.
In the following, the vector field $X_\mu$ denotes either the U(1)$_D$ or U(1)$_{B-L}$ gauge boson in the interaction basis. The covariant derivative for fermions and scalars is defined by
\begin{align}
    D_\mu = \partial_\mu - i g_1 Y B_\mu - i g_2 \tau^a W^a_\mu - i g' Q' X_\mu,
\end{align}
where $B_\mu$ and $W^a_\mu~(a=1,2,3)$ are the U(1)$_Y$ and SU(2)$_L$ gauge bosons, and $Y$ and $Q'$ are the hypercharge and the extra U(1) gauge charge, respectively. 
The matrix $\tau^a$ is defined by $\tau^a \equiv \sigma^a/2$ with $\sigma^a$ being the Pauli matrices. 
The gauge couplings of U(1)$_Y$, SU(2)$_L$, and U(1)$_D$ are denoted by $g_1$, $g_2$, and $g'$, respectively. For the U(1)$_{B-L}$ model, $g'$ should be replaced with $g_{B-L}$.
Throughout this paper, the SU(3) color interaction is omitted because it is irrelevant to our study.
 
\subsection{Dark photon model} \label{subsec:dp-model}
In the dark photon model, all SM particles are assumed to be singlet under the secluded U(1)$_D$ symmetry. However, the kinetic mixing between $X_\mu$ and $B_\mu$ is allowed by the symmetries. Meanwhile, $\Phi$ is assigned to have a secluded U(1)$_D$ 
charge $Q^D_\phi = +1$ without loss of generality and can form a quartic coupling with $H$. 

The Lagrangian of the dark photon model with the dark Higgs field is given by 
\begin{align}
\mathcal{L} = \mathcal{L}_{\mathrm{SM}} - \frac{1}{4} X^{\mu \nu} X_{\mu \nu}  + \frac{\epsilon}{2} B^{\mu\nu} X_{\mu\nu} 
+ |D_\mu \Phi|^2 - V(H,\Phi), \label{eq:dp-lag}
\end{align}
where $\mathcal{L}_{\mathrm{SM}}$ represents the SM Lagrangian without the scalar potential. The second term is the kinetic term of $X_\mu$, and 
the third one is the kinetic mixing term with a constant parameter $\epsilon$. The field strength tensors of $X_\mu$ and $B_\mu$ are given by 
\begin{align}
X_{\mu\nu} = \partial_\mu X_\nu - \partial_\nu X_\mu,~~~B_{\mu\nu} = \partial_\mu B_\nu - \partial_\nu B_\mu,
\end{align}
respectively. The fourth term is the kinetic term of $\Phi$, and 
the last term $V(H,\Phi)$ is the scalar potential including $H$. The concrete form of $V(H, \Phi)$ is given by
\begin{align}
    V(H, \Phi) = -\mu_H^2 H^\dagger H -\mu_\Phi^2 \Phi^\dagger \Phi + \frac{\lambda_H}{2} (H^\dagger H)^2 
    + \frac{\lambda_\Phi}{2} (\Phi^\dagger \Phi)^2 + \lambda_{H\Phi} (H^\dagger H) (\Phi^\dagger \Phi), \label{eq:dp-scalar-pot}
\end{align}
where $\mu^2_H$ and $\mu^2_\Phi$ as well as the quartic couplings $\lambda_H$, $\lambda_\Phi$, and $\lambda_{H\Phi}$ are taken to 
be positive. The values of these parameters are assumed to be chosen appropriately so that the electroweak and the secluded symmetry can be spontaneously broken. After the symmetry breaking, the scalar bosons mix with each other through the quartic term. 

Based on the Lagrangian \eqref{eq:dp-lag} with the vevs \eqref{eq:dp-iggs-vev}, the interaction Lagrangian can be expressed in the mass basis of the dark photon $A'$, the dark Higgs boson $\phi$, and the SM Higgs boson $h$, respectively.
The interaction Lagrangian relevant to $A'$ production via the dark Higgs boson decays can be read as~\cite{Araki:2024uad}
\begin{align}
    \mathcal{L}_{\mathrm{int}} \supset g' Q^D_\phi m_{A'}\cos\alpha \phi A'_{\mu}A'^{\mu} + \varepsilon eA'_{\mu}J_{\mathrm{EM}}^\mu - \sin\alpha \sum_f\frac{m_f}{v}\phi\bar{f}f, \label{eq:dp-int-lag}
\end{align}
where $\varepsilon = \epsilon \cos\theta_W$ is the gauge kinetic mixing with the Weinberg angle $\theta_W$, and $\alpha$ is the mixing angle between the SM Higgs and the dark Higgs boson. 
The masses of $A'$ and the SM fermion $f$ are denoted as $m_{A'}$ and $m_f$, respectively. 
The electromagnetic current $J^\mu_{\mathrm{EM}}$ is defined by
\begin{align}
J^\mu_{\mathrm{EM}} &= \sum_f Q_f \bar{f} \gamma^\mu f,
\end{align} 
where $Q_f$ is the electric charge of $f$, with $e$ being the elementary charge.

\subsection{Gauged U(1)$_{B-L}$ model} \label{subsec:bl-model}
In the gauged U(1)$_{B-L}$ model, the SM quarks and leptons have the U(1)$_{B-L}$ gauge charge $+1/3$ and $-1$, respectively. Since the model becomes anomalous only with the SM particle contents, three SM singlet fermions are introduced to cancel the gauge anomaly. We identify these fermions as right-handed neutrinos $\nu_R$ and assign the U(1)$_{B-L}$ charge $-1$. The gauge charge of the dark Higgs field is assigned to $Q^{B-L}_\phi = +2$ 
so that the right-handed neutrino can acquire Majorana masses after the spontaneous breaking of U(1)$_{B-L}$. The Lagrangian of the U(1)$_{B-L}$ model is given by
\begin{align}
\mathcal{L} = \mathcal{L}_{\mathrm{SM}} +  \mathcal{L}_{\nu_R}- \frac{1}{4} X^{\mu \nu} X_{\mu \nu}  + \frac{\epsilon}{2} B^{\mu\nu} X_{\mu\nu} 
+ |D_\mu \Phi|^2 - V(H,\Phi), \label{eq:bl-lag}
\end{align}
where the second term represents the Lagrangian of the right-handed neutrinos. For the sake of minimality, we set the kinetic mixing parameter to be negligibly small and omit this term throughout this paper. The Lagrangian in the mass basis of the U(1)$_{B-L}$ gauge boson $Z'$ and the dark Higgs boson $\phi$ can be found in refs.~\cite{Eijima:2022dec,Seto:2024lik}. 

The Lagrangian relevant to $Z'$ production via the dark Higgs boson decays is given by 
\begin{align}
    \mathcal{L}_{\mathrm{int}} \supset g_{B-L} Q^{B-L}_\phi m_{Z'}\cos\alpha \, \phi Z'_{\mu} Z'^{\mu} + g_{B-L} Z'_{\mu}J_{B-L}^\mu - \sin\alpha \sum_f\frac{m_f}{v}\phi\bar{f}f, \label{eq:bl-int-lag}
\end{align}
where the U(1)$_{B-L}$ gauge current $J^{\mu}_{B-L}$ is given by
\begin{align}
J_{B-L}^\mu = \sum_f Q_f^{B-L} \bar{f} \gamma^\mu f,
\end{align}
with $Q_f^{B-L}$ being the U(1)$_{B-L}$ charge of fermion $f$. The second term originates from the gauge interaction and plays a similar role of the second one in eq.~\eqref{eq:dp-int-lag}. 
The interaction Lagrangian can be obtained from eq.~\eqref{eq:dp-int-lag} by the replacement of
\begin{align}
    g' \to g_{B-L},~~~ m_{A'} \to m_{Z'},~~~ \varepsilon \to \frac{g_{B-L}}{e},~~~Q_f \to Q_f^{B-L}.~~~Q^D_\phi \to Q^{B-L}_\phi. \label{eq:coupling-replace}
\end{align}

\section{Production via dark Higgs boson decays} \label{sec:production}
\begin{figure}[t]
\begin{center}
\begin{tabular}{ccc}
    \includegraphics[scale=0.5]{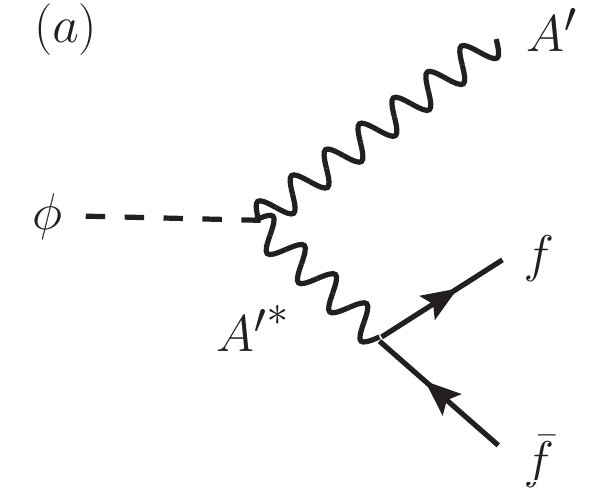}~~ &
    \includegraphics[scale=0.5]{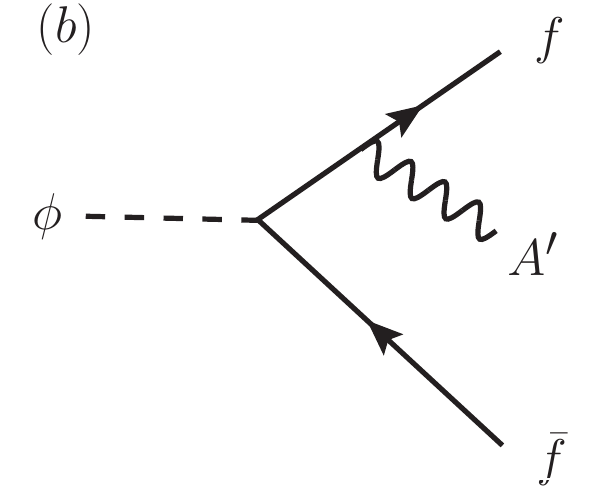}~~ &
    \includegraphics[scale=0.5]{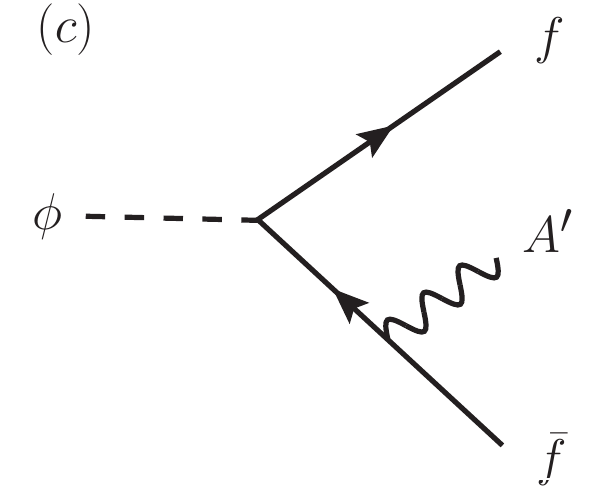} 
\end{tabular}
\end{center}
    \caption{
    Feynman diagrams of $\phi \to A' f \bar{f}$
    }
\label{fig:off-shell-dp-diagram}
\end{figure}
In this paper, we study three production processes of the dark photon or the U(1)$_{B-L}$ gauge boson through dark Higgs boson decays : 
\begin{enumerate}
\setlength{\parskip}{0mm} %
\setlength{\itemsep}{1mm} %
    \item[(A)] the on-shell $\phi$ decay into an $A'$ pair, $\phi \to A'A'$~\cite{Araki:2020wkq}~, 
    \item[(B)] the on-shell $\phi$ decay into single $A'$ and SM particles, $\phi \to A' + {\rm SM}$~,
    \item[(C)] the off-shell $\phi$ decay into an $A'$ pair, $\phi^\ast \to A' A'$~\cite{Araki:2024uad},
\end{enumerate}
where $A'$ should be replaced with $Z'$ in the U(1)$_{B-L}$ model.
The off-shell $\phi$ decays appear in meson decays, e.g. $B \to K + \phi^\ast \to K A'A'$. 
Each production process can be dominant for (A) $2m_{A'} < m_\phi$, (B)  $m_{A'} \leq m_\phi \leq 2m_{A'}$ and large $\varepsilon$, and (C) either $m_\phi < 2 m_{A'}$ or $m_B-m_K< m_\phi$. 
The productions from (A) and (C) have already been analyzed, and these decay widths can be found in refs.~\cite{Araki:2020wkq} and \cite{Araki:2024uad}, respectively. The production from (B) is firstly analyzed as the source of light gauge bosons in the context of long-lived particle search such as FASER and SHiP.
Figure \ref{fig:off-shell-dp-diagram} shows the Feynman diagrams of (B) $\phi \to A' f\bar{f}$. The diagram (a) is the off-shell $A'$ decay. It occurs through the $\phi$-$A'$-$A'$ interaction and, hence, is related to the origin of the gauge boson mass.
The amplitude of (a) is proportional to $g'\varepsilon$ in the dark photon model or $g_{B-L}^2$ in the U(1)$_{B-L}$ model and unsuppressed by the scalar mixing. On the other hand, the amplitudes of diagrams (b) and (c) are proportional to $\alpha Y_f \varepsilon$ or $\alpha Y_f g_{B-L}$ where $Y_f$ is the Yukawa coupling constant of $f$. 
Thus, unless $g'$ or $g_{B-L}$ is very small, the amplitude of (a) always dominates over those of (b) and (c). When $g'$ or $g_{B-L}$ is tiny, three production processes can become comparable. 
In that case, however, $\phi$ predominantly decays into the SM particles instead of (a), (b), and (c), because the amplitude of $\phi \to \mathrm{SMs}$ is only suppressed by $\alpha Y_f$.
Thus, it is enough to consider diagram (a) for our purpose.
We note that the diagram (a) can be an important source of $A'$ in the dark photon model since there is no strong constraint on $g'$, while it is not for the U(1)$_{B-L}$ model since $g_{B-L}$ is tightly constrained by experiments\footnote{This situation can be evaded when one considers a large kinetic mixing parameter. In that case, the results will be similar to the dark photon model.}.

For the dark photon model, the differential decay width of $\phi \to A' f \bar{f}$ with respect to the momentum transfer $q^2$ is given by
\begin{align}
    \frac{d}{dq^2}\Gamma(\phi \to A' f \bar{f}) &= \frac{(\varepsilon g' Q_f \cos\alpha)^2 \alpha_{\mathrm{EM}}}{4 \pi^2} \frac{m_{A'}^2}{m_\phi} 
    \left( \frac{1}{(q^2 - m_{A'}^2)^2 + m_{A'}^2 \Gamma_{A'}^2} \right) \nn \\
    &\quad \times
    (q^2 + 2 m_f^2)\left( 1 + \frac{m_\phi^4}{12 m_{A'}^2 q^2} \lambda^2_{A'} \right) \lambda_{A'} \beta, \label{eq:diff-decay-width2}
\end{align}
where $\Gamma_{A'}$ is the total decay width of $A'$, and $\alpha_{\mathrm{EM}} = e^2/(4 \pi)$ is the fine structure constant. The range of $q^2$ is from $4m_f^2$ to $(m_\phi - m_{A'})^2$. The functions $\lambda_A$ and $\beta$ in eq.~\eqref{eq:diff-decay-width2} are defined by
\begin{subequations}
\begin{align}
    \lambda_{A'} &= \lambda\left(\frac{m_{A'}}{m_\phi}, \frac{q}{m_\phi}\right)
    \equiv \sqrt{1 - 2 \frac{(m_{A'}^2 + q^2)}{m_\phi^2} + \frac{(m_{A'}^2 - q^2)^2}{m_\phi^4}}, \\
    \beta &= \sqrt{1 - \frac{4 m_f^2}{q^2}}.
\end{align}
\end{subequations}
For the gauged U(1)$_{B-L}$ model, the differential decay width can be simply obtained from eq.~\eqref{eq:diff-decay-width2} by making the replacement of eq.~\eqref{eq:coupling-replace}. 

The differential decay branching ratio into $A'$ with all the SM particles can be expressed approximately by using eq.~\eqref{eq:diff-decay-width2} as 
\begin{align}
    \frac{d}{dq^2}\Gamma(\phi \to A' + \mathrm{SM}) = \frac{1}{\mathrm{Br}({A'}^\ast \to f \bar{f})} \frac{d}{dq^2}\Gamma(\phi \to A' f \bar{f}), \label{eq:diff-decay-width-sm}
\end{align}
where Br$({A'}^\ast \to f\bar{f})$ is the branching ratio of $A'$ into $f \bar{f}$ with mass $m_{{A'}^\ast}^2=q^2$. This approximation is valid since the decay width of $A'$ is very narrow compared to its mass due to the small coupling constants $\varepsilon$.

\section{Dark Higgs decays} \label{sec:higgs-decay}
\begin{figure}[t]
\begin{center}
\begin{tabular}{cc}
    \includegraphics[scale=0.35]{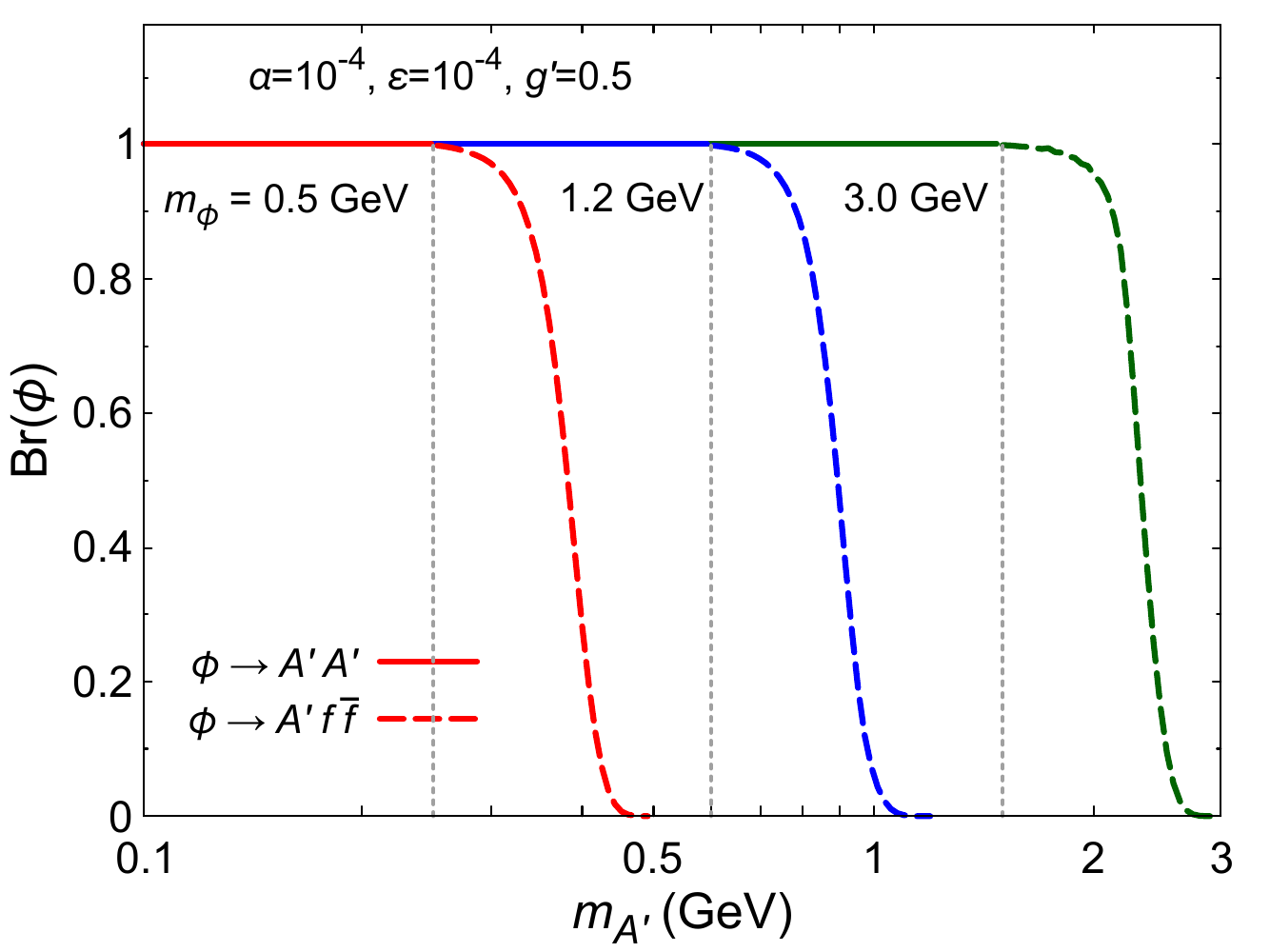} &
    \includegraphics[scale=0.35]{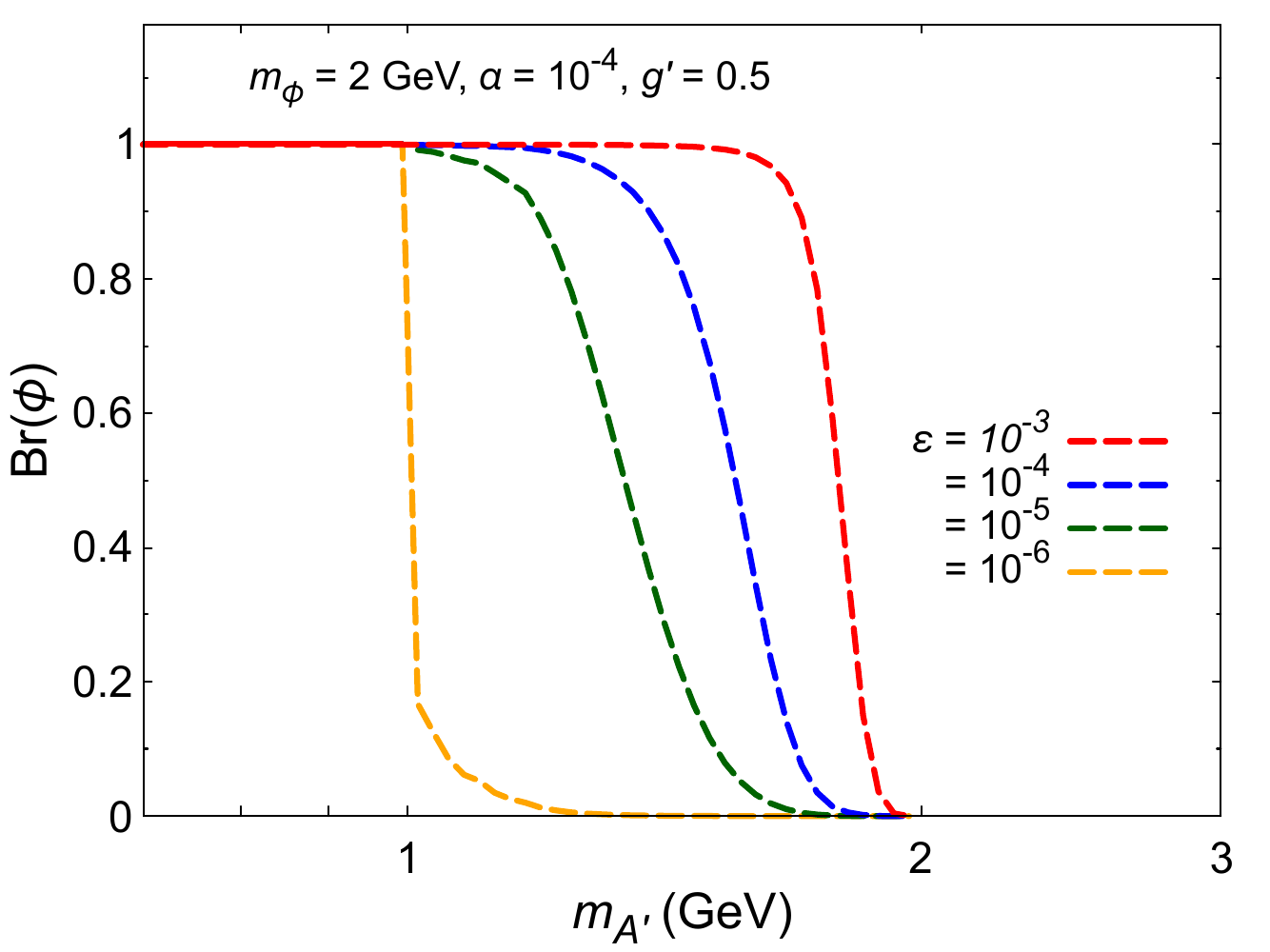}
\end{tabular}
\end{center}
    \caption{
    The decay branching ratios of the dark Higgs boson in the dark photon model. Left: the branching ratios of $\phi \rightarrow A'A'$ (solid) and $\phi \rightarrow A' + \mathrm{SM}$ (dashed) for $m_\phi = 0.5$ GeV (red), $1.2$ GeV (blue), and $3.0$ GeV (green) with $\varepsilon = 10^{-4}$. The vertical dotted lines represent the kinematical thresholds for each $m_\phi$.
    Right: the branching ratios for $\varepsilon = 10^{-3}$ (red), $10^{-4}$ (blue), $10^{-5}$ (green), and $10^{-6}$ (orange) with $m_\phi = 2.0$ GeV. 
    In both panels, the scalar mixing and the gauge coupling constant are fixed to $\alpha = 10^{-4}$ and $g' = 0.5$, respectively.
    }
\label{fig:dh-br}
\end{figure}
The decay of the dark Higgs boson is modified by the new decay channels, that is, (A) $\phi \rightarrow A'A'$ and (B) $\phi \rightarrow A' + \mathrm{SM}$ discussed in the previous section.
The total decay width of the dark Higgs boson is given by
\begin{align}
    \Gamma_\phi = \Gamma(\phi \to \mathrm{SM}) + \Gamma(\phi \to A'A') + \Gamma(\phi \to A' + \mathrm{SM}),
\end{align}
where $\Gamma(\phi \to \mathrm{SM})$ denotes the all possible final states containing only SM particles above thresholds.

Figure \ref{fig:dh-br} shows the decay branching ratios into the $A'A'$ (solid) and the $A' + \mathrm{SM}$ (dashed) final state. The left panel shows the dependence on the dark Higgs boson mass by varying $m_\phi = 0.5$ GeV (red), $1.2$ GeV (blue), and $3.0$ GeV (green) with $\varepsilon = 10^{-4}$. 
The gray dotted vertical lines represent the kinematical thresholds of $\phi \to A'A'$ decay. 
The other parameters are fixed to $\alpha = 10^{-4}$ and $g' = 0.5$. 
The right panel shows the dependence on $\varepsilon$ for $10^{-3}$ (red), $10^{-4}$ (blue), $10^{-5}$ (green), and $10^{-6}$ (orange) with $m_\phi = 2.0$ GeV. 
In both panels, one can see that, below the threshold of $\phi \to A'A'$, the decay into $A'A'$ dominates the dark Higgs boson decay, while above the threshold, the decay into $A' + \mathrm{SM}$ can become significant depending on $\varepsilon$.

\begin{figure}[t]
\begin{center}
\begin{tabular}{cc}
    \includegraphics[scale=0.35]{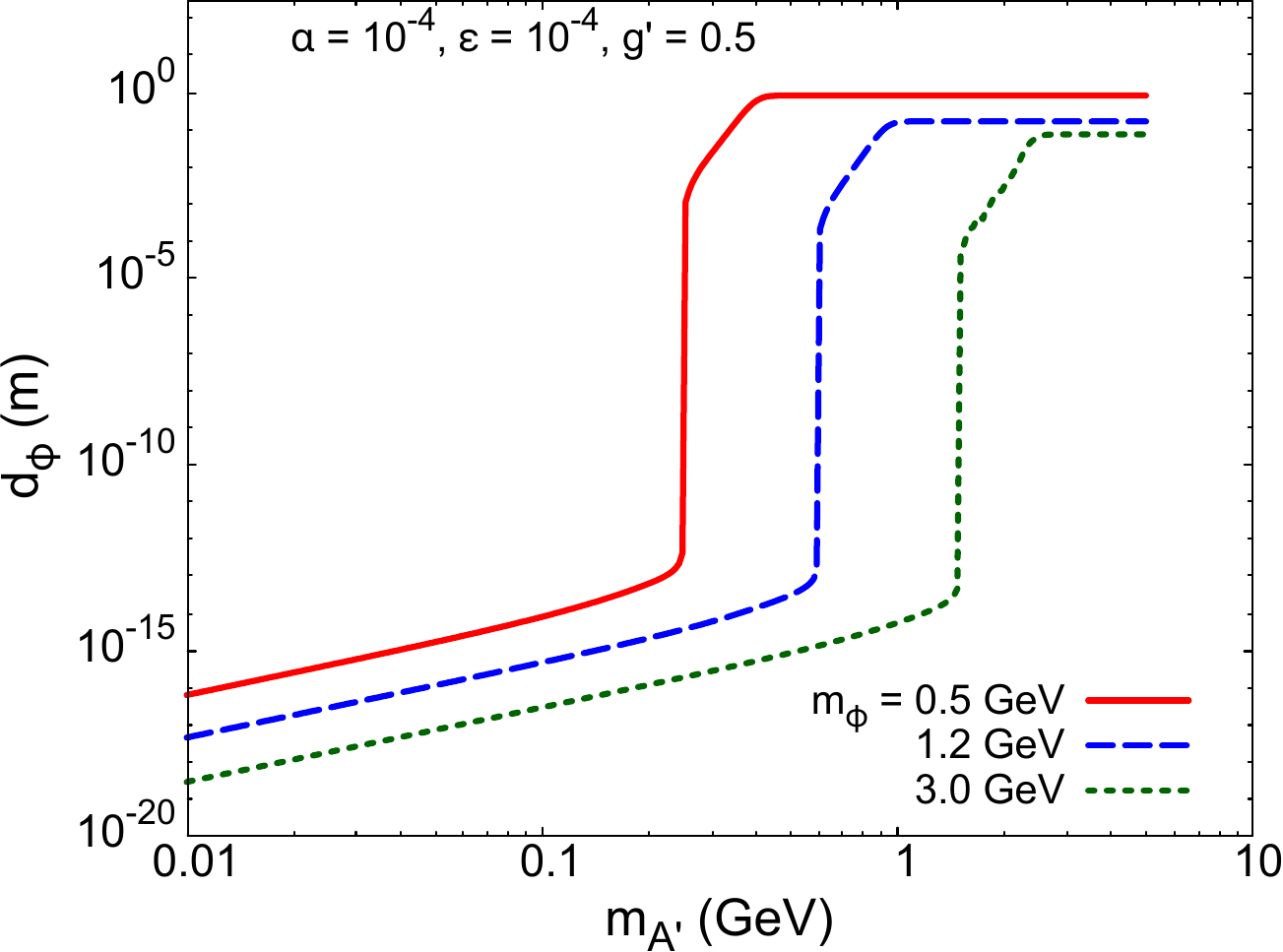} &
    \includegraphics[scale=0.35]{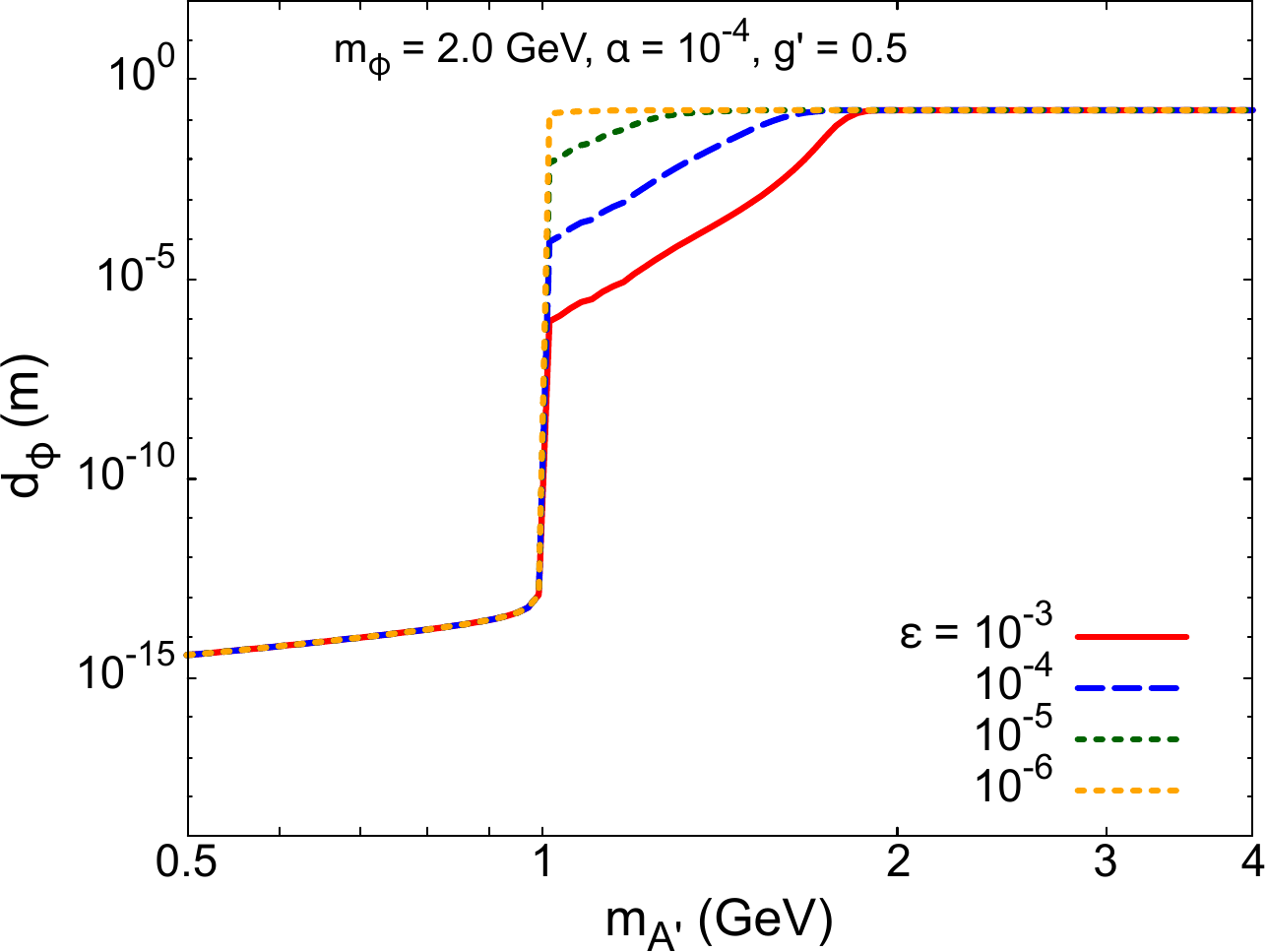}
\end{tabular}
\end{center}
    \caption{
    The decay lengths of the dark Higgs boson in the dark photon model.
    Left: the decay lengths for $m_\phi = 0.5$ GeV (red solid), $1.2$ GeV (blue dashed), and $3.0$ GeV (green dotted) with $\varepsilon = 10^{-4}$.
    Right: the decay lengths for $\varepsilon = 10^{-3}$ (red solid), $10^{-4}$ (blue dashed), $10^{-5}$ (green dotted), and $10^{-6}$ (orange dashed-dotted) with $m_\phi = 2.0$ GeV. 
    In both panels, the scalar mixing and the gauge coupling constant are fixed to $\alpha = 10^{-4}$ and $g' = 0.5$, respectively.
    }
\label{fig:dh-decay-length}
\end{figure}
Figure \ref{fig:dh-decay-length} shows the decay length of the dark Higgs boson, $d_\phi$. The left panel shows the decay lengths for $m_\phi = 0.5$ GeV (red solid), $1.2$ GeV (blue dashed), and $3.0$ GeV (green dotted) with $\varepsilon = 10^{-4}$. The decay length becomes shorter as $m_\phi$ becomes larger.
Below the threshold of $\phi \to A'A'$, the dark Higgs boson instantly decays into a pair of $A'$, and its decay length is very short. 
Above the threshold, the decay length is much longer and macroscopic. The dark Higgs boson decays into $A'+\mathrm{SM}$ until $m_{A'} < m_\phi - 2 m_e$ in the dark photon model ($m_{A'} < m_\phi$ in the U(1)$_{B-L}$ model). 
Once the $\phi \to A'+\mathrm{SM}$ decay channel is closed, the dark Higgs boson predominantly decays into the SM particles. Even if $m_{A'} > m_\phi - 2 m_e$, however, as we will see later, dark photons can still be produced from the off-shell dark Higgs decays: (C) $\phi^\ast \to A'A'$. 
The right panel focuses on the contributions to the decay length by 
$\phi \to A' +\mathrm{SM}$ for $m_\phi = 2.0$ GeV. The red (solid), blue (dashed), green (dotted), and orange (dashed-dotted) curves correspond to $\varepsilon = 10^{-3}$, $10^{-4}$, $10^{-5}$, and $10^{-6}$, respectively. The other parameters are taken $\alpha=10^{-4}$ and $g'=0.5$. The partial decay width of $\phi \to A' +\mathrm{SM}$ is proportional to $\varepsilon^2$, and thus the decay length becomes shorter 
as $\varepsilon$ becomes larger. This decay can be significant for $\varepsilon > 10^{-5}$. From the figure, the decay length is roughly from $10^{-5}$\,m to $1$\,m. When the dark Higgs boson is boosted with the Lorentz factor $\beta\gamma > 10^2$, it can reach detectors located several $100$\,m away and then decay into one dark photon with a charged particle pair. In this case, the dark photon is not necessarily long-lived. This implies that short-lived dark photons or larger $\varepsilon$ regions can be explored at long-lived particle search experiments through this new production process.

For the U(1)$_{B-L}$ model, the decay branching ratio and the decay length are calculated by making the replacement of eq.~\eqref{eq:coupling-replace}.
The decay width of $\phi \to Z' + \mathrm{SM}$ scales as $g_{B-L}^4$ instead of $g'^2 \varepsilon^2$. 
As will be shown later, in the parameter space we are interested in, the gauge coupling constant is rather small.
For instance, $g_{B-L} = 10^{-4}$ corresponds to $\varepsilon \simeq 10^{-7}$ in the dark photon model. The decay width with such a small $\varepsilon$ is negligibly small 
as can be seen from figure~\ref{fig:dh-br}. 
We note in passing that the decay width of $\phi^\ast \to Z'Z'$ in the U(1)$_{B-L}$ model is also negligibly small since it scales as $g_{B-L}^2 \alpha^2$.

\section{Distribution of gauge bosons}
The distributions of the dark photons produced via the on-shell dark Higgs boson decays (A) and (B) are shown in this section. As an illustrative example, we take $m_{A'} = 0.1,~
0.8$, and $1.2$\,GeV for $m_\phi = 2.0$\,GeV. The first two correspond to light and heavy dark photon mass compared to $m_\phi/2$  respectively for the production (A), and the last one is to the case for the production (B). Other parameters are fixed to $g'=0.05$ and $\alpha = 10^{-3}$ for FASER and $g'=0.5$ and $\alpha=10^{-4}$ for FASER2 and SHiP. For the off-shell dark photon decays, $\varepsilon$ is taken to be $10^{-4}$. For simplicity, we only consider $B$ meson decays which are the main production channel of the dark Higgs boson. 
The $B$ meson distribution with respect to its momentum and angle to the beam axis can be generated by Pythia 8~\cite{Sjostrand:2014zea,Sjostrand:2006za}. The Monash tune~\cite{Skands:2014pea} is additionally applied for proton-proton collision at LHC. For the FASER and FASER2 experiment, we use the data sets implemented in \texttt{FORESEE package}~\cite{Kling:2021fwx} while we generated the data sets for SHiP assuming the target consists of pure molybdenum.\footnote{In the actual SHiP setup, the target is composed of titanium-zirconium-doped molybdenum alloy~\cite{ShipECN3}. However, for simplicity, we model it as a pure molybdenum target in the present analysis.}
The dark Higgs boson distributions are calculated by using the $B$ meson distributions. Then the dark photon distributions are obtained by calculating the production processes (A) and (B) using the dark Higgs boson distributions.

\begin{figure}[t]
\tabcolsep = -0.1cm
\centering
\begin{tabular}{ccc}
    \includegraphics[scale=0.27]{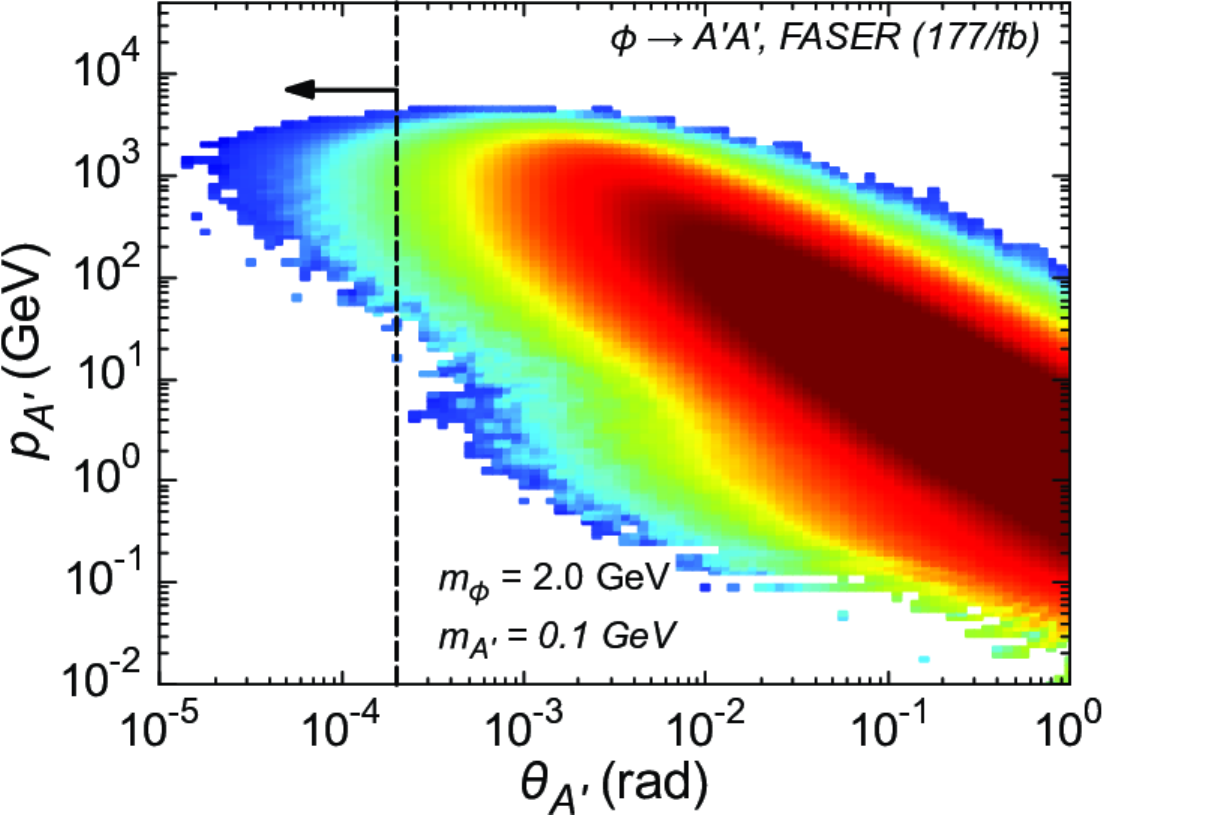} &
    \includegraphics[scale=0.27]{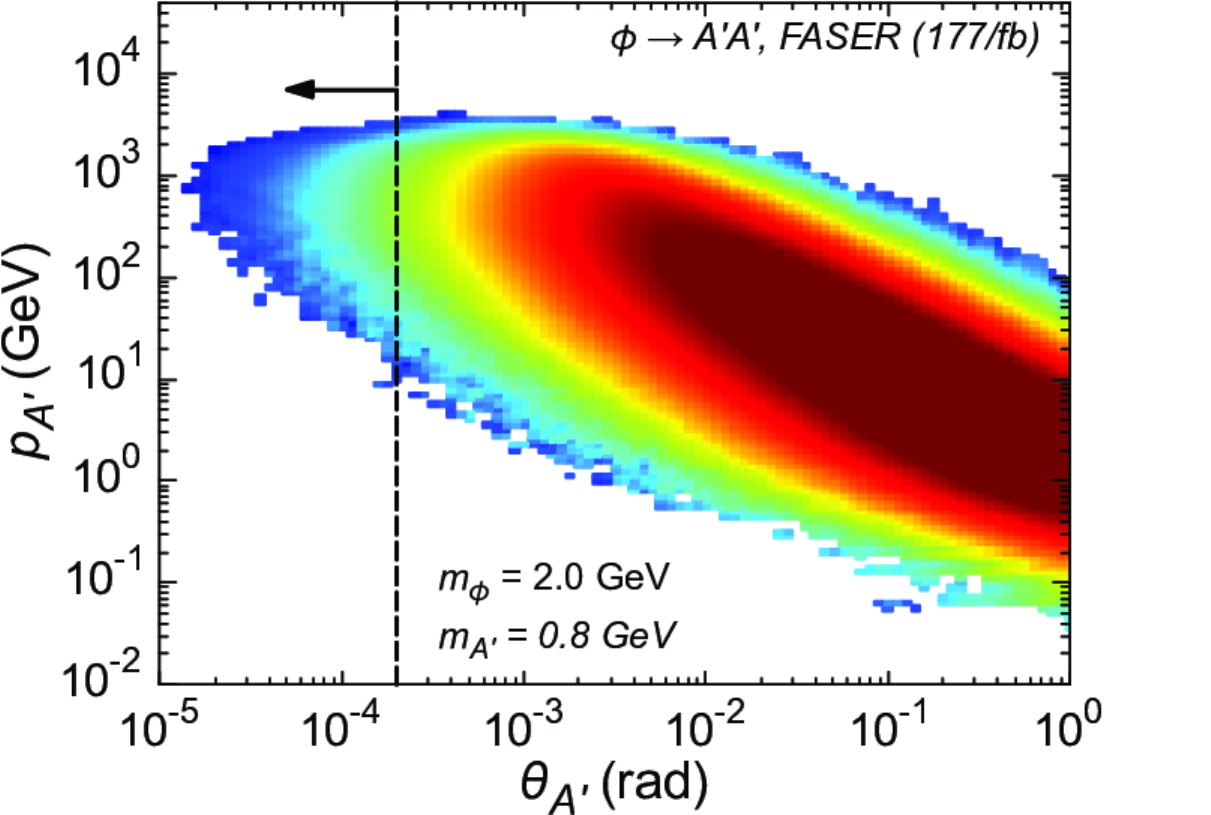} &
    \includegraphics[scale=0.27]{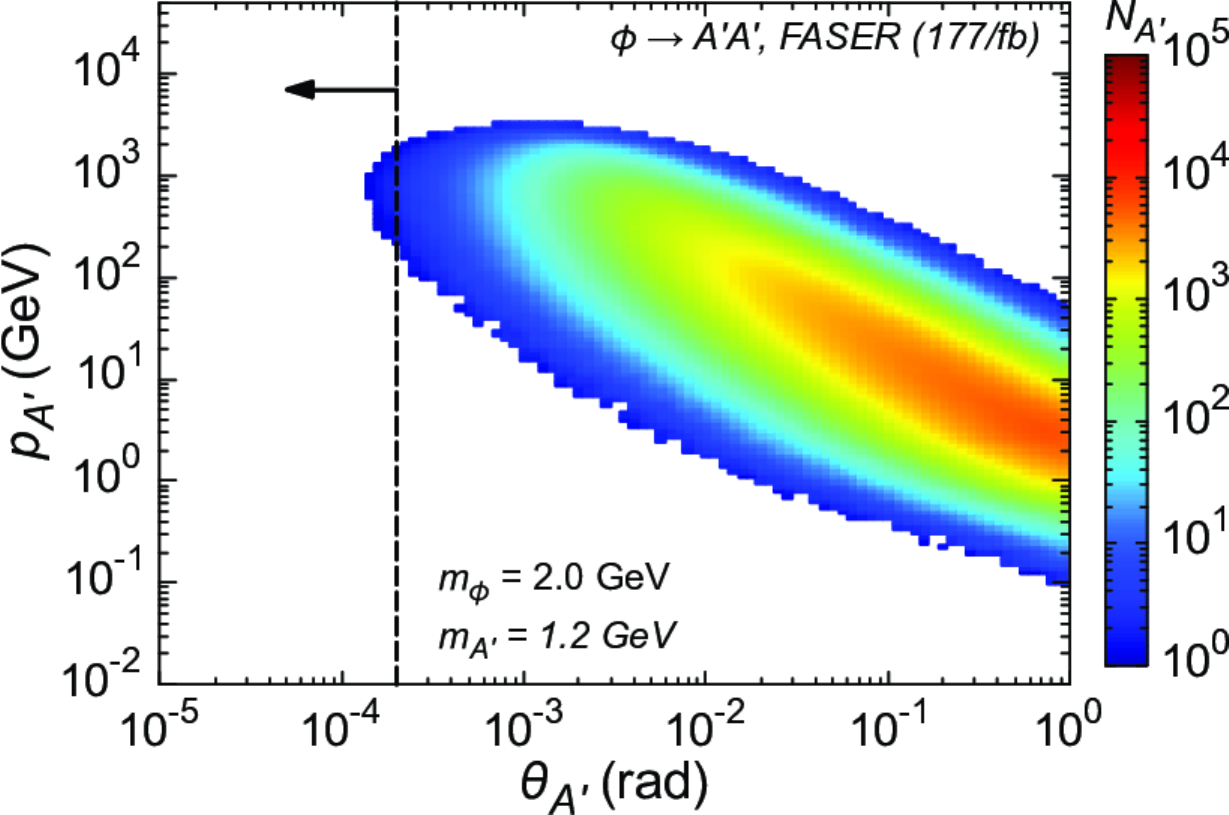} \\
    \includegraphics[scale=0.27]{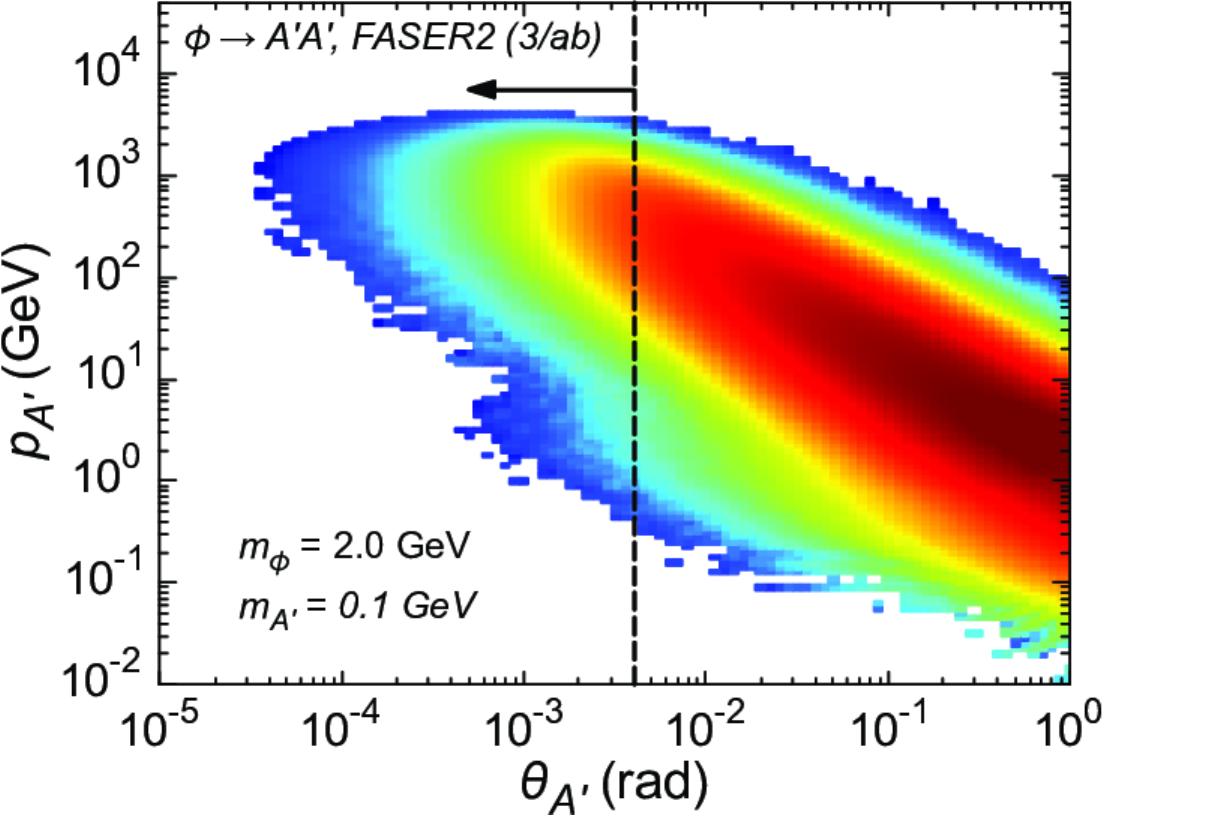} &
    \includegraphics[scale=0.27]{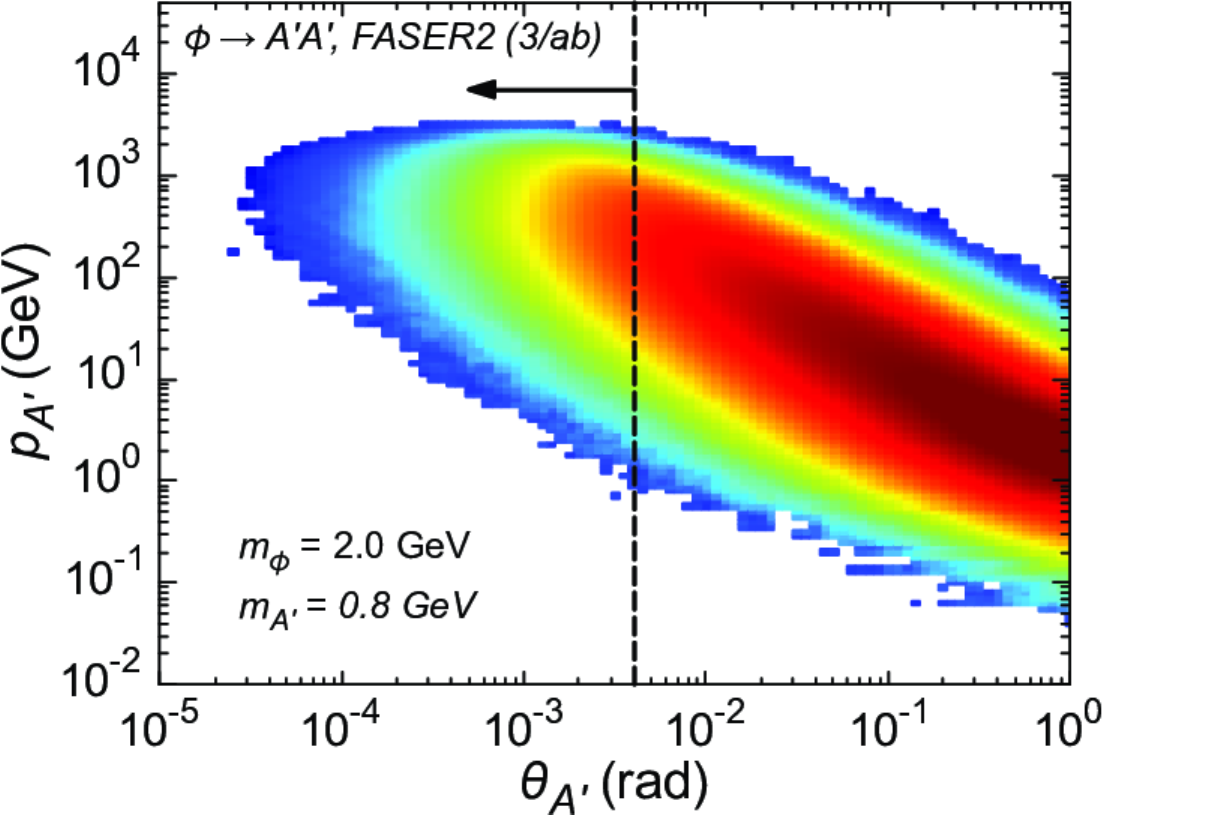} &
    \includegraphics[scale=0.27]{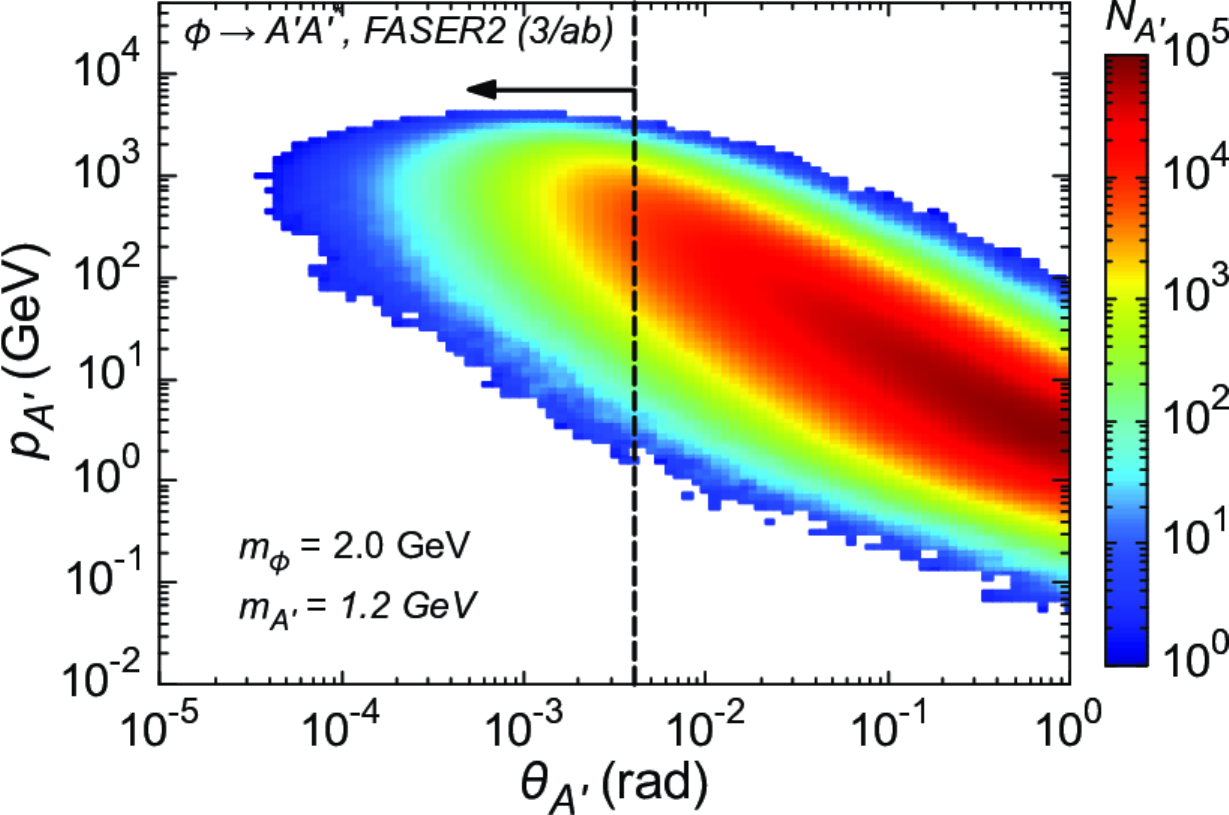} \\
    \includegraphics[scale=0.27]{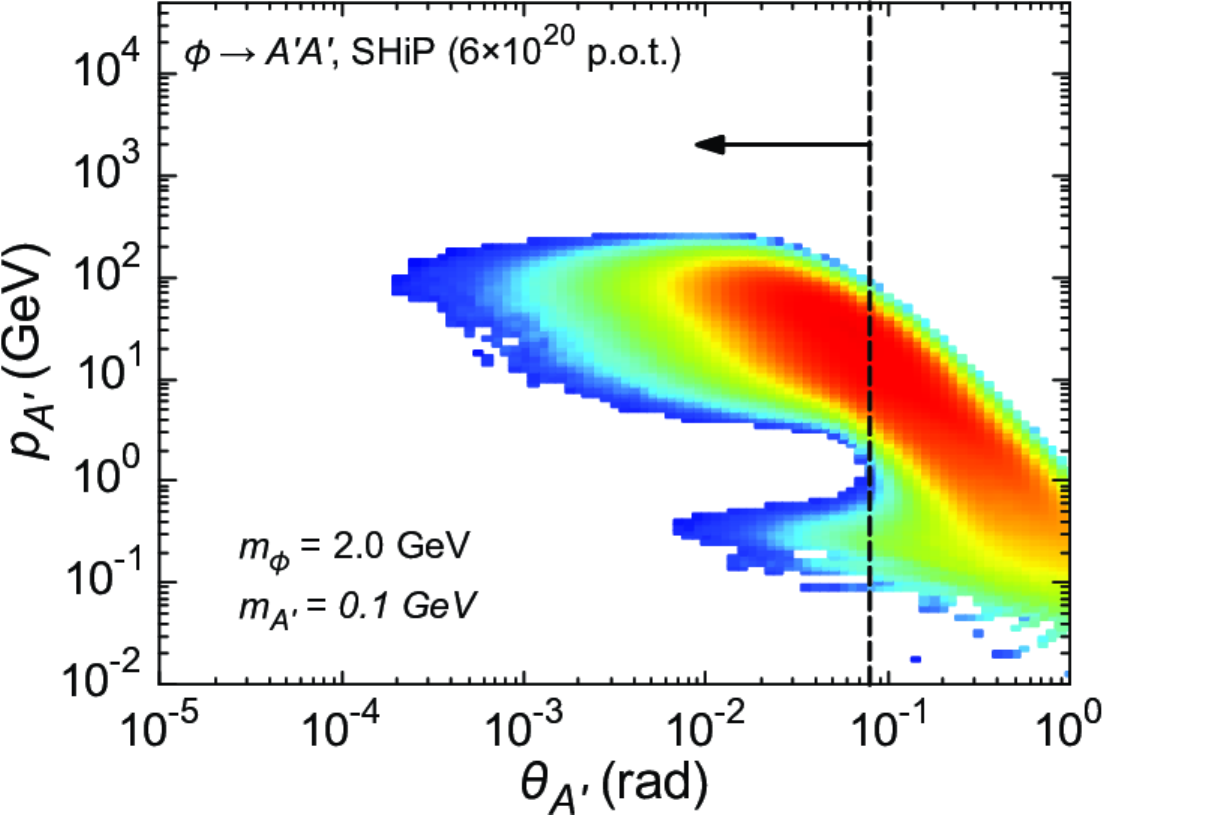}  &
    \includegraphics[scale=0.27]{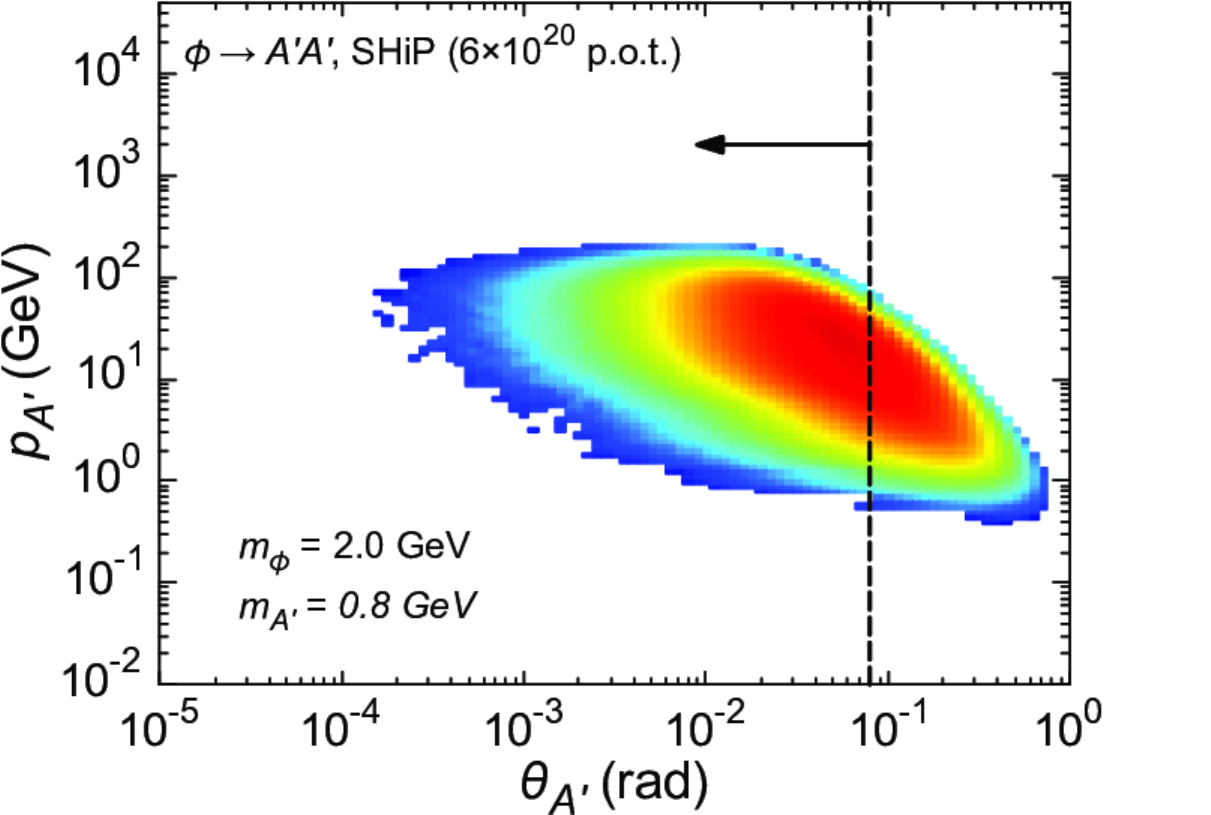}  &
    \includegraphics[scale=0.27]{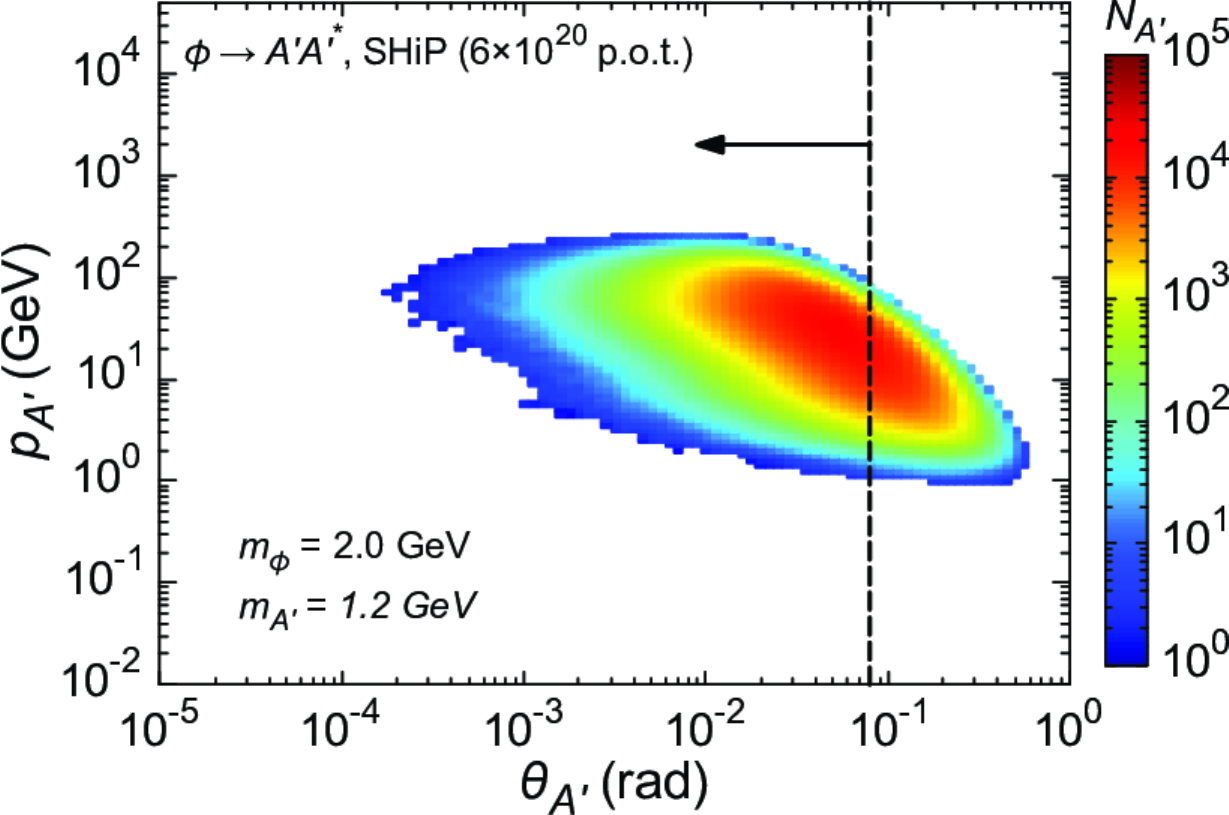} \\
\end{tabular}
    \caption{
    The distribution of the dark photons produced by the on-shell dark Higgs decays at FASER (upper), FASER2 (middle) and SHiP (lower) in $(\theta_{A'},~p_{A'})$ plane. The masses of the dark Higgs and dark photon are indicated in panels. The parameters are fixed to $g' = 0.05$ and $\alpha=10^{-3}$ for FASER, and $g' = 0.5$ and $\alpha=10^{-4}$ for FASER2 and SHiP, respectively. The kinetic mixing is taken to $\epsilon=10^{-4}$ for the off-shell dark photon decay.
    }
\label{fig:dark-photon-dist}
\end{figure}

\begin{table}[t]
\centering
\begin{tabular}{|c|c|c|c|c|c|c|c|} \hline
 \hspace{2cm} & ~~$L_{\rm min}$~~ & ~~$L_{\rm max}$~~ & ~~$R$~~ & ~~~~ & ~~~~ & ~~~~ & ~~$\mathcal{L}$/$N_{\mathrm{pot}}$~~ \\ \hline 
 FASER  & 478.5 & 480 & 0.1 & ~~~ &  ~~~ & ~~~ & 177\,fb$^{-1}$ \\ \hline \hline
 \hspace{2cm} & ~~$L_{\rm min}$~~ & ~~$L_{\rm max}$~~ & ~~$W_{\mathrm{min}}$~~ & ~~$W_{\mathrm{max}}$~~ & ~~$H_{\mathrm{min}}$~~ & ~~$H_{\mathrm{max}}$~~ & ~~$\mathcal{L}$/$N_{\mathrm{pot}}$~~ \\ \hline 
 FASER2  & 650 & 660 & 2.6 & 2.6 & 1.0 & 1.0 & 3\,ab$^{-1}$ \\ \hline
 SHiP     & 33.5 & 83.5 & 1.0 & 4.0 & 2.7 & 6.0 & $6 \times 10^{20}$ \\ \hline
\end{tabular}
\caption{
Dimensions of the FASER, FASER2~\cite{Adhikary:2024nlv,Feng:2025adw,FPF:2025bor}, and SHiP~\cite{SHiP:2015vad,SHiP:2020vbd,ShipECN3} detectors and the integrated luminosity ($\mathcal{L}$)/proton on target ($N_{\mathrm{pot}}$) used in this study. 
$L_{\rm min}$ and $L_{\rm max}$ are the distances to the front and rear end of the detector from the interaction point, respectively. $R$ is the radius of cylindrical shape of the FASER detector while $W_{\mathrm{min/max}}$ and  $H_{\mathrm{min/max}}$ are the width and height of the rectangular detectors for FASER2 and SHiP at the front/rear face, respectively. The distance and length in this figure are in meters.
}
\label{tab:detector-dimension}
\end{table}

\begin{table}[t]
\centering
\begin{tabular}{|c|c|c|c|} \hline
 \hspace{2cm} & ~~$0.1$\,GeV~~ & ~~$0.8$\,GeV~~ & ~~$1.2$\,GeV~~ \\ \hline 
 FASER & ~~$2.4 \times 10^4~(4.9 \times 10^8)$~~ & ~~$3.7 \times 10^4~(4.9 \times 10^8)$~~ & ~~$2.3 \times 10^2~(3.0 \times 10^6)$~~ \\ \hline
 FASER2  & ~~$1.1 \times 10^6~ (8.4 \times 10^7)$~~ & ~~$1.6 \times 10^6~ (8.4 \times 10^7)$~~ & ~~$8.4 \times 10^5~ (4.1 \times 10^7)$~~ \\ \hline
 SHiP     & ~~$1.4 \times 10^6~ (2.8 \times 10^6)$~~ & ~~$2.0 \times 10^6~ (2.8 \times 10^6)$~~ & ~~$1.1 \times 10^6~ (1.4 \times 10^6)$~~ \\ \hline
\end{tabular}
\caption{
The number of dark photons produced from the on-shell dark Higgs decays within the angle coverage shown in figure~\ref{fig:dark-photon-dist}. The number in parenthesis is the total number of the dark photon produced via the dark Higgs boson decays.
}
\label{tab:dp-numbers}
\end{table}

Figure~\ref{fig:dark-photon-dist} shows the distributions of the dark photon in $(\theta_{A'},~p_{A'})$ plane, where $\theta_{A'}$ is the angle of dark photon momentum with respect to the beam direction, and $p_{A'}$ is the absolute value of dark photon momentum. The upper, middle, and lower panels represent the distributions at the FASER, FASER2, and SHiP experiments, respectively. The vertical dashed lines indicate the angular coverage of each detectors. The geometry of the detectors in each experiment is shown in table~\ref{tab:detector-dimension}, in which we assume $L_{\rm min}=650$\,m for the FASER2 experiment~\cite{Feng:2025adw}. We employ the integrated luminosity is 177\,fb$^{-1}$ for FASER and 3\,ab$^{-1}$ for FASER2 while the total proton on target (p.o.t.) is $6 \times 10^{20}$ for SHiP. 
In the FASER and FASER2 experiment, the on-shell dark Higgs decay can produce about $10^6$\,-\,$10^8$ dark photons depending on $m_{A'}$. Most of the dark photons concentrate in large angles and low momentum region. 
However, some fraction of dark Higgs boson are very energetic along the beam direction. A substantial number of dark photons from such dark Higgs boson decays are boosted to the detector direction with high momentum. These explanations also hold to the off-shell dark photon decays. 
In the SHiP experiment, the on-shell dark Higgs decay can produce about $10^6$ dark photons. The dark photon distribution is separated into high and low momentum regions in the light dark photon case. The high momentum dark photons originate from those emitted along the detector direction in the dark Higgs boson rest frame. Such dark photons can be more energetic due to the boost. On the other hand, the low momentum dark photons originate from those emitted in the opposite direction to the detector. Since the beam energy in the SHiP experiment is $400$\,GeV, those dark photons remain relatively less energetic even after the boost. 
As the dark photon mass increases, this separation gradually disappears. 

The number of dark photons within the angle coverage of each detector is summarized in table~\ref{tab:dp-numbers}. The numbers in parenthesis are the total dark photons produced from the dark Higgs boson decays.
The total number of dark photons at SHiP is less than that at FASER2. However, the number of dark photons entering the detector is of the same order or larger due to the large angle coverage of the SHiP detector.

\section{Number of signal events}
In our analyses, we define the signal as a decay into a pair of charged particles or photons inside the detector. 
In the FASER, FASER2, and SHiP experiments, charged particles passing through the detectors can be identified, and their momenta can be measured.
For these decays, three production processes of the dark photon or the U(1)$_{B-L}$ gauge boson from the dark Higgs boson decays are considered:(A) $\phi \to A'A'$, (B) $\phi \to A' + \mathrm{SM}$, and (C) $\phi^\ast \to A'A'$, where $A'$ should be replaced with $Z'$ in the U(1)$_{B-L}$ model.
In process (B), we regard a pair of SM particles from the dark Higgs boson decay as the signal. As for the production of the dark Higgs boson, we consider $B$ meson decays, which are assumed to occur at the interaction point. From these dark Higgs bosons, the total number of expected signal events is calculated. In the following, we present the formulae for the signal events in each process.

The expected number of signal events for the production from (A) $\phi \rightarrow A'A'$ is derived in ref.~\cite{Araki:2020wkq} and expressed as
\begin{align}
\label{eq:noe_on}
   N_{\rm A}
   &= \mathcal{L} \int dp_B d\theta_B d\varphi_B
   \frac{d\sigma_{pp \to B}}{dp_B d\theta_B d\varphi_B}~
   {\rm Br}(B \to X_s \phi)~
   {\rm Br}(\phi \to A'_1 A'_2)
   \nonumber \\
   & \hspace{6cm} \times
   {\rm Br}(A' \to {\rm signal})~
   \sum_{i=1}^2 ~
   \mathcal{P}_{A'_i}^{\rm det}~,
\end{align}
where the integrated luminosity $\mathcal{L}$ should be replaced with the proton on target $N_{\rm pot}$ for the SHiP experiment.
The momentum of $B$ mesons is denoted as $p_B$, and the polar and the azimuthal angles of $p_B$ with respect to the beam direction are defined by $\theta_B$ and $\varphi_B$, respectively.
The branching ratio of $B$ meson is given in ref.~\cite{Feng:2017vli}, see also refs.~\cite{Chivukula:1988gp,Grinstein:1988yu}, which is proportional to $\alpha^2$.
As for the decay branching ratio of the gauge boson, we assume ${\rm Br}(A' \to {\rm signal})=1$ for the dark photon model except for the FASER experiment; we set ${\rm Br}(A' \to {\rm signal})={\rm Br}(A' \to e\bar{e})$ when we study the dark photon model at FASER.
For the U(1)$_{B-L}$ model, we assume ${\rm Br}(Z' \to \textrm{signal})=1-{\rm Br}(Z' \to \textrm{invisible})$, where the left- and right-handed neutrinos are regarded as invisible final states.
The probability that the produced dark photons decay inside the detector is written by $\mathcal{P}_{A'_i}^{\rm det}$ for both $A'_1$ and $A'_2$, and we apply the calculation procedure given in ref.~\cite{Araki:2022xqp} to the cases of a rectangular and a pyramidal frustum detector for FASER2 and SHiP, respectively. 

For the production from (B) $\phi \to A' + {\rm SM}_i$, where ${\rm SM}_i$ means possible SM particles, the expected number of signal events is given by
\begin{align}
\label{eq:noe_off-dp}
   N_{\rm B}
   &= \mathcal{L} \int dp_B d\theta_B d\varphi_B
   \frac{d\sigma_{pp \to B}}{dp_B d\theta_B d\varphi_B}~
   {\rm Br}(B \to X_s \phi)
   \nonumber \\
   & \hspace{1cm}
   \times \int^{(m_\phi-m_{A'})^2}_{4m_\textrm{SM}^2} dq^2
   \left[
        \sum_{i} \frac{d}{d q^2}
        {\rm Br}(\phi \to A' + \textrm{SM}_i)~
        {\rm Br}(A' \to {\rm signal})~
        \mathcal{P}_{A'}^{\rm det}(q^2) 
    \right.
   \nonumber \\
   & \hspace{8cm}
   \left. +
        \sum_{j} \frac{d}{d q^2} 
        {\rm Br}(\phi \to A' + \textrm{SM}_j)~
        \mathcal{P}_\phi^{\rm det}
   \right]~,
\end{align}
where $m_\textrm{SM}$ is the mass of the final state SM particle.
In the square bracket, the subscript $i$ runs over all possible final states, and the probability $\mathcal{P}_{A'}^{\rm det}(q^2)$ is calculated in the same manner as that in eq.~\eqref{eq:noe_on}.
The second term is added to take into account ${\rm SM}_j$ as the signal.
The subscript $j$ does not include neutrinos in the case of the U(1)$_{B-L}$ model, and $\mathcal{P}_\phi^{\rm det}$ is the probability that the dark Higgs boson decays inside the detector.
For the FASER detector, we use
\begin{align}
 \mathcal{P}_{\phi}^{\rm det} =
 \left( 
    e^{-\frac{L_{\mathrm{min}}}{\beta\gamma d_\phi}} - e^{-\frac{L_{\mathrm{max}}}{\beta\gamma d_\phi}}
 \right) 
 \Theta 
 \left( 
    R - L_\mathrm{max}~ \frac{\sqrt{p_{\phi, x}^2 + p_{\phi, y}^2}}{p_{\phi, z}} 
 \right)~,
 \label{eq:P_phi_clyd}
\end{align}
while for the FASER2 and the SHiP detector 
\begin{align}
 \mathcal{P}_{\phi}^{\rm det} =
 \left( 
    e^{-\frac{L_{\mathrm{min}}}{\beta\gamma d_\phi}} - e^{-\frac{L_{\mathrm{max}}}{\beta\gamma d_\phi}}
 \right) 
 \Theta_x 
 \left( 
    \frac{H_{\rm max}}{2} - L_\mathrm{max}~ \left| \frac{p_{\phi, x}}{p_{\phi, z}} \right|
 \right)
 \Theta_y 
 \left( 
    \frac{W_{\rm max}}{2} - L_\mathrm{max}~ \left|\frac{p_{\phi, y}}{p_{\phi, z}} \right|
 \right)~.
 \label{eq:P_phi_rect}
\end{align}
The decay length of the dark Higgs boson is denoted as $d_\phi$ with $\beta\gamma$ being the Lorentz factor.
The step functions $\Theta$, $\Theta_x$ and $\Theta_y$ are multiplied to restrict the momentum of a dark Higgs boson, $(p_{\phi, x}$, $p_{\phi, y}$, $p_{\phi, z})$, in order for the dark Higgs boson to pass through the detector. The distance $L$, the radius $R$, the height $H$ and the width $W$ of the detector are given in table~\ref{tab:detector-dimension}. For the SHiP detector, $H_\mathrm{max},~W_\mathrm{max}$, and $L_\mathrm{max}$ in the step functions should be replaced with $H_\mathrm{min},~W_\mathrm{min}$, and $L_\mathrm{min}$, respectively.

Lastly, for the production from (C) $\phi^\ast \rightarrow A'A'$, we follow the calculation discussed in ref.~\cite{Araki:2024uad}; the expected number of signal events is expressed as
\begin{align}
\label{eq:noe_off-dh}
   N_{\rm C}
   &= \mathcal{L} \int dp_B d\theta_B d\varphi_B
   \frac{d\sigma_{pp \to B}}{dp_B d\theta_B d\varphi_B}~
   \int_{4m_{A'}^2}^{m_b^2} d q^2
   \left[ \frac{d}{d q^2} {\rm Br}(B \to X_s A'_1 A'_2) \right]
   \nonumber \\
   & \hspace{6cm} \times
   {\rm Br}(A' \to {\rm signal})~
   \sum_{i=1}^2 ~
   \mathcal{P}_{A'_i}^{\rm det}(q^2)~,
\end{align}
where the expressions of ${\rm Br}(B \to X_s A'_1 A'_2)$ and $\mathcal{P}_{A'_i}^{\rm det}(q^2)$ are given in ref.~\cite{Araki:2024uad}, and $m_b$ is the mass of the b-quark.

\section{Exclusion limit and expected sensitivity} \label{sec:numerical-results}
We now present numerical results of the exclusion limit at FASER and the expected number of signal events at the FASER2 and the SHiP experiment. 
The dimensions of the detectors, the luminosity, and the proton on target are given in table~\ref{tab:detector-dimension}. 
In this study, we assume that these experiments are background free and derive 95\% C.L. exclusion regions which correspond to regions where more than 3 events are expected. The discussion on backgrounds at the FASER2 experiment can be found in ref.~\cite{Adhikary:2024nlv}. 
Moreover, in order to avoid unexpected background, we set a lower bound on momentum of the gauge boson: $p>500$ GeV, $100$ GeV, and $40$ GeV for the FASER, the FASER2, and the SHiP experiment, respectively.

Light new particles have been extensively searched for in beam dump, collider, and neutrino experiments as well as in astrophysical observations. Tight constraints on the parameter space of the dark photon $(m_{A'}, \varepsilon)$, the U(1)$_{B-L}$ gauge boson $(m_{Z'}, g_{B-L})$ and the dark Higgs boson $(m_\phi, \alpha)$ have been placed independently so far. Such constraints are summarized in literature (for recent reviews, see refs.~\cite{Fabbrichesi:2020wbt,Feng:2022inv,Ferber:2023iso} and references therein). 
Furthermore, in the dark photon model studied with dark Higgs boson, new constraints from perturbative unitarity and SM Higgs invisible decay have been derived in ref.~\cite{Araki:2024uad}. 
Another new constraints on the dark photon and U(1)$_{B-L}$ gauge boson was derived in ref.~\cite{Araki:2023xgb} using the results of the ND280 detector at the T2K experiment and the recent results from rare meson decays ($B \to K \ell^+ \ell^-,~K \to \pi \nu \bar{\nu}$) in ref.~\cite{Seto:2025mte}. 
We apply these constraints to our results.
%

\subsection{FASER}
\begin{figure}[t]
\begin{tabular}{cc}
\centering
    \includegraphics[scale=0.35]{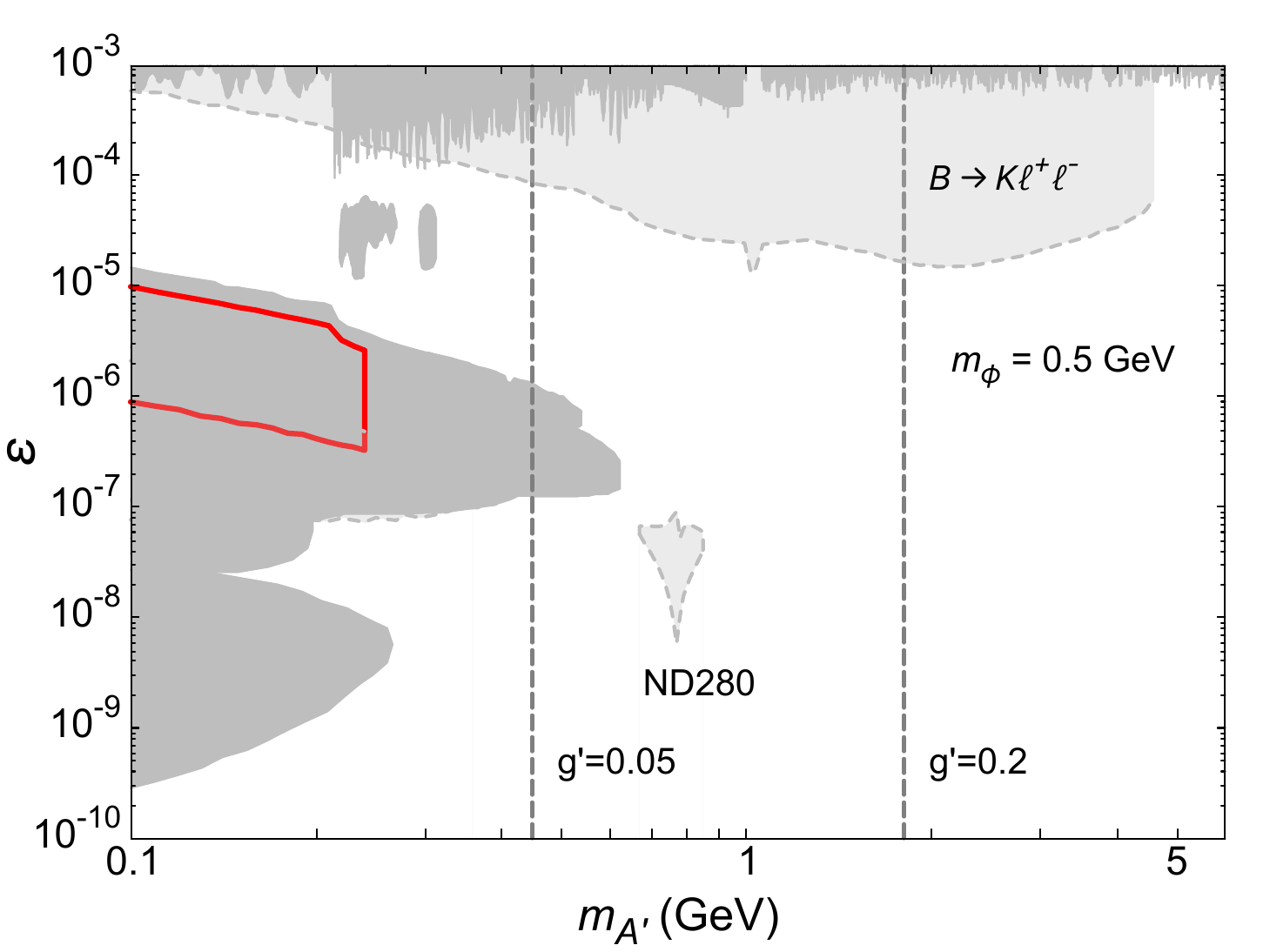} &
    \includegraphics[scale=0.35]{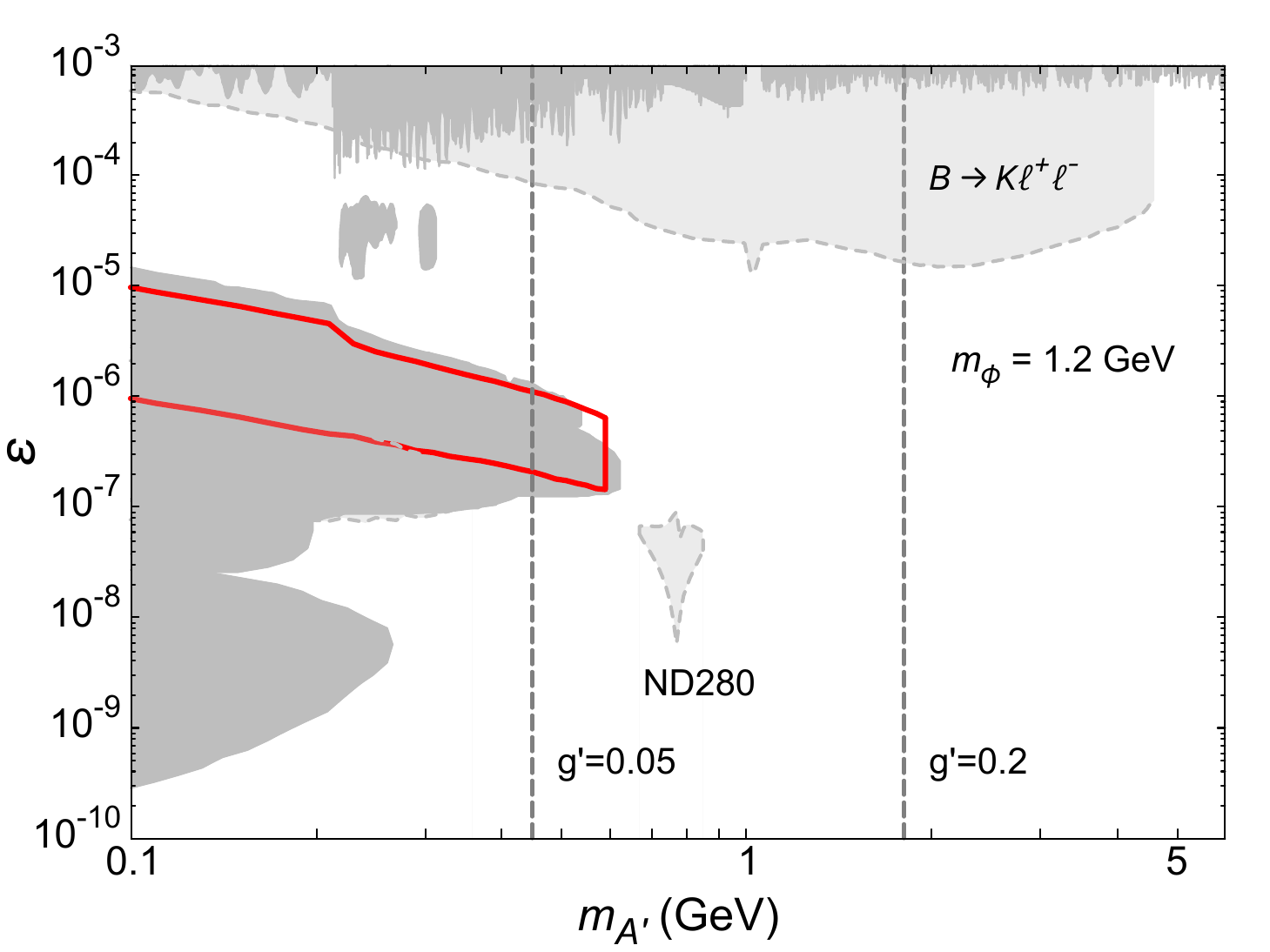} \\
    \includegraphics[scale=0.35]{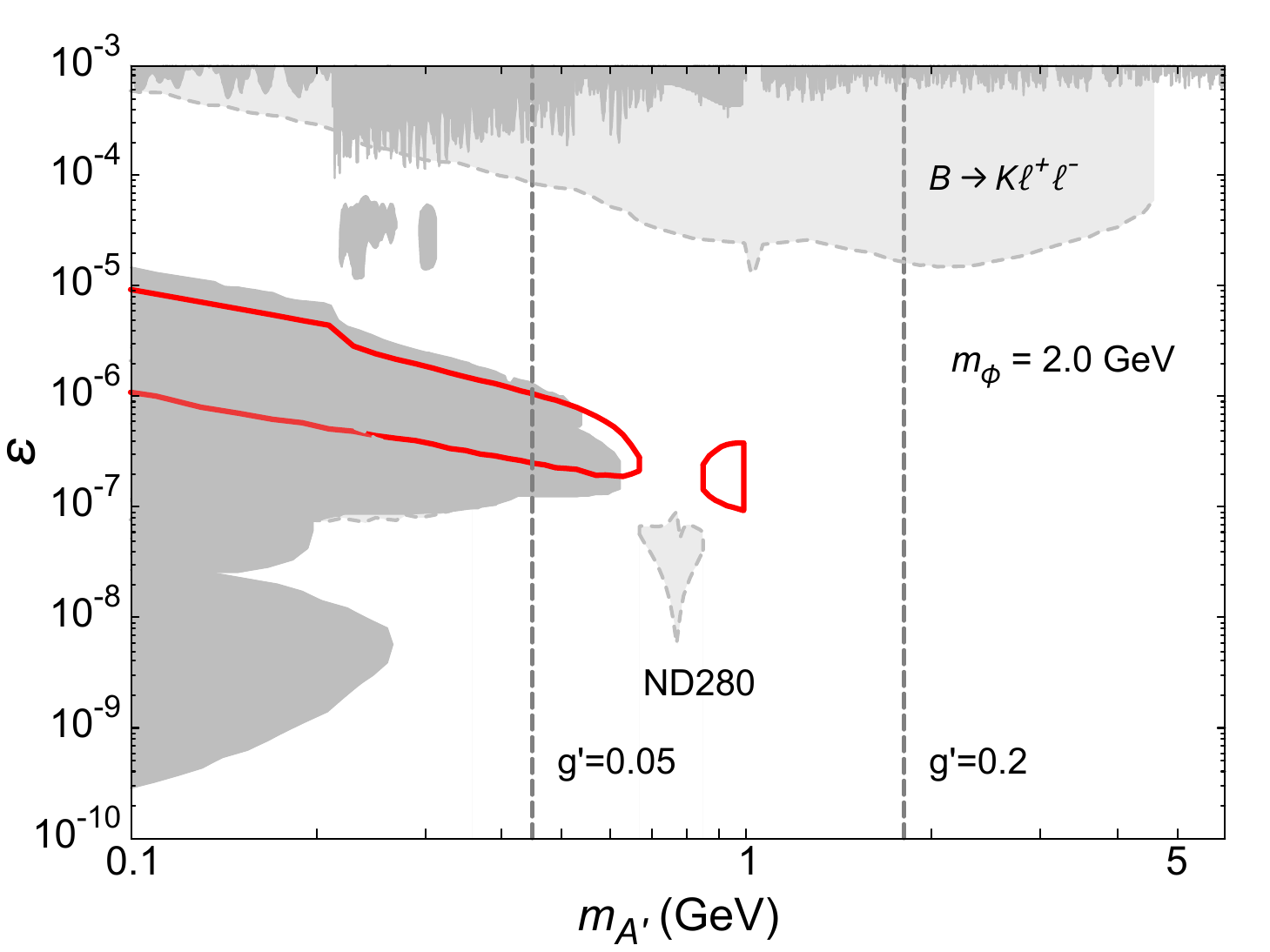} &
    \includegraphics[scale=0.35]{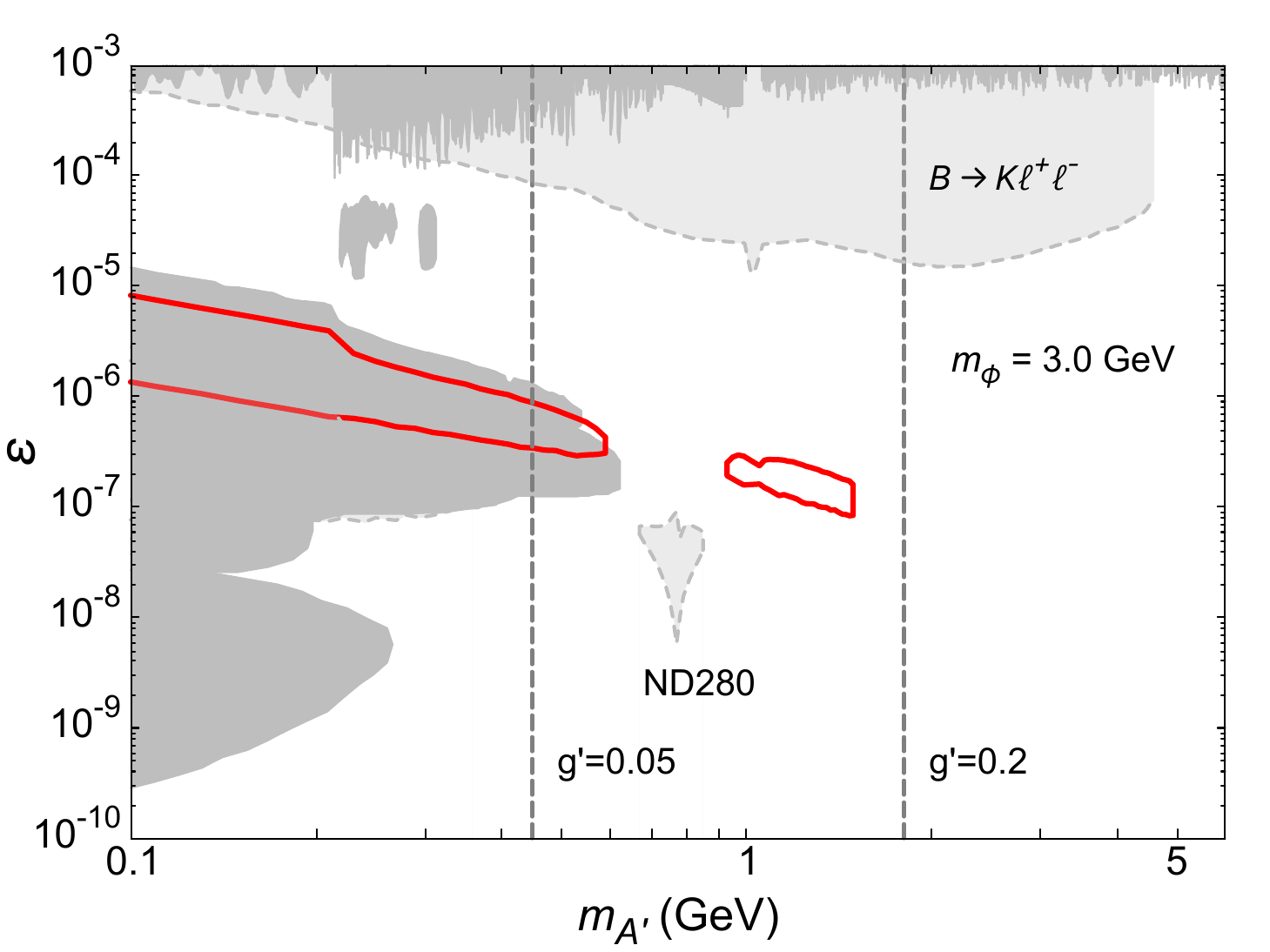}
\end{tabular}
    \caption{
    The $95$\% C.L. excluded regions by FASER for the dark photon decays for $m_\phi=0.5$ (top left), $1.2$ (top right), $2.0$ (bottom left), and $3.0$ (bottom right) GeV. The gauge coupling and the scalar mixing are taken to $g'=0.05$ and $\alpha=10^{-3}$, respectively
    }
\label{fig:sensitivity-faser-DP}
\end{figure}
Firstly, we derive excluded regions for the dark photon model using the latest FASER result with the integrated luminosity $177$ fb$^{-1}$~\cite{moriond}.  
The regions surrounded by the red curves in figure~\ref{fig:sensitivity-faser-DP} are excluded by the null result of the dark photon search. 
The scalar mass is taken to $m_\phi=0.5$ (top-left), $1.2$ (top-right), $2.0$ (bottom-left), and $3.0$ (bottom-right) GeV.
The gauge coupling constant and the scalar mixing angle are fixed to $g' = 0.05$ and $\alpha=10^{-3}$, respectively. 
The left-side of dashed vertical lines are excluded for $g'=0.2$ and $0.05$ as denoted in the figures. 
Dark gray regions are the exclusion limit from the terrestrial experiments taken from the latest version of ref.~\cite{Kling:2021fwx}. 
In the calculations, we take into account only the $e^+ e^-$ final state, that is, ${\rm Br}(A' \to {\rm signal})={\rm Br}(A' \to e^+ e^-)$, and place a lower cut of $500$ GeV on the dark photon momentum.
The excluded regions are obtained solely by the production (A) $\phi \to A'A'$, and there are no contributions from (B) $\phi \to A' + {\rm SM}$ and (C) $\phi^\ast \to A'A'$.
In the bottom panels, there is a gap between 0.6 GeV and 0.9 GeV.
This is because the dark photon resonantly decays into the SM hadrons in that mass region.

We find that a scalar mixing angle of the order of $\mathcal{O}(10^{-3})$ is necessary to impose constraints in the unexplored parameter space.
It should be noted that the dark Higgs boson almost completely decays into a pair of dark photons as long as $g' > 10^{-5}$ and $m_\phi > 2m_{A'}$; the experimental constraints on such an invisibly decaying dark Higgs boson are not so strict, and $\alpha=10^{-3}$ is still allowed \cite{Ferber:2023iso}. 
Given $\alpha = 10^{-3}$, on the other hand, the constraint from the invisible decay of the SM Higgs boson places a strong constraint on $g'$ and $m_{A'}$.
In the figures, we show the lower limits on $m_{A'}$ from the invisible decay for $g' = 0.05$ and $g' = 0.2$.
We note that the limit shifts to the left for a smaller $g'$. 
Note also that the excluded regions remain almost unchanged until $g' > 10^{-5}$ and start to narrow for $g' < 10^{-5}$.
Regarding the dependence on $m_\phi$, we cannot find any new excluded regions for $m_\phi > 3.3$ GeV.

The limit can be evaded in any cases of $\alpha < 10^{-3}$, $g' < 10^{-5}$ or $m_\phi < 2 m_{A'}$. Since small $g'$ results in no signals from the productions (B) and (C), we focus the following analyses on $\alpha < 10^{-3}$ for the FASER2 and SHiP experiments. 

\subsection{FASER2}
\begin{figure}[t]
\centering
    \includegraphics[scale=0.45]{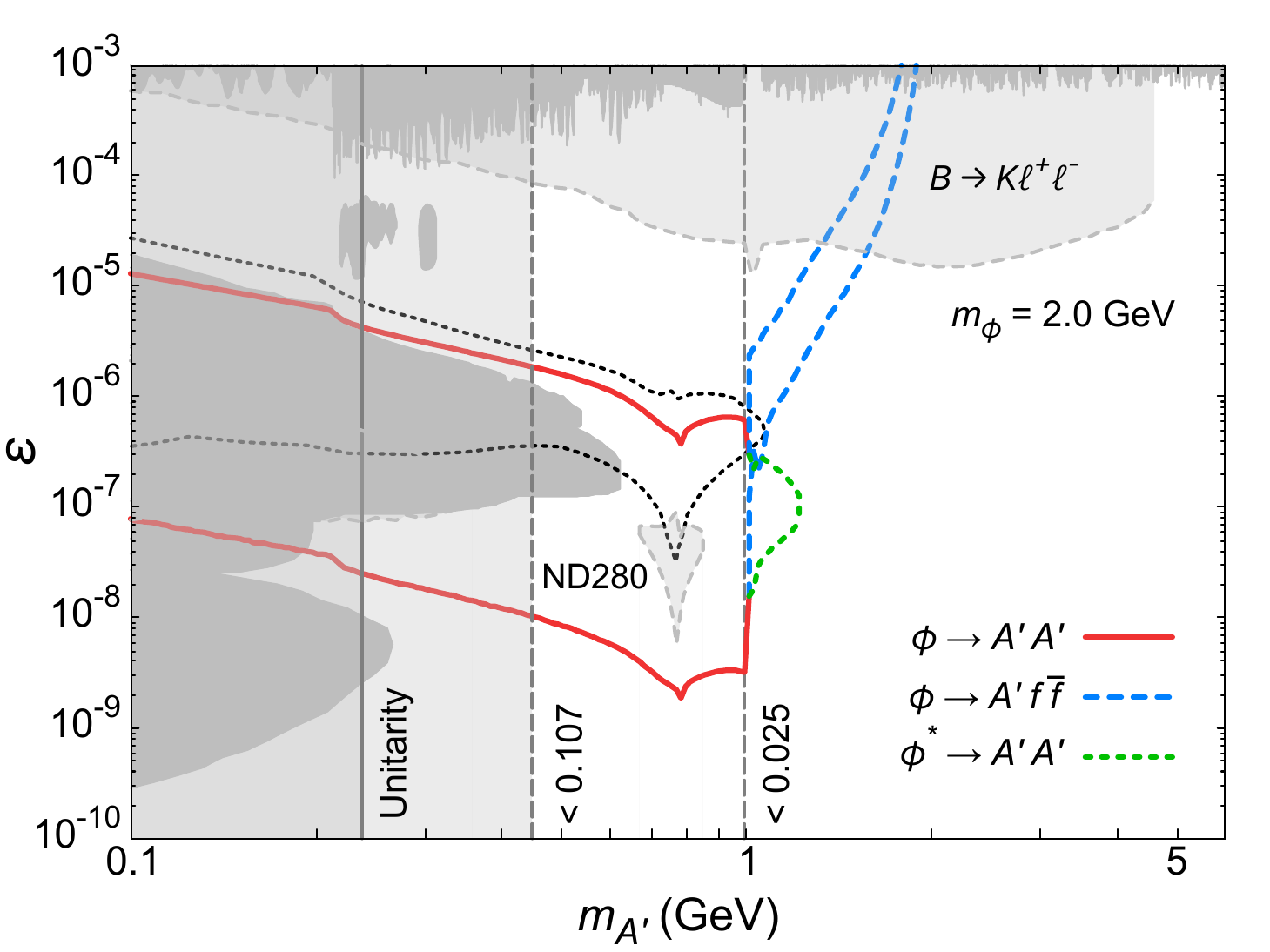} 
    \caption{
    The $95$\% C.L. sensitivity contours at FASER2 for the dark photon decays for $m_\phi=2.0$. The red, blue and green curves represent the sensitivity region from $\phi \to A'A'$, $\phi \to A' + \mathrm{SM}$ and $\phi^\ast \to A'A'$, respectively. The gauge coupling and scalar mixing are taken to $0.5$ and $10^{-4}$, respectively
    }
\label{fig:sensitivity-faser2-DP-separate}
\end{figure}
Figure~\ref{fig:sensitivity-faser2-DP-separate} shows the expected sensitivity region in the dark photon model in $m_{A'}$-$\varepsilon$ plane from each production processes: (A) $\phi \to A'A'$ (red), (B) $\phi \to A' + \mathrm{SM}$ (blue), and (C) $\phi^\ast \to A'A'$ (green). The sensitivity region to the vanilla dark photon for FASER2 is shown by the black dotted curve~\cite{Kling:2021fwx}. Here we have included the latest result from FASER~\cite{moriond}.
The dark Higgs mass and scalar mixing are taken to $m_\phi = 2.0$ GeV and $\alpha = 10^{-4}$, respectively. 
The light gray regions with the vertical solid and dashed line represent the bounds from the perturbative unitarity and the SM Higgs invisible decay Br$(h \to \mathrm{invisible}) < 0.107$, respectively. The future expected limit of the SM Higgs invisible decay at HL-LHC (Br$(h \to \mathrm{invisible}) < 0.025$) is also indicated by the dashed line. The light gray regions bounded by the dashed curves are the exclusion limits from the rare meson decays \cite{Seto:2025mte} and the T2K experiment \cite{Araki:2023xgb} as indicated in the figure. 
For the dark photon model, the excluded regions are derived with eqs.~\eqref{eq:noe_off-dp} and \eqref{eq:noe_off-dh}.
Below the kinematical threshold $m_{A'} < m_\phi/2$, the production from the on-shell decays $\phi \to A'A'$ dominates the signal, while above the threshold the exclusion regions are obtained by $\phi^\ast \to A'A'$ and $\phi \to A' + {\rm SM}$. Comparing with the cylindrical detector \cite{Araki:2024uad}, these processes provide the sensitivity regions to small $\varepsilon$ even for the far and rectangular detector at FASER2. As can be seen from figure~\ref{fig:sensitivity-faser2-DP-separate}, there also exists region that extends upward along $\varepsilon$, which is derived from $\phi \to A' + {\rm SM}$.
As shown in section~\ref{sec:higgs-decay}, the kinetic mixing needs to be $\varepsilon > 10^{-5}$ for $g'=0.5$ in order for $\phi \to A' + {\rm SM}$ to dominate the decay.
Though such a large kinetic mixing results in a short decay length for the dark photon, the dark Higgs boson can travel over long distance in this case.
Even for $\varepsilon \sim 10^{-3}$, some of the dark Higgs bosons can reach the detector.
Nevertheless, $\varepsilon$ cannot arbitrarily be large, since the dark Higgs boson also becomes short lived if $\varepsilon$ is too large.
In figure~\ref{fig:sensitivity-faser2-DP-separate}, an ${\mathcal O}$(1) gauge coupling is assumed to enhance the contributions of $\phi^\ast \to A'A'$ and $\phi \to A' + {\rm SM}$.
A large gauge coupling, on the other hand, tightens the bounds from the invisible decay of the SM Higgs boson and the perturbative unitarity. 

\begin{figure}[t]
\begin{tabular}{cc}
\centering
    \includegraphics[scale=0.35]{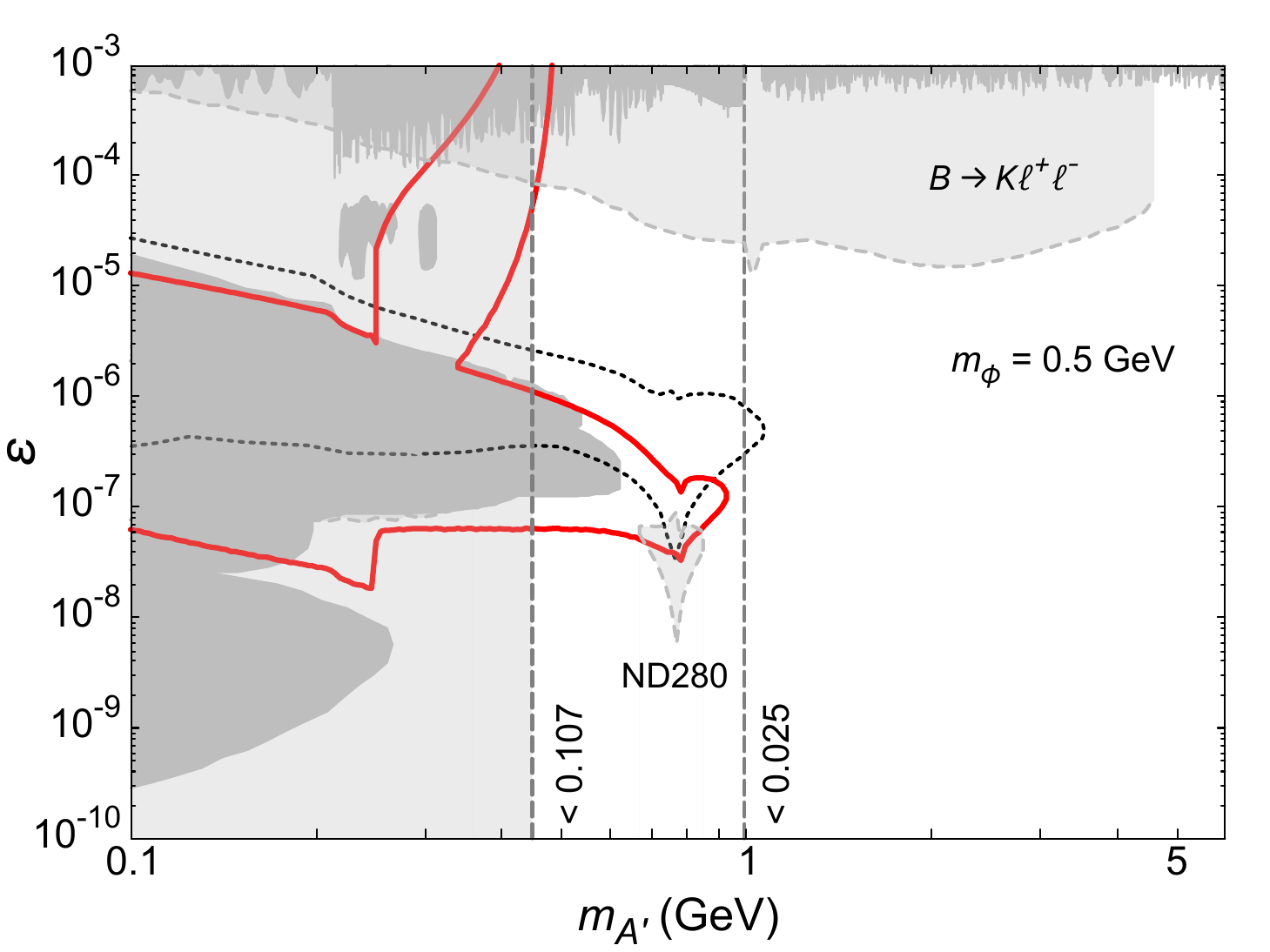} &
    \includegraphics[scale=0.35]{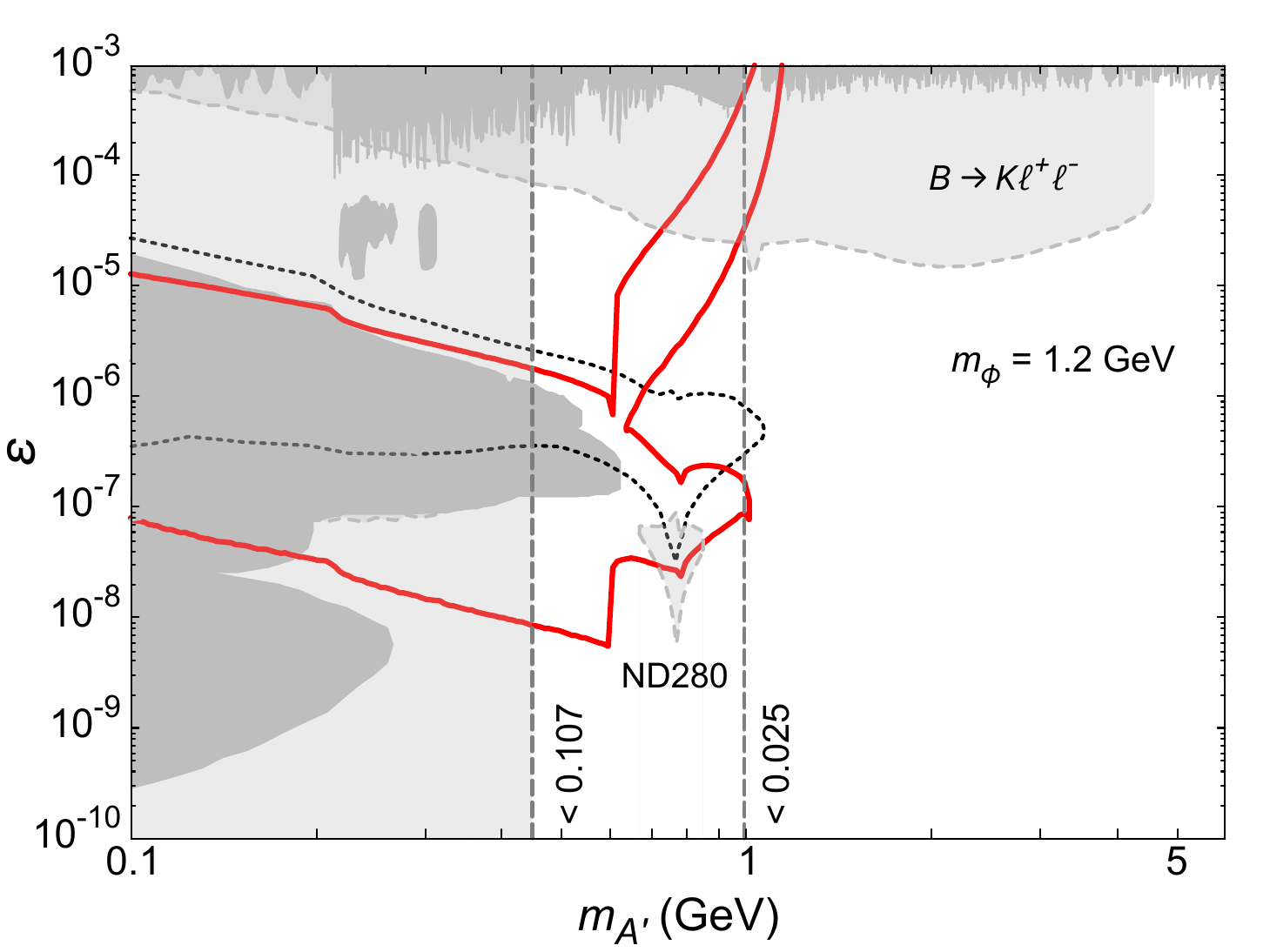} \\
    \includegraphics[scale=0.35]{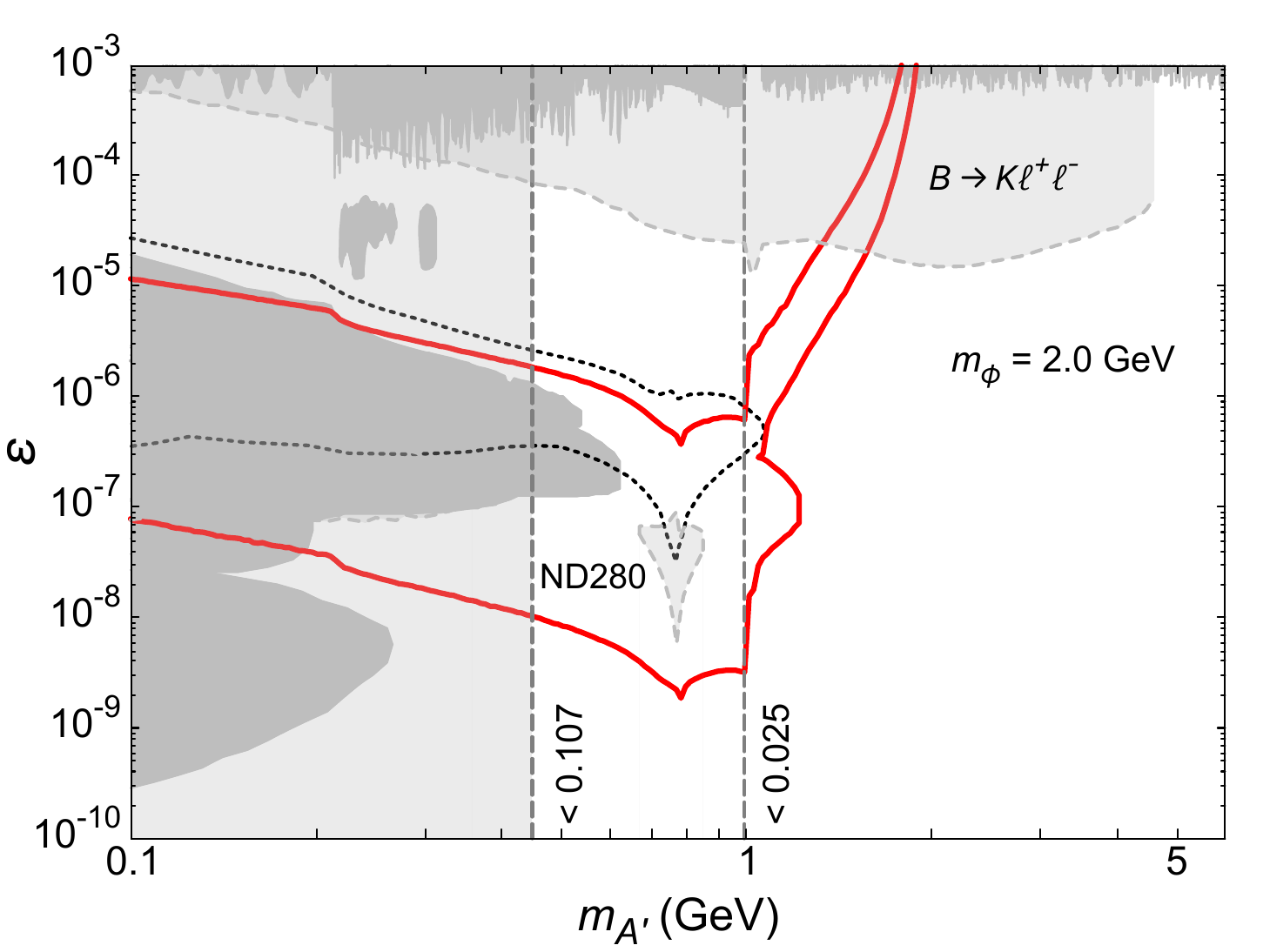} &
    \includegraphics[scale=0.35]{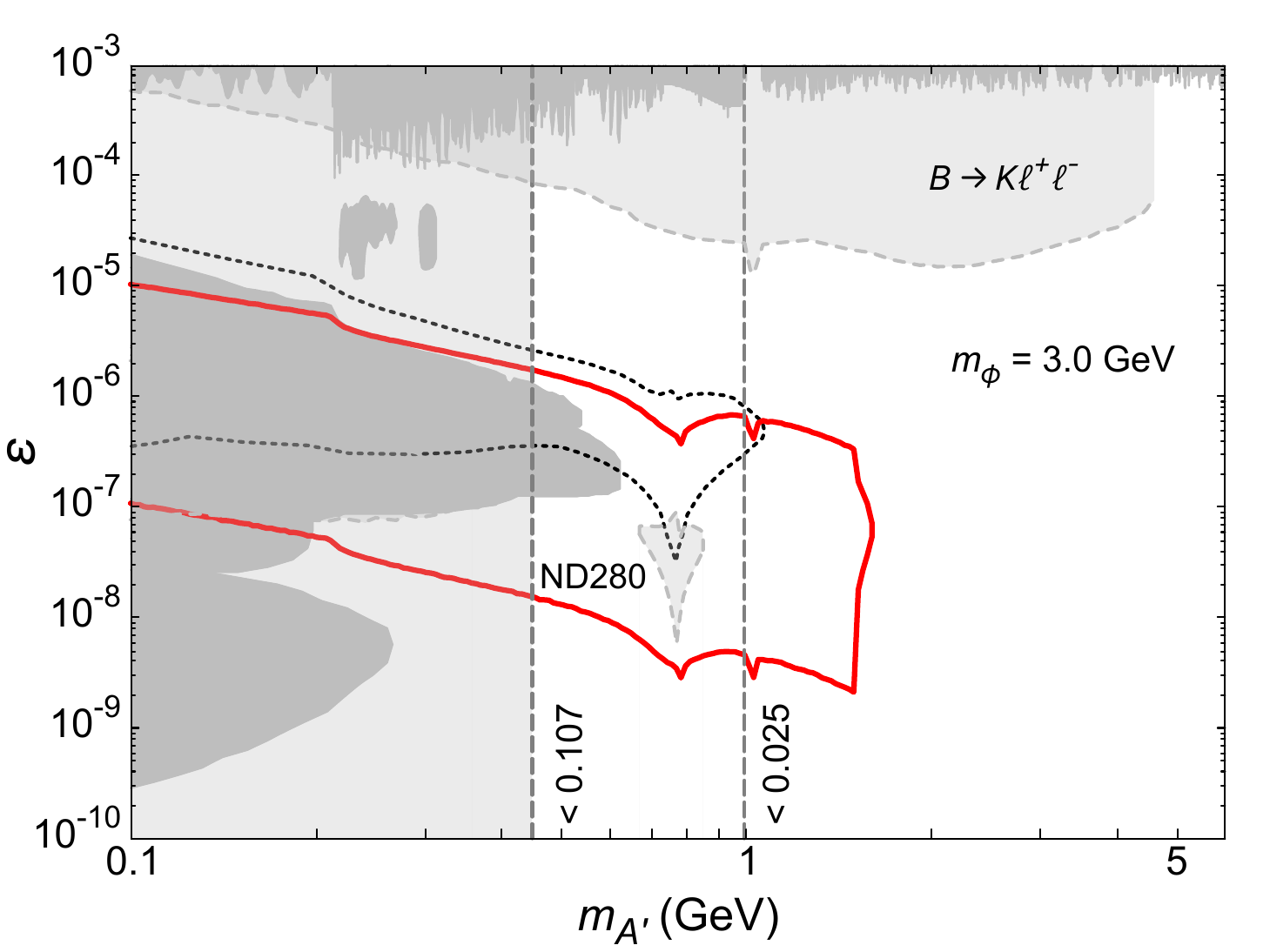}
\end{tabular}
    \caption{
    The $95$\% C.L. sensitivity contours at FASER2 for the dark photon decays for $m_\phi=0.5$ (top left), $1.2$ (top right), $2.0$ (bottom left) and $3.0$ (bottom right) GeV. The gauge coupling and scalar mixing are taken to $g'=0.5$ and $\alpha=10^{-4}$, respectively
    }
\label{fig:sensitivity-faser2-DP}
\end{figure}
\begin{figure}[t]
\begin{tabular}{cc}
\centering
    \includegraphics[scale=0.35]{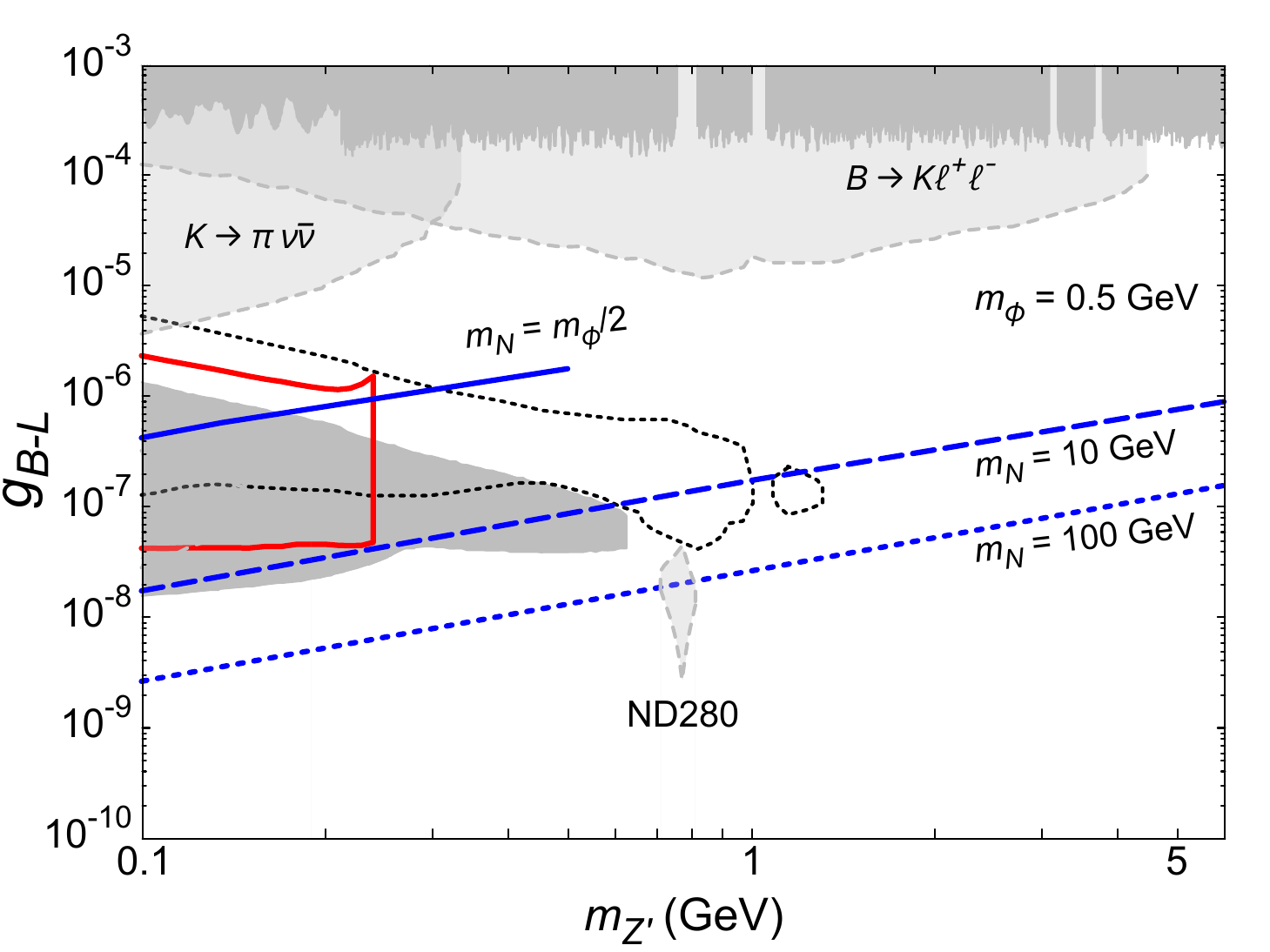} &
    \includegraphics[scale=0.35]{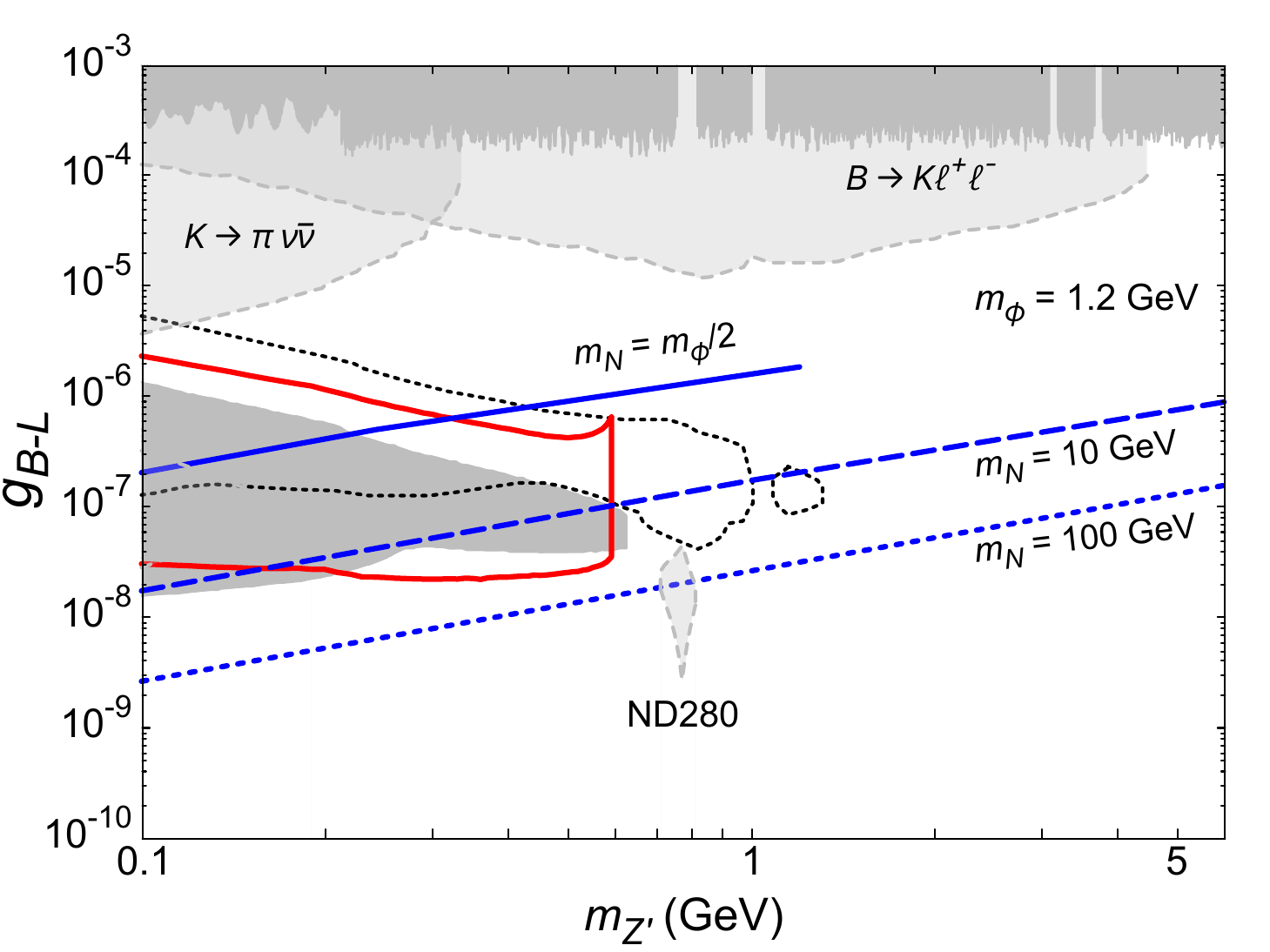} \\
    \includegraphics[scale=0.35]{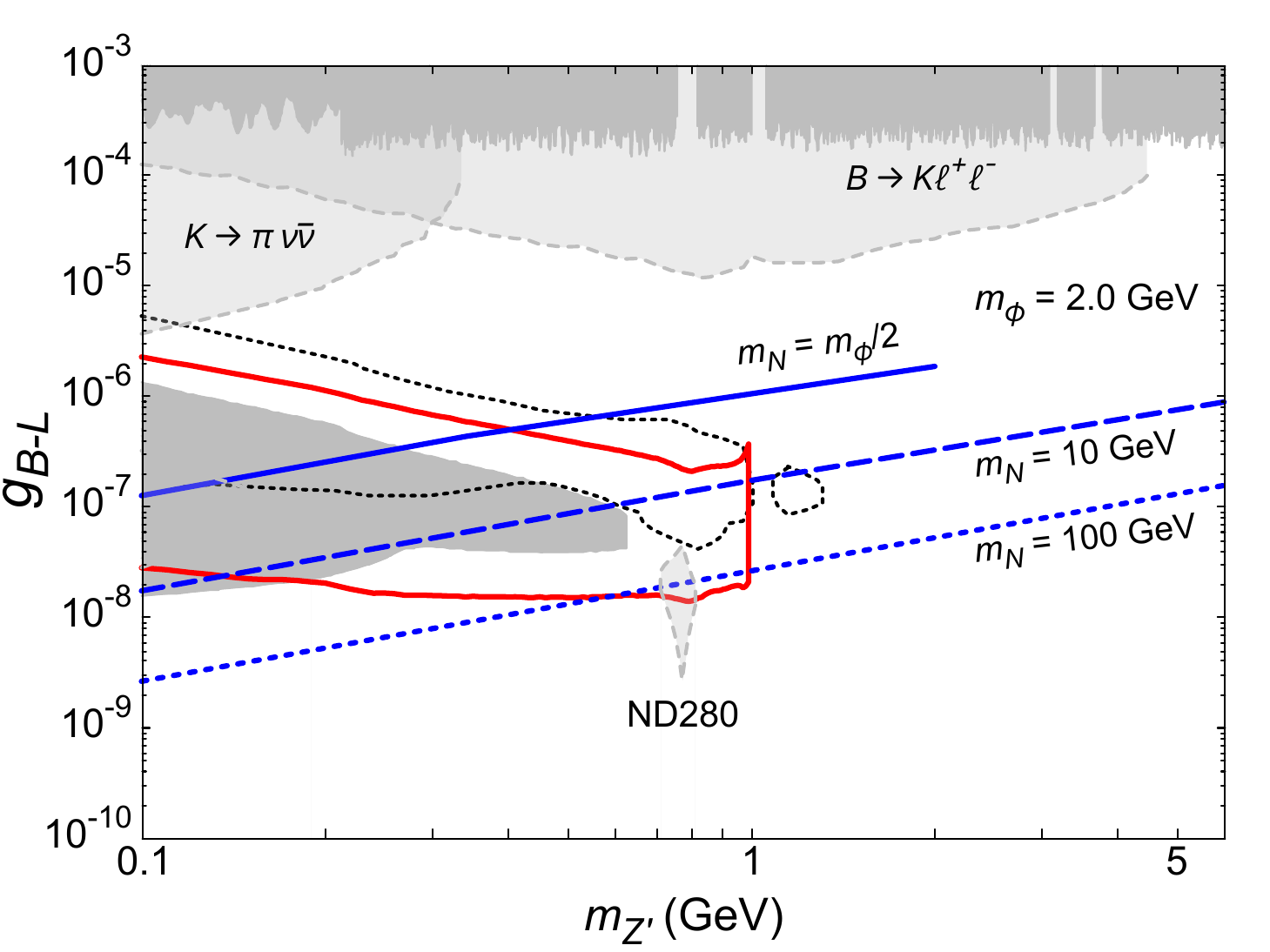} &
    \includegraphics[scale=0.35]{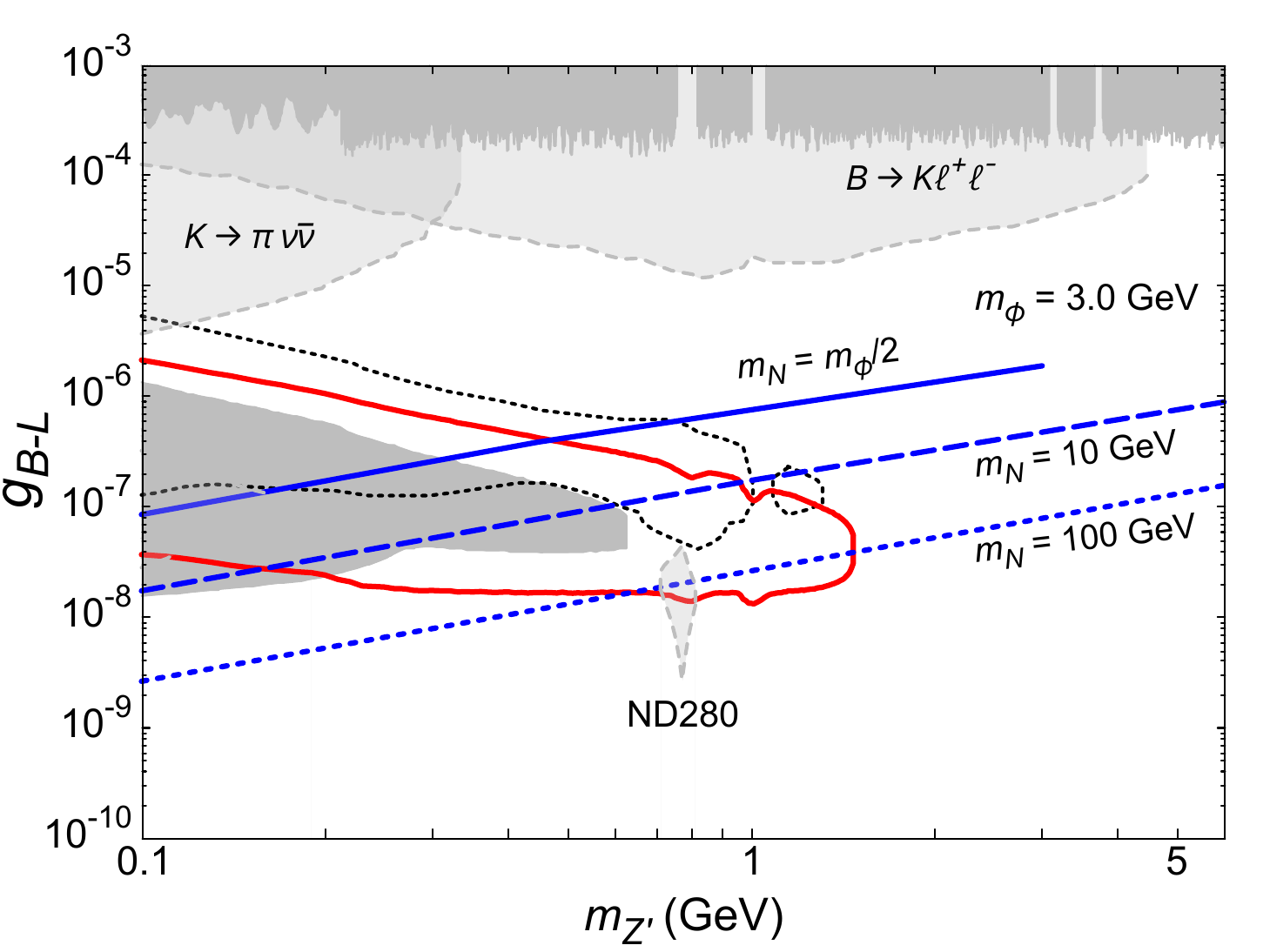}
\end{tabular}
    \caption{
    The $95$\% C.L. sensitivity contours at FASER2 for the U(1)$_{B-L}$ gauge boson decays for $m_\phi=0.5$ (top left), $1.2$ (top right), $2.0$ (bottom left) and $3.0$ (bottom right) GeV. The scalar mixing is taken to $\alpha=10^{-4}$. On the blue solid, dashed and dotted lines, the relic abundance of the DM can be explained. 
    The solid line is shown for $m_{Z'}< 2m_N$.
    }
\label{fig:sensitivity-faser2-BL}
\end{figure}

In figures~\ref{fig:sensitivity-faser2-DP} and \ref{fig:sensitivity-faser2-BL}, we show 95\% C.L. expected sensitivity regions (red solid) for the dark photon and the U(1)$_{B-L}$ model, respectively, at the FASER2 experiment. 
In the top-left, top-right, bottom-left, and bottom-right panels, we set $m_\phi = 0.5$ GeV, $1.2$ GeV, $2.0$ GeV, and $3.0$ GeV, respectively. The scalar mixing and gauge coupling constant are taken as the same values of figure~\ref{fig:sensitivity-faser2-DP-separate}. In the dark photon model, the sensitivity to the vanilla dark photon model is shown by dotted curve \cite{Kling:2021fwx}. 
In the U(1)$_{B-L}$ model, dotted curve represents the sensitivity region obtained by light meson decays and bremsstrahlung process, derived in ref.~\cite{Feng:2022inv}.

In the U(1)$_{B-L}$ model, the expected number of signal events from eqs.~\eqref{eq:noe_off-dp} and \eqref{eq:noe_off-dh} are negligibly small.
This is because the U(1)$_{B-L}$ model contains only the gauge coupling constant, and it is constrained to be $g_{B-L} < 10^{-4} \,\mathchar`-\, 10^{-3}$.
Thus, we use only eq.~\eqref{eq:noe_on} for the U(1)$_{B-L}$ model.
The bounds from perturbative unitarity and invisible Higgs decays are very weak due to the small gauge coupling. Those constrain much lighter mass region than $m_{Z'}=0.1$ GeV. 
As was shown in ref.~\cite{Seto:2024lik}, the sterile neutrino DM, $N$, in the IR freeze-in scenario can be realized in the gauged U(1)$_{B-L}$ model. In the IR freeze-in scenario, the DM production is most efficient at the temperature of the order of the DM mass if the initial states are
thermalized as our $Z'$ boson. In our calculation of the DM relic abundance, production cross sections are evaluated
with the interaction vertices on the the electroweak symmetry breaking vacuum, which is valid for the temperature $\leq \mathcal{O}(100)$ GeV.
Thus, the DM mass should be taken below $100$ GeV in our results.
In figure \ref{fig:sensitivity-faser2-BL}, the blue solid, dashed and dotted lines represent the parameters which reproduce the DM relic abundance for $2m_N > m_{Z'}$ with $m_N=m_\phi /2$ and $10,~100$ GeV, respectively.
In ref.~\cite{Seto:2024lik}, it was pointed out that the freeze-in scenario of the sterile neutrino DM in the gauged U(1)$_{B-L}$ model can be explored by FASER2 and SHiP for $m_\phi < 2m_N$.
With this spectrum, the sterile neutrino DM can be mainly produced via the scatterings of $Z'$, $Z'Z' \to NN$. Such scattering processes can be enhanced by the longitudinal modes of $Z'$. Although $\phi$ cannot produce the sterile neutrinos directly, it contributes to the DM production through the s-channel of the scattering. These facts result in relatively large $g_{B-L}$ depending on $m_\phi$, which can be explored by the long-lived particle search  experiments.
The solid line is truncated at $2m_N = m_{Z'}$, where the DM production channel $Z' \to 2N$ becomes kinematically forbidden.
One can see in the figures that, for $m_\phi \gtrsim 1~\mathrm{GeV}$, the parameter space for $m_N \gtrsim 10~\mathrm{GeV}$ can be probed by FASER2. It should be noticed that the sensitivity region by the dark Higgs boson decays covers small $g_{B-L}$ region, which cannot be reached by the normal $Z'$ productions.

\subsection{SHiP}

\begin{figure}[t]
\begin{tabular}{cc}
\centering
    \includegraphics[scale=0.35]{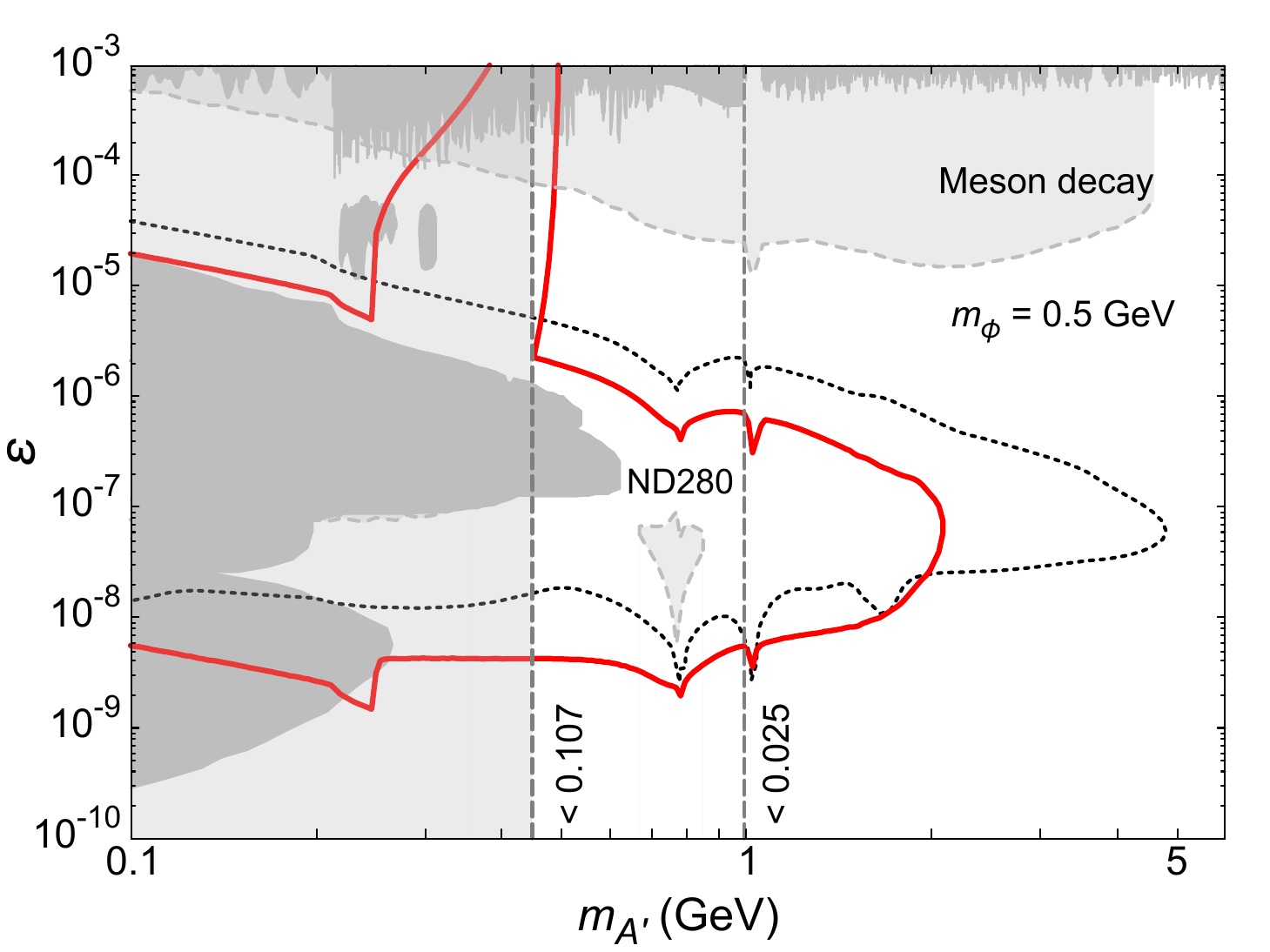} &
    \includegraphics[scale=0.35]{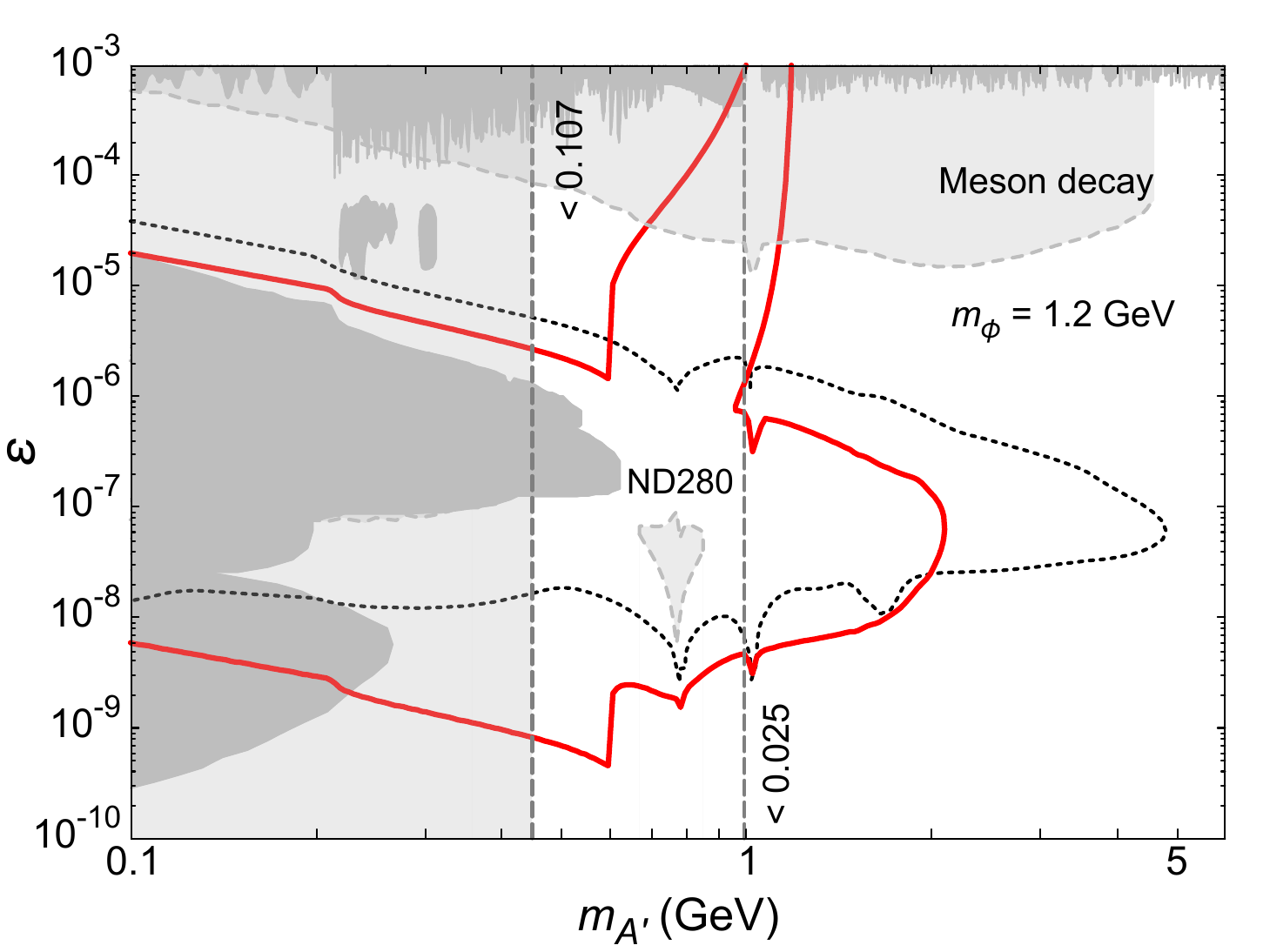} \\
    \centering
    \includegraphics[scale=0.35]{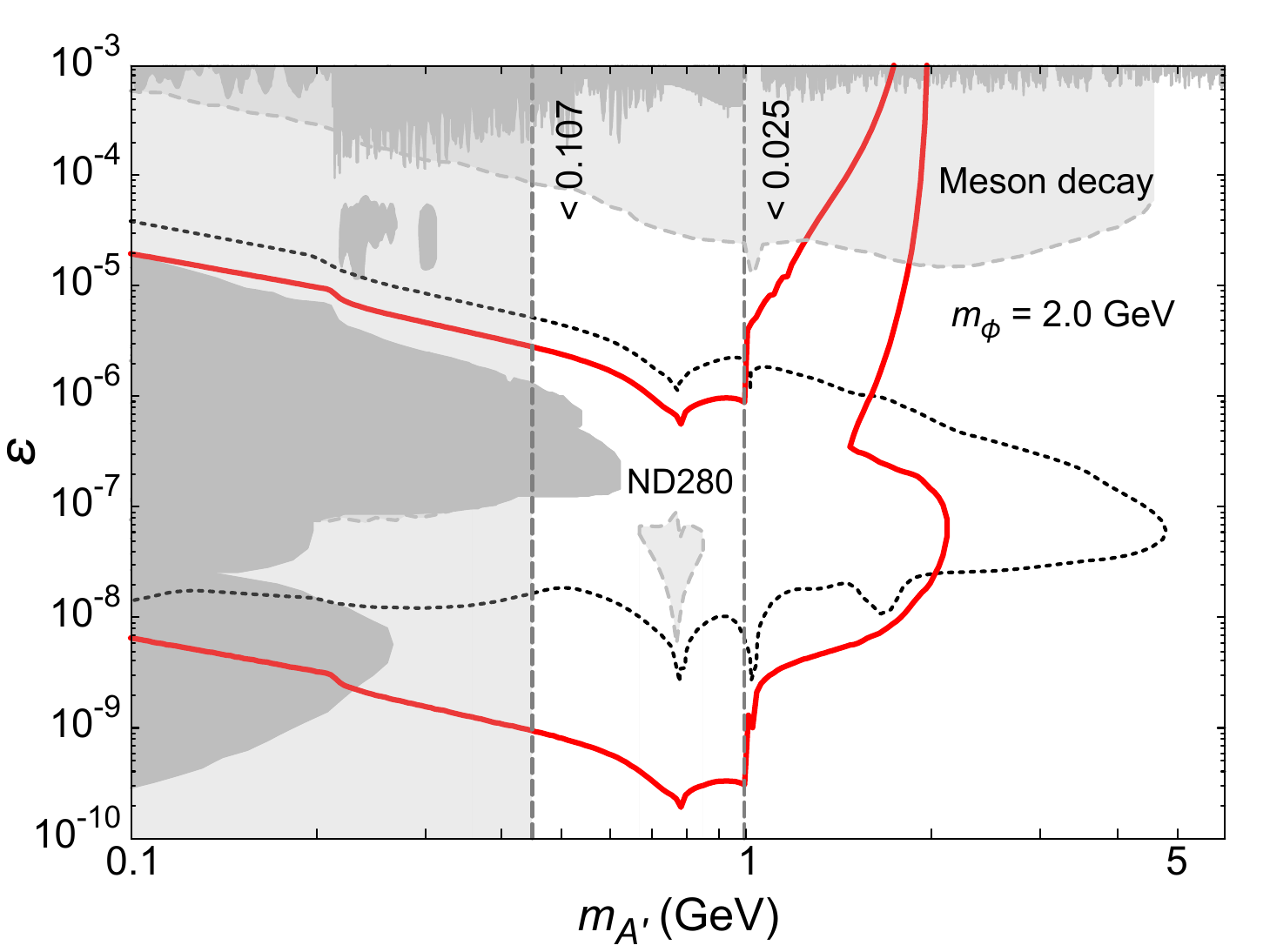} &
    \includegraphics[scale=0.35]{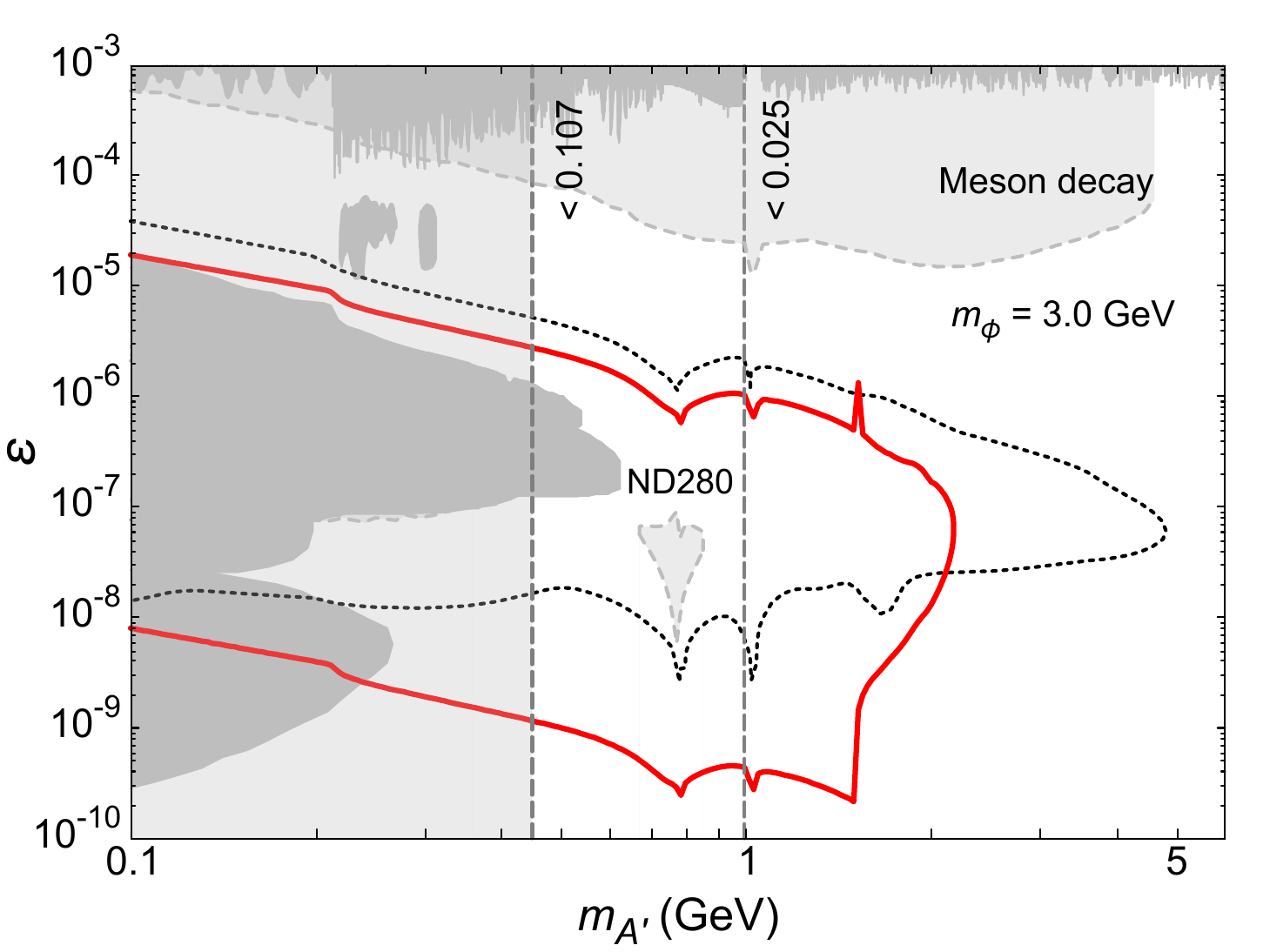}
\end{tabular}
    \caption{
    The $95$\% C.L. sensitivity contours at SHiP for the dark photon decays. The parameters are taken as the same as figure~\ref{fig:sensitivity-faser2-DP}.
    }
\label{fig:sensitivity-ship-DP}
\end{figure}

\begin{figure}[t]
\begin{tabular}{cc}
\centering
    \includegraphics[scale=0.35]{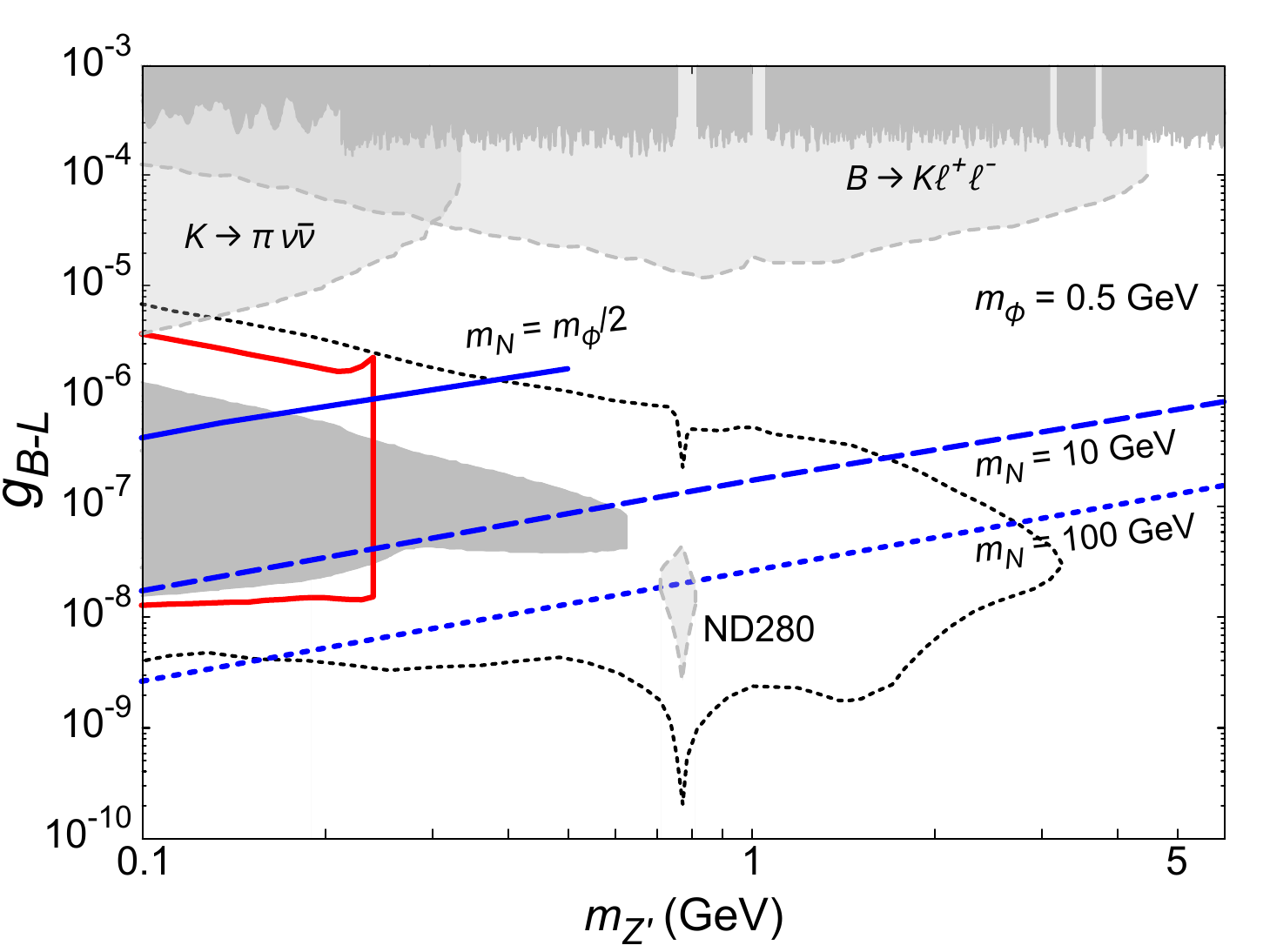} &
    \includegraphics[scale=0.35]{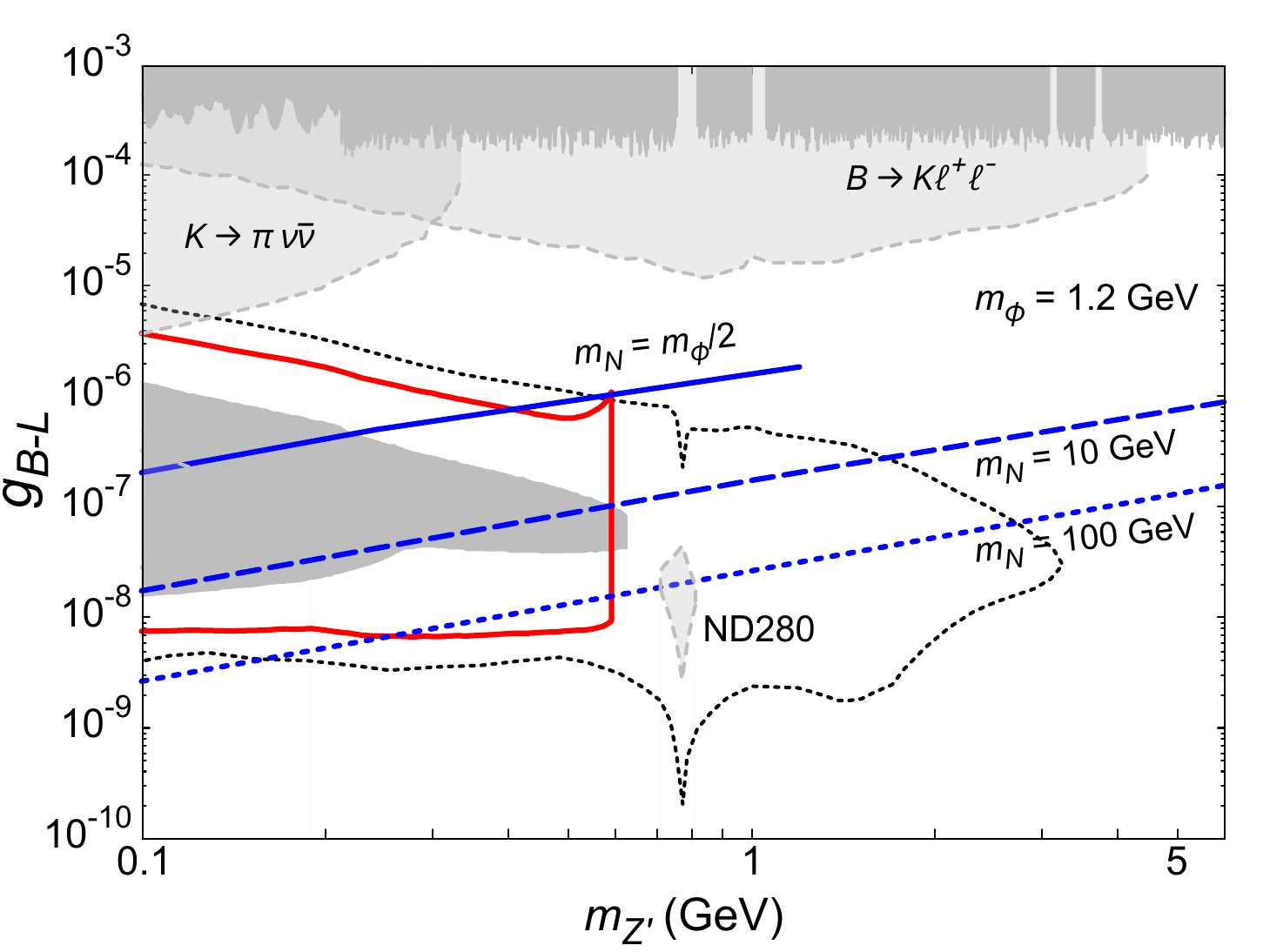} \\
    \centering
    \includegraphics[scale=0.35]{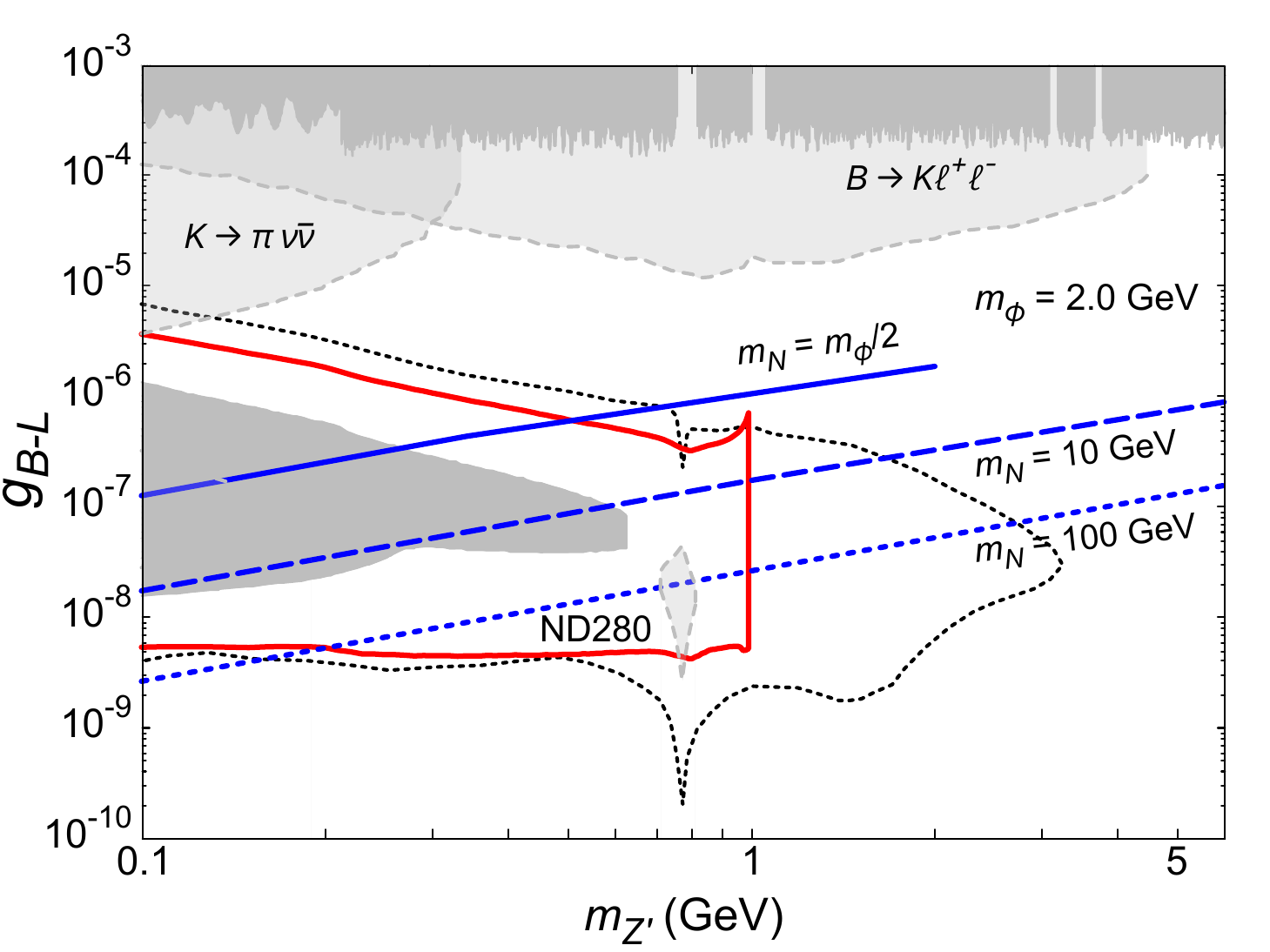} &
    \includegraphics[scale=0.35]{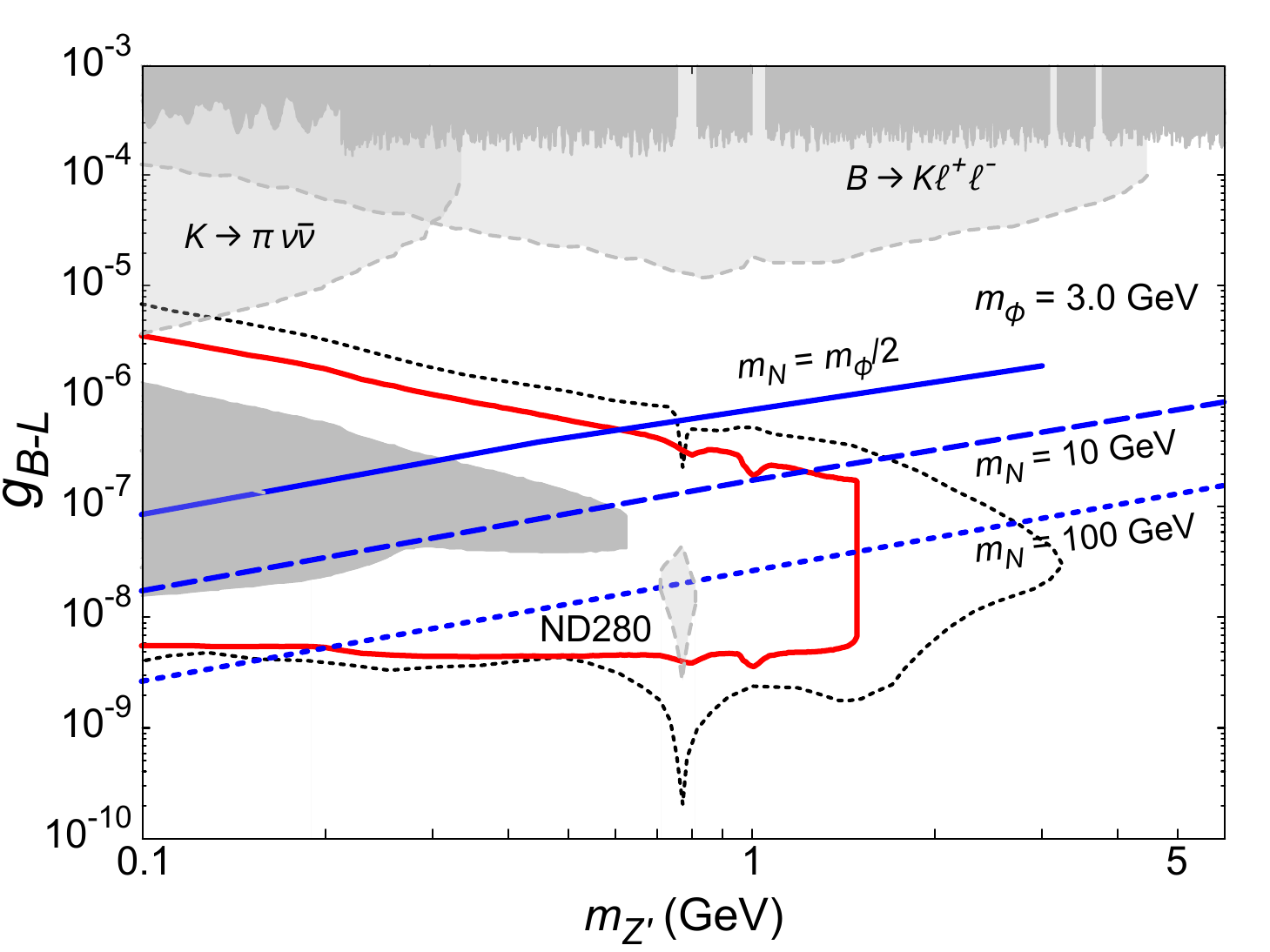}
\end{tabular}
    \caption{
    The $95$\% C.L. sensitivity contours at SHiP for the U(1)$_{B-L}$ gauge boson. The parameters are taken as the same as figure~\ref{fig:sensitivity-faser2-BL}.
    }
\label{fig:sensitivity-ship-BL}
\end{figure}
Figures~\ref{fig:sensitivity-ship-DP} and \ref{fig:sensitivity-ship-BL} are the same plots as figures~\ref{fig:sensitivity-faser2-DP} and \ref{fig:sensitivity-faser2-BL} for the dark photon and the U(1)$_{B-L}$ model, respectively, at the SHiP experiment. 
In both figures, dotted curves represent the sensitivity regions by the light meson decays and the bremsstrahlung production~\cite{ShipECN3}. 
The expected sensitivity region at SHiP covers wider parameter space than that at FASER2 due to its larger decay volume. 
Before closing this section, we comment on the reconstruction efficiency, which is assumed to be $100$\% in our analyses. In ref.~\cite{SHiP:2020vbd}, the reconstruction efficiency having two good tracks from dark photon decays was estimated. It was shown that those for proton bremsstrahlung and Drell-Yan like QCD processes are above $80$\% in most of the parameter space, while that for meson decay process can be below $50$\% as $\varepsilon$ decreases. Estimating the efficiency for the dark photons from the dark Higgs decays is beyond the scope of this paper. Instead, we calculated the sensitivity by assuming the efficiency of $10$\%. We found that the lower edges of the sensitivity regions become narrower, corresponding to values of $\varepsilon$ that are larger by one order of magnitude.

\section{Summary and Discussion}

We have studied the dark Higgs boson decays as sources of dark photon and U(1)$_{B-L}$ gauge bosons at the FASER, FASER2, and SHiP experiments. The on-shell dark Higgs decay into $A' f \bar{f}$ is newly taken into account in addition to both the on-shell and off-shell dark Higgs decays into $A'A'$. 
We derived the exclusion limit for the dark photon from the latest results from the FASER experiment. Then, we showed the expected sensitivity regions to the parameter space of the gauge bosons produced from the dark Higgs boson decays at the future FASER2 and SHiP experiments. 

Applying the latest results from the FASER experiment to the dark photon model, we found that the parameter space shown in figure~\ref{fig:sensitivity-faser-DP} can be excluded for the range of $10^{-5} < g' < 0.05$ with $\alpha > 10^{-3}$ where the dark Higgs dominantly decays into $A'A'$. 

Based on the bounds obtained from FASER, we analyzed the expected sensitivity regions at the FASER2 and SHiP experiments, respectively. For FASER2, we updated our previous analysis by taking into account the decay of $\phi \to A' f \bar{f}$ and the rectangular detector shape. In the dark photon model, it was found that the the single $A'$ production process provided a possibility in exploring larger $\varepsilon$ region. Such regions cannot be probed in the vanilla dark photon model. We also showed that FASER2 can be sensitive to smaller $\varepsilon$ region even for the rectangular detector through the on-shell dark Higgs decays. Similarly in the U(1)$_{B-L}$ model, the small gauge coupling region $g_{B-L} < 10^{-7}$ can be explored through the on-shell dark Higgs decay into $Z'Z'$. Such a parameter region can reproduce the observed relic abundance by the sterile neutrino dark matter in the freeze-in scenario. Thus the FASER2 and SHiP experiments will examine the scenario.

For the SHiP experiment, we derived the sensitivity region for the same parameter space as the FASER2 case in the dark photon and U(1)$_{B-L}$ models. It was shown that wider parameter space of the dark photon and U(1)$_{B-L}$ gauge bosons can be explored.

\section*{Acknowledgements}
This work was partially supported by JSPS KAKENHI Grant Numbers JP24K07024 (T.A.), JP23K13097 (K.A.), JP25KJ0401 (K.A.), JP23K03402 (O.S.), JP22K03622 (T.S.), JP23H01189 (T.A. and T.S.), K2-SPRING program grant number JPMJSP2136 (Y.N.), and National Natural Science Foundation of China Grant
Number NSFC-12347112 (Y.U.).

\appendix
\section{Distribution of $B$ meson and dark Higgs boson}
\begin{figure}[t]
\tabcolsep = -0.1cm
\centering
\begin{tabular}{ccc}
    \includegraphics[scale=0.27]{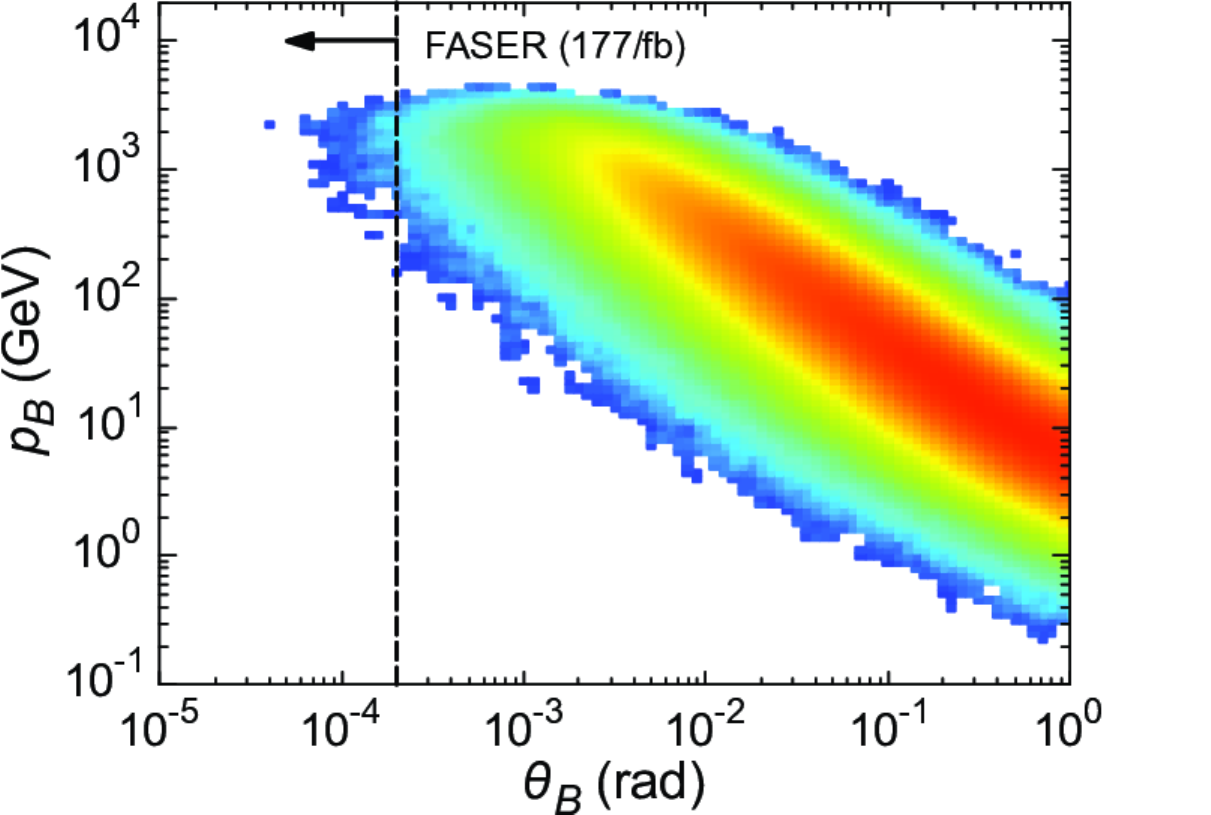} &
    \includegraphics[scale=0.27]{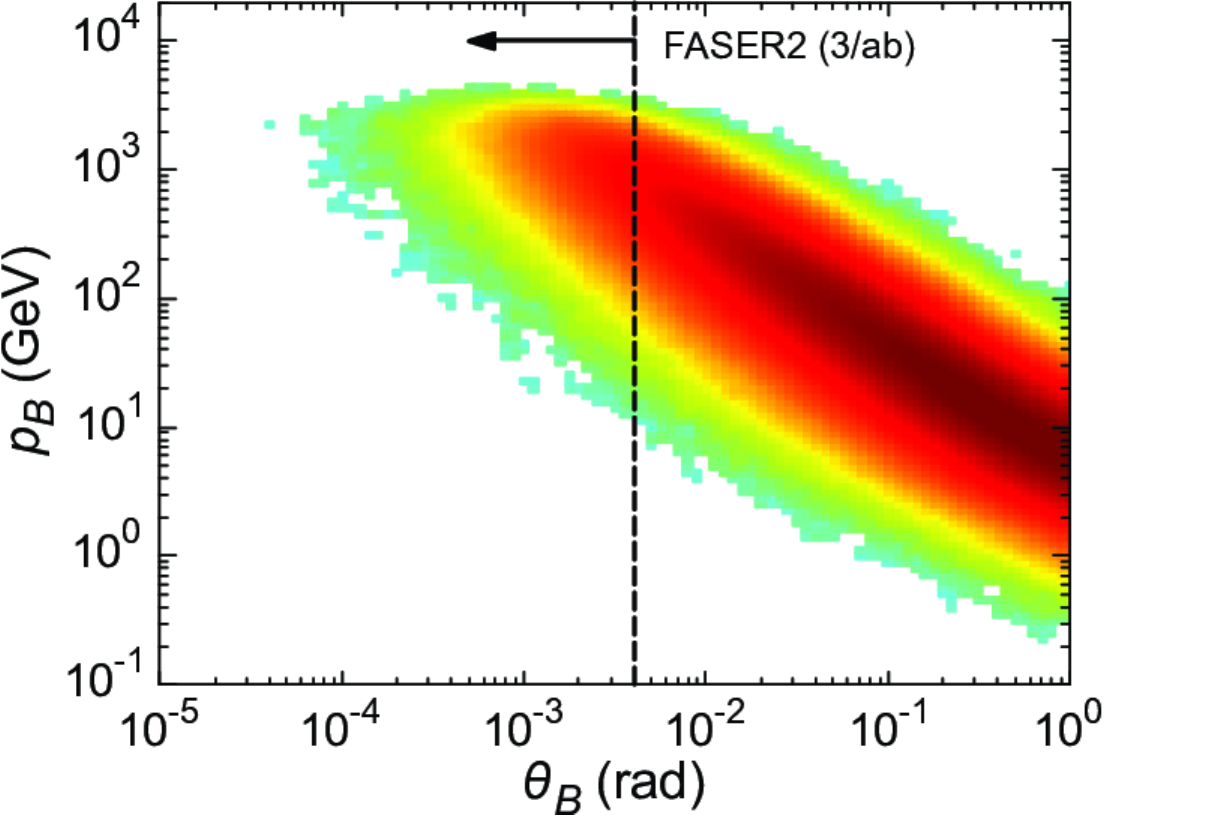} &
    \includegraphics[scale=0.27]{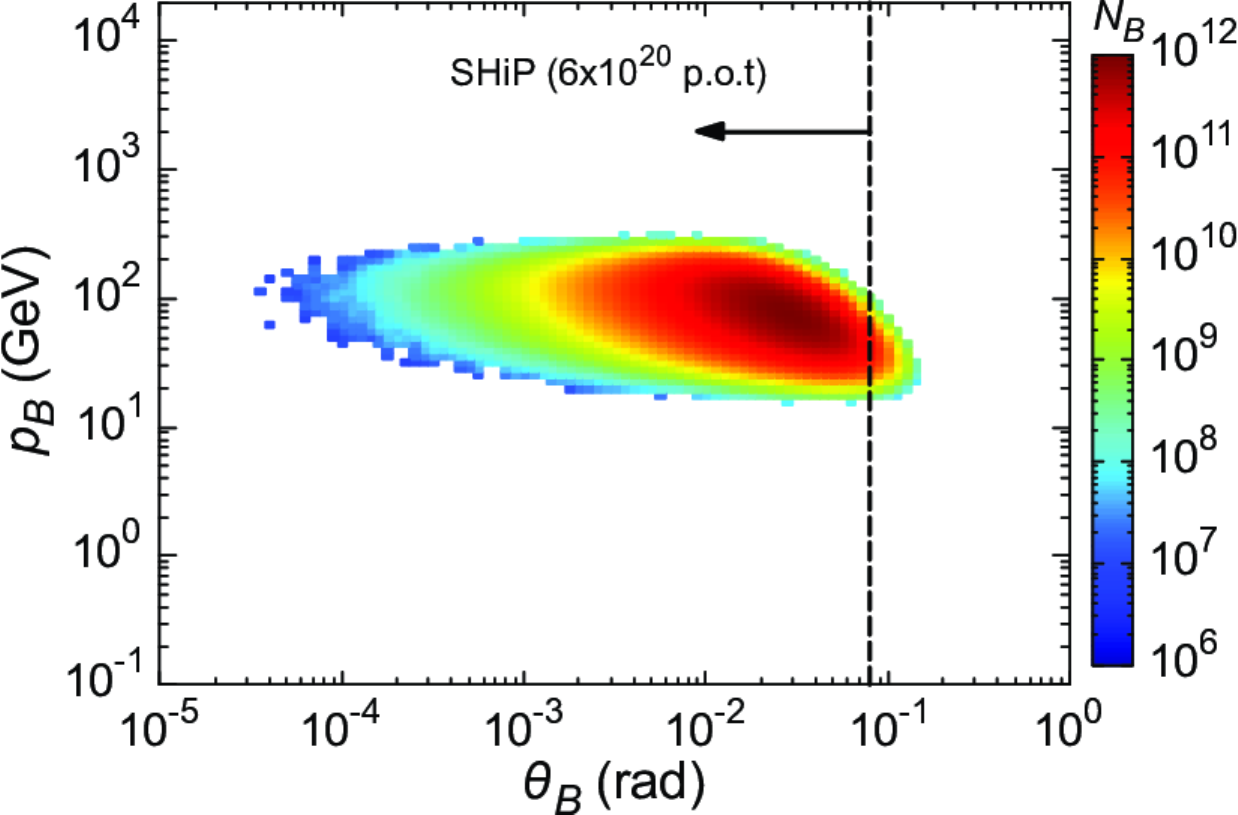} \\
    \includegraphics[scale=0.27]{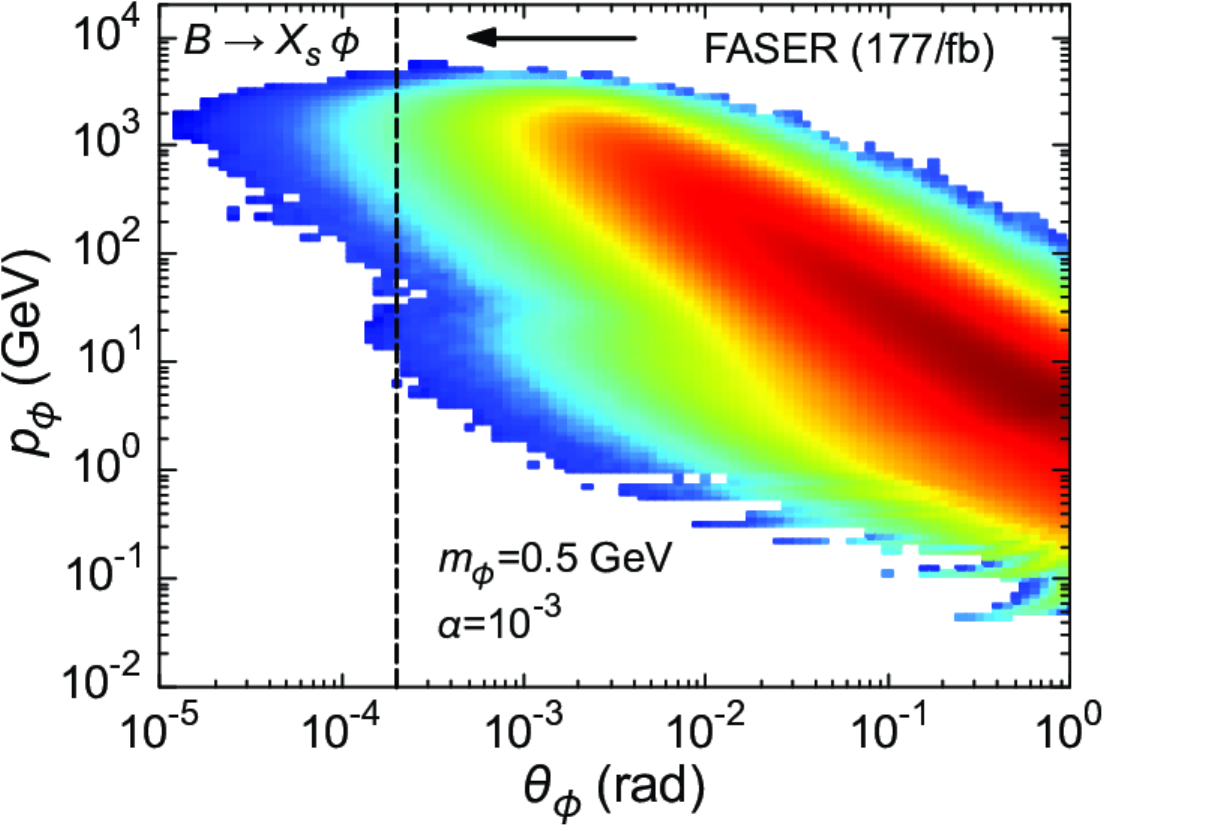} &
    \includegraphics[scale=0.27]{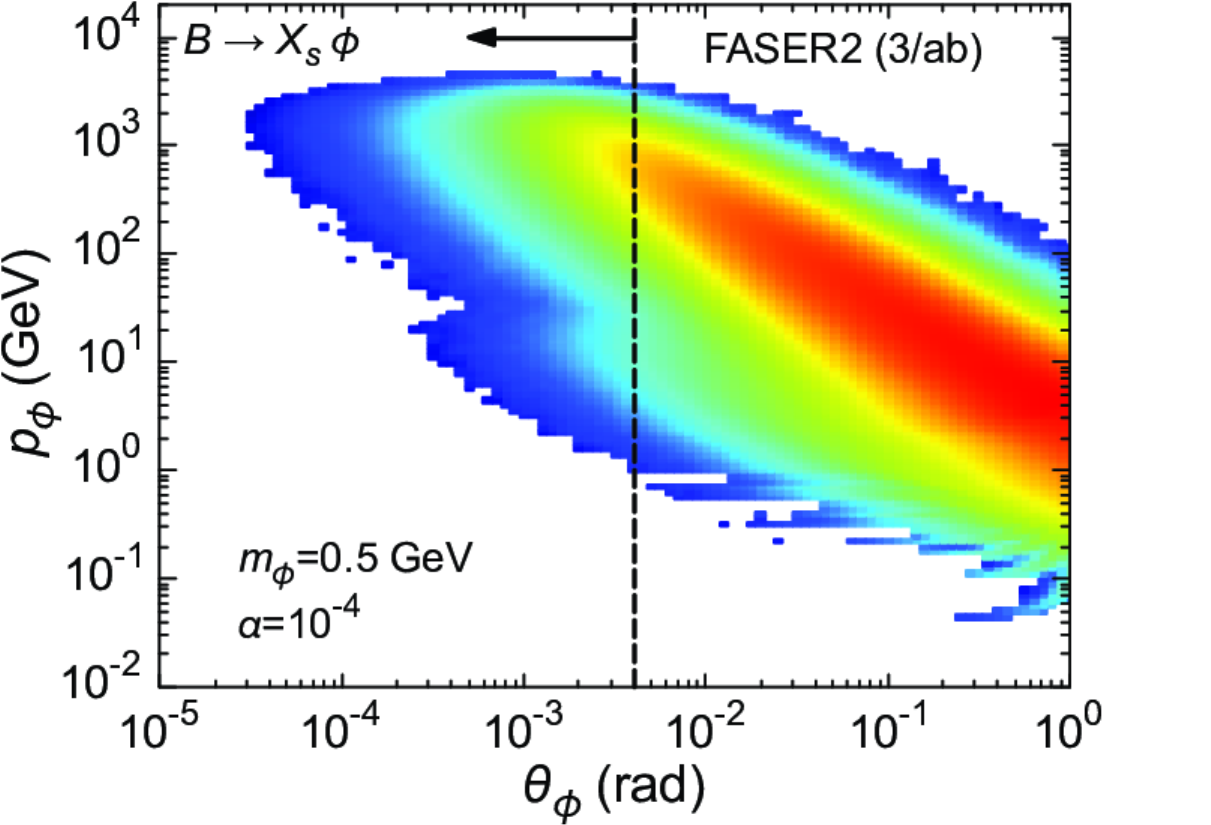} &
    \includegraphics[scale=0.27]{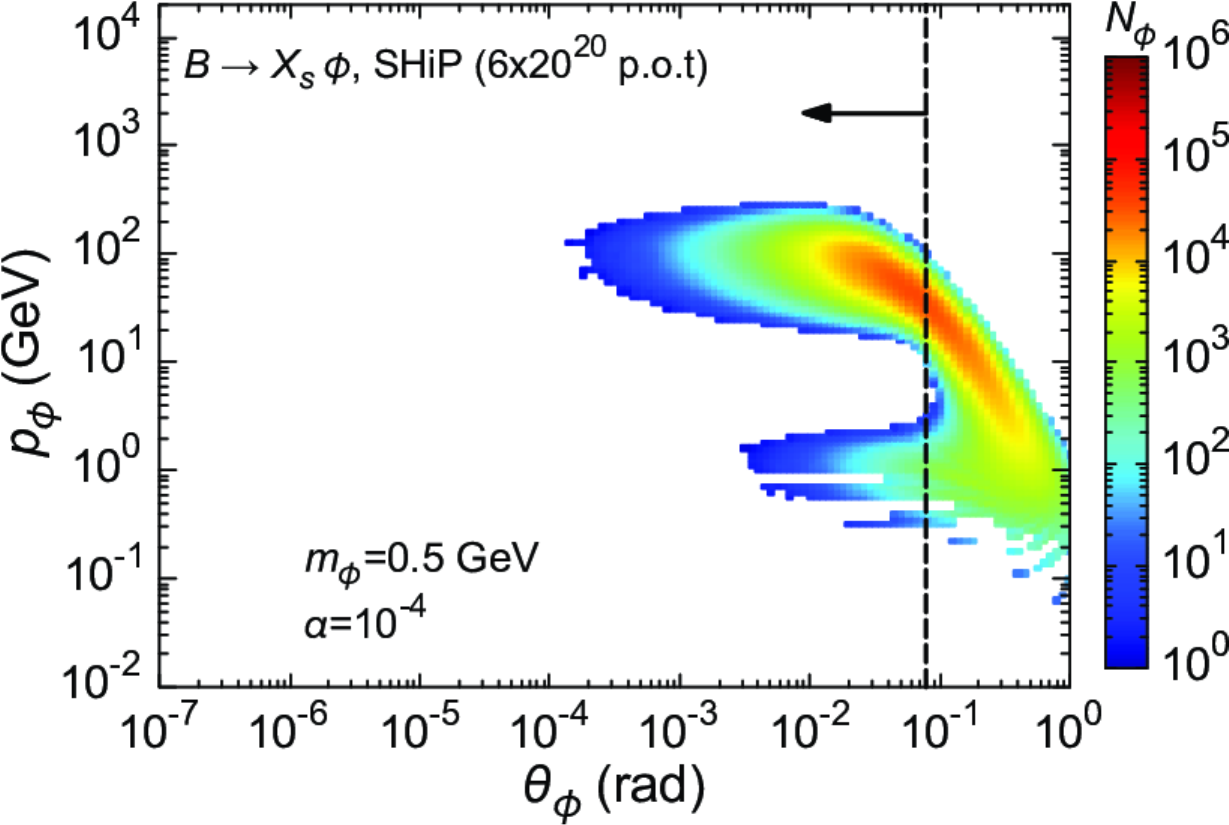} \\
    \includegraphics[scale=0.27]{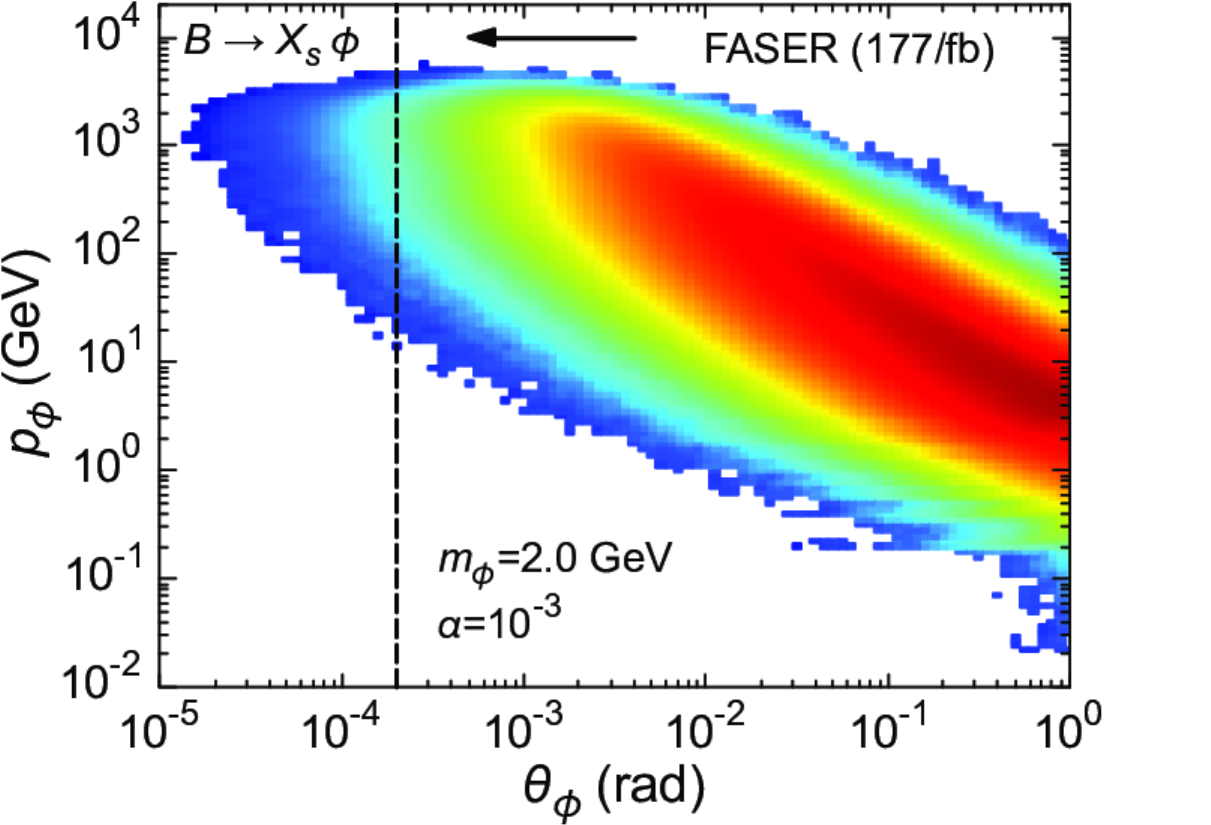} &
    \includegraphics[scale=0.27]{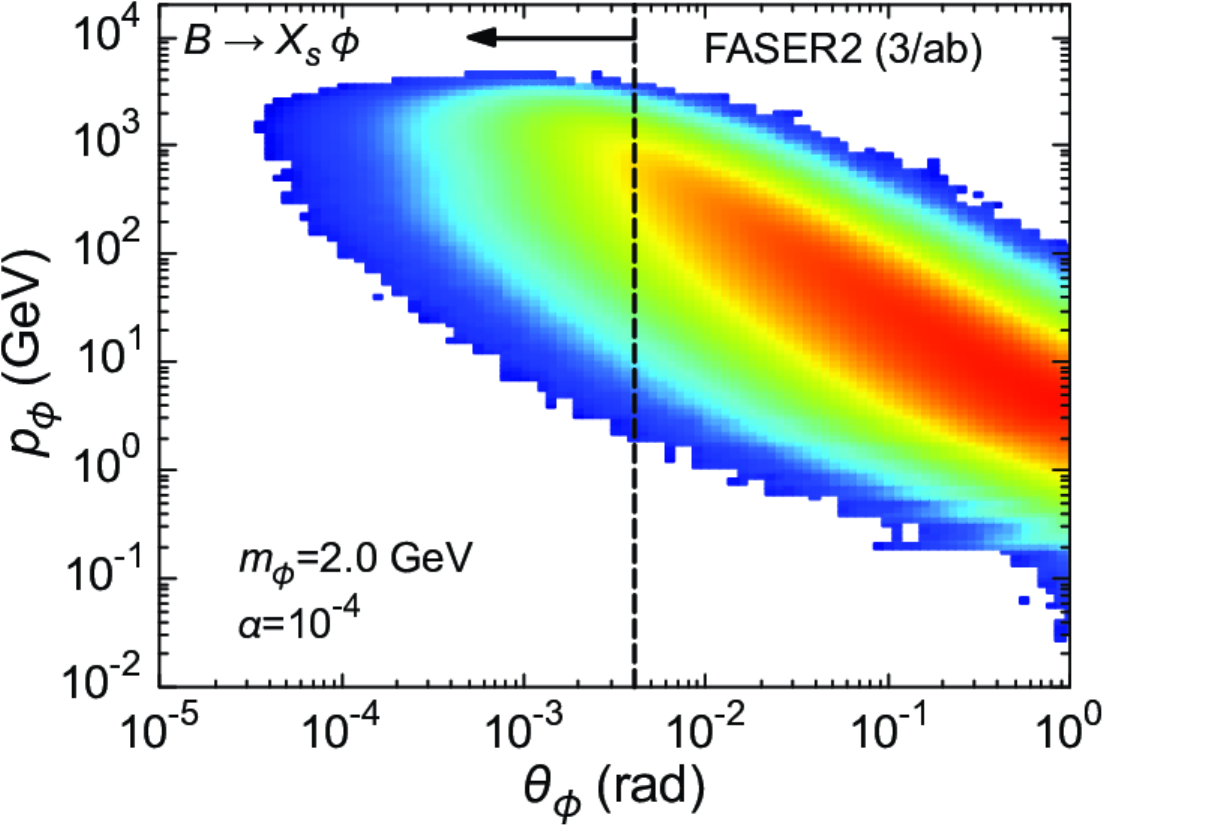} &
    \includegraphics[scale=0.27]{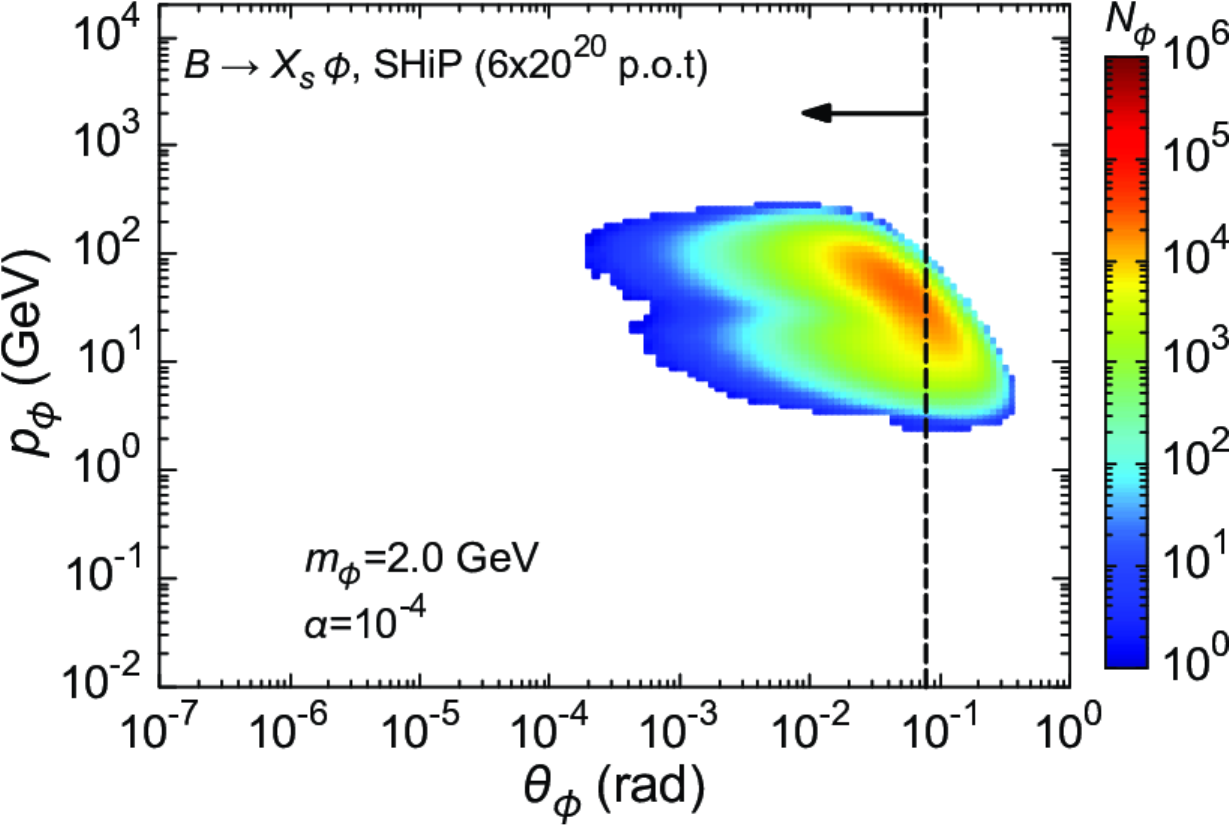} \\
\end{tabular}
    \caption{
        Upper panels: $B$ meson distribution produced at ATLAS interaction point for $177$ fb$^{-1}$ (left), $3$ ab$^{-1}$ (center), and SHiP target (right) . Vertical dashed lines represent the angle coverage of the FASER, FASER2, and SHiP detectors, respectively.
        Middle and Lower panels: Dark Higgs boson distributions at FASER (left), FASER2 (center), and SHiP (right). The dark Higgs mass is taken to $m_\phi=0.5$ GeV(middle), and $2.0$ GeV(bottom), respectively. The scalar mixing is fixed to $\alpha=10^{-3}$ for FASER, and $\alpha=10^{-4}$ for FASER2, and SHiP, respectively.
    }
\label{fig:B-dark-hippgs-dist}
\end{figure}
As supplemental material, we show the distributions of $B$ meson and dark Higgs boson produced at the ATLAS and SHiP, respectively.

Top panels of figure~\ref{fig:B-dark-hippgs-dist} show the $B$ meson distribution at ATLAS for $177$ fb$^{-1}$ (left), $3$ ab$^{-1}$ (middle) and SHiP (right). The vertical dashed line represent the angle coverage of the FASER, FASER2, and SHiP detectors. Using these distributions, the distributions of the dark Higgs boson with $m_\phi = 0.5$ and $2.0$ GeV are calculated, which are shown in the middle and bottom panels. The distributions for $m_\phi = 1.2$ GeV are similar to those with $m_\phi = 2.0$ GeV.

\bibliographystyle{apsrev}
\bibliography{ref}

\end{document}